%% file: spectral_density_arxiv_v3.tex
\documentclass[11pt]{article}
\pdfoutput=1

\usepackage{jcappub}
\usepackage{mathrsfs,amssymb,amsmath,verbatim,mathtools,graphicx,bm,booktabs,tabularx,multirow,relsize,framed,orcidlink}
\usepackage{xcolor}
\usepackage[T1]{fontenc}
\usepackage[utf8]{inputenc}

\usepackage[noabbrev]{cleveref} 
\crefname{section}{section}{sections}
\Crefname{section}{Section}{Sections}
\crefname{appendix}{appendix}{appendices}
\Crefname{appendix}{Appendix}{Appendices}
\crefname{footnote}{footnote}{footnotes}
\Crefname{footnote}{Footnote}{Footnotes}

\numberwithin{equation}{section}

\input{mydefs}

\newcommand{\unige}{D\'epartement de Physique Th\'eorique, Universit\'e de Gen\`eve, 24 quai Ernest Ansermet, 1211 Gen\`eve 4, Switzerland}
\newcommand{\gwsc}{Gravitational Wave Science Center (GWSC), Universit\'e de Gen\`eve, CH-1211 Geneva, Switzerland}
\newcommand{\mrs}{Aix-Marseille Universit\'e, Universit\'e de Toulon, CNRS, CPT, Marseille, France}

\graphicspath{figures/} 

\begin{document}

\title{The spectral density of astrophysical stochastic backgrounds}

\author[1,2]{Enis Belgacem\,\orcidlink{0000-0003-4920-0911},}
\author[1,2]{Francesco Iacovelli\,\orcidlink{0000-0002-4875-5862},}
\author[1,2]{Michele Maggiore\,\orcidlink{0000-0001-7348-047X},}
\author[3]{Michele Mancarella\,\orcidlink{0000-0002-0675-508X},}
\author[1,2]{Niccol\`o Muttoni\,\orcidlink{0000-0002-4214-2344}\,}

\affiliation[1]{\unige}
\affiliation[2]{\gwsc}
\affiliation[3]{\mrs}

\emailAdd{Enis.Belgacem@unige.ch}
\emailAdd{Francesco.Iacovelli@unige.ch}
\emailAdd{Michele.Maggiore@unige.ch}
\emailAdd{mancarella@cpt.univ-mrs.fr}
\emailAdd{Niccolo.Muttoni@unige.ch}

\abstract{
We provide a detailed derivation of the spectral density of the stochastic background generated by the superposition of coalescing compact binaries. We show how the expression often used in the literature emerges from an average over the extrinsic parameters of the binaries (times of arrival, polarization angles,  orbit inclinations and arrival directions) and how the Stokes parameters related to circular and linear polarization are set to zero by such averaging procedure.  We then consider the effect of shot noise, i.e. the fact that for the superposition of a finite number of sources
these averages are only approximate, and we show how it  generates circular and  linear polarizations (even for isotropic backgrounds) as well as spatial anisotropies, and we compute them explicitly for a realistic population of binary black holes and binary neutron stars.
}

\maketitle
\flushbottom

\section{Introduction}\label{intro}

Stochastic backgrounds  of gravitational waves (GWs) are among the main targets of present and future GW experiments. Such backgrounds can have an astrophysical origin, or being generated by cosmological processes in the early Universe (see refs.~\cite{Maggiore:1999vm,Maggiore:2007ulw,Regimbau:2011rp,Romano:2016dpx,Maggiore:2018sht,Caprini:2018mtu,Christensen:2018iqi,Lawrence:2023buo} for reviews).
Pulsar timing arrays have already provided evidence for a stochastic GW background which  has a possible  explanations as  due to the superposition of signals from supermassive black hole binaries~\cite{NANOGrav:2023gor,EPTA:2023fyk,Reardon:2023gzh,Xu:2023wog}.
The astrophysical background due to the superposition of  compact binaries  coalescences (CBCs) might already be detectable in the most advanced stage of current (second generation) ground-based GW detectors~\cite{LIGOScientific:2019vic,Perigois:2020ymr,Regimbau:2022mdu}. 
At third-generation (3G) detectors, such as Einstein Telescope (ET)~\cite{Hild:2008ng,Punturo:2010zz,Hild:2010id,Maggiore:2019uih} and  Cosmic Explorer (CE)~\cite{Reitze:2019iox,Evans:2021gyd,Evans:2023euw,Gupta:2023lga}, the astrophysical stochastic background from CBCs will be detectable with a very high signal-to-noise ratio~\cite{Regimbau:2022mdu,Branchesi:2023mws,Iacovelli:2024mjy}.
Beside carrying important astrophysical information by itself, the astrophysical GW  background  can be a foreground to stochastic backgrounds of cosmological origin, and also from this point of view is therefore important to characterize it accurately, in view of subtracting it~\cite{Cutler:2005qq,Harms:2008xv,Regimbau:2016ike,Pan:2019uyn,Sachdev:2020bkk,Sharma:2020btq,Lewicki:2021kmu,Perigois:2021ovr,Zhou:2022otw,Zhou:2022nmt,Zhong:2022ylh,Pan:2023naq,Zhong:2024dss,Li:2024iua,Belgacem:2024ntv}. 

A first  aim of this paper is to put on a firmer theoretical ground the derivation of a basic quantity used in the description of the astrophysical  background from CBCs, namely its spectral density, presenting an accurate derivation of its expression. We will see in particular how it emerges from an average over the times of arrival, polarization angles,  orbit inclinations and arrival directions  of an ensemble of CBCs and how, in the process,  the Stokes parameters related to circular and linear polarization are  averaged to zero. We will then use our formalism to  take into account the effect of shot noise, i.e. of the fact that any given realization of the astrophysical background, with a finite number of events, only corresponds to an imperfect average over these parameters, so  it leaves a residual in quantities that otherwise would be averaged to zero~\cite{Jenkins:2019uzp,ValbusaDallArmi:2023ydl}.  

The paper is organized as follows. In \cref{sect:StatChar} we  recall standard definitions on discrete stochastic backgrounds and we  define the averages over extrinsic and intrinsic parameters of an ensemble of CBCs. In \cref{sect:compSastro} we  compute  the spectral density of the astrophysical background, by performing explicitly the averages over the extrinsic parameters.
In \cref{sect:energy} we  compute the energy density of a generic (anisotropic and polarized) stochastic background, showing that there is no energy density associated to the Stokes parameters describing linear and circular polarizations (a fact that sometimes gives rise to confusion in the literature).
In \cref{sect:shot} we use our formalism to describe the generation of  polarization and anisotropies from shot noise. We will see, in particular, that shot noise can generate  linear polarization  even for an isotropic distribution  of sources (contrary to statements in the literature). We will then estimate numerically these shot-noise effects on a realistic population of BBHs and BNSs. \Cref{sect:concl} contains our conclusions. Some technical material is presented in appedices~\ref{app:extractpsi}-\ref{app:finiteT}.

\section{Statistical characterization of discrete stochastic backgrounds}\label{sect:StatChar}

In this section we begin by introducing standard definitions for the astrophysical stochastic background, in order to fix the notation, and we will then discuss the averages that are involved in its characterization, distinguishing between the averages over extrinsic and intrinsic parameters.\footnote{This section has some overlap with the corresponding introductory section of the companion paper~\cite{Belgacem:2024ntv}, where we use this formalism to study the subtraction of the astrophysical background in a two-detector correlation.}

\subsection{Basic definitions}\label{sect:basic}

A superposition of GWs coming from all directions, with propagation directions labeled by a unit vector $\hatn$,  can be written as 
\be\label{snrhab}
h_{ab}(t,\vx)
=\int_{-\infty}^{\infty} df \int_{S^2} d^2\hatn\,\sum_{A=+,\times}  \hti_A(f,\hatn) e^A_{ab}(\hatn) \, \,   e^{-2\pi i f(t-\hatn\cdot\vx /c )}
\, ,
\ee
where $a,b=1,2,3$ are spatial indices, $A=+,\times$ labels the two polarizations, and $e_{ab}^{A}$ are the polarization tensors, normalized as 
\be\label{normaee}
e_{ab}^{A}(\hatn)e_{ab}^{A'}(\hatn)=2\delta^{AA'}\, ,
\ee
where the sum over the repeated spatial indices is understood
(we follow the notation and conventions in ref.~\cite{Maggiore:2007ulw}). 
A stochastic GW background can only be characterized statistically, through an ensemble average. Conceptually,
this ensemble average is an average over many different realizations of the Universe. Of course, in practice we only have a single realization of the Universe, and what we observe in a given observation may or may not be well represented by the ideal mathematical operation of averaging over Universe realizations. In particular,
cosmological backgrounds (such as, for instance, the background produced by  the amplification of vacuum fluctuations in the early Universe), can be considered
as due to the superposition of an effectively infinite number of signals, so they are fully characterized by such ensemble averages, through the corresponding correlators (or just the two-point correlator, for a Gaussian background). 
In contrast,  the superposition of astrophysical signals will retain some  dependence on the specific realization. For instance, a stochastic background due to the superposition of $10^4$ BBHs (that correspond to the BBH coalescences taking place in the Universe in about 1 month, given our current knowledge of their rate~\cite{KAGRA:2021duu}) will retain some dependence on the specific sample of sources that coalesced during that period; a subsequent data stretch of one more month will effectively correspond to a different realization of the same discrete stochastic process. In the following, we will use the notation $\langle \ldots \rangle_U$
for the average over different realizations of the Universe. In \cref{sect:def_averages_ext_int} we will  discuss how to perform the ``Universe'' averages for an ensemble of CBCs, separating it into averages over extrinsic and intrinsic parameters.

If a stochastic background of GWs is stationary,  isotropic and unpolarized, its two-point correlator  can be written as
\be\label{ave}
\langle \tilde{h}_A^*(f,\hatn)
\tilde{h}_{A'}(f',\hatn')\rangle_U =
\delta (f-f')\, \frac{\delta(\hatn,\hatn')}{4\pi}\,
 \delta_{AA'}\, \frac{1}{2}S_h(f),
\ee
where  $\delta(\hatn,\hatn')$ is a Dirac delta over the two-sphere,  
\be
\delta(\hatn,\hatn')=\delta (\phi -\phi ') \delta (\cos\theta -\cos\theta ')\, ,
\ee
and  $(\theta ,\phi)$ are the polar angles that define $\hatn$. The function $S_h(f)$, defined by this expression,  is the spectral density of the stochastic background. 

Actually, \eq{ave} assumes an infinite observation time. Any realistic observation will only last for a finite observation time $T$. In this case, the assumption of stationarity is to some extent violated by the fact that we have an explicit reference to a given   time segment of finite duration $T$, and is recovered only for $fT\gg 1$.  In section~\ref{sect:compSastro} we will compute explicitly the result for $fT$ generic, and we will see how \eq{ave} is modified.
For ground-based detectors, even of third generation, that operates at $f\, >\, (5-10)$~Hz, the condition $fT\gg 1$ is  well satisfied already for stretches of data of about $60$~s (as typically used to estimate the PSD of a detector). However, the expression for $fT$ (and $f'T$) generic can be particularly useful for detectors working at lower frequencies. At 
a  frequency $f\simeq 10^{-3}$~Hz typical of LISA, the condition $fT\gg 1$ is  satisfied only for an observation time $T\gg 10^3$~s;   pulsar timing arrays (PTA), operating in the nHz region,  are sensitive by construction to GWs with frequencies of the order of the inverse of the observation time  so, after observing for a time $T$,   they access frequencies down   precisely to $1/T$, and therefore at the lower edge of the bandwidth accessible to PTA, the condition
$fT\gg 1$ by definition is not satisfied (e.g., at a frequency $f=1$~nHz the condition $fT\gg 1$ becomes $T\gg 30~{\rm yr}$). So, in particular for PTA, it is important to write the result in a form which is valid in general, and not only in the limit $fT\gg 1$.

The introduction of a finite observation time $T$ introduces deviations from stationarity which are controlled by the parameter $fT$ and become smaller and smaller as $fT$ grows, so in particular can be negligible at ground-based detectors, for realistic values of $T$. This is due to the fact that 
the statistical properties of a population of astrophysical sources only change on cosmological timescales (say, millions of years), which are huge compared to the observation time $T$, and this is even more true for a background of cosmological origin; so, the underlying process
is statistically invariant under time translations (over a timescale of the order of the observation time, and in fact even much bigger than it) and the only deviation from a distribution of events uniform in time will be due to the specific  realization in a finite stretch of data. In contrast, a stochastic GW background can be intrinsically anisotropic, or polarized. In this case, the most general form of the two-point correlator (assuming again for the moment stationarity) is then given by~\cite{Seto:2008sr} (see also refs.~\cite{Romano:2016dpx,Renzini:2020rjw,Renzini:2022alw} for reviews) 
\be\label{Stokes1}
\langle\tilde{h}_A^*(f,\hat{\bf{n}})\tilde{h}_{A'}(f',\hat{\bf{n}}')\rangle_{U}=\delta(f-f')\,\frac{\delta(\hat{\bf{n}},\hat{\bf{n}}')}{4\pi}
\frac12 H_{A A'}(f,\hat{\bf{n}})\, .
\ee
Once again, we will see explicitly how this expression   is modified  for $fT$ generic, which  is important in particular for PTA.

Note that in \eq{Stokes1} we still assume that signals coming from different directions are not correlated, i.e. the correlator is still proportional to $\delta(\hat{\bf{n}},\hat{\bf{n}}')$; however, the possibility of an anisotropy is encoded in the fact that the function $H_{A A'}(f,\hat{\bf{n}})$ could depend on $\hatn$; the possibility of a polarization of the background, instead, is encoded in the fact that  $H_{A A'}(f,\hat{\bf{n}})$ is  a  $2\times 2$  matrix in the polarization indices, not necessarily proportional to $\delta_{AA'}$.
Taking the complex conjugate of \eq{Stokes1} we see that  $H_{A A'}(f,\hat{\bf{n}})$  is a Hermitian matrix, so it can be decomposed (with real coefficients) into the basis made by the identity matrix  and the Pauli matrices, 
\be\label{Stokes2}
H_{A A'}(f,\hat{\bf{n}})=I(f,\hat{\bf{n}})\,\delta_{A A'}+U(f,\hat{\bf{n}})\,\sigma^{1}_{A A'}+V(f,\hat{\bf{n}})\,\sigma^{2}_{A A'}+Q(f,\hat{\bf{n}})\,\sigma^{3}_{A A'}\, ,
\ee
where the three Pauli matrices are given by
\begin{equation}\label{Pauli}
\sigma^{1}=\,
\begin{pmatrix}
0       & 1 \\
1       & 0
\end{pmatrix}\,,\qquad 
\sigma^{2}=\,
\begin{pmatrix}
0       & -i \\
i       & 0
\end{pmatrix}\,,\qquad
\sigma^{3}=\,
\begin{pmatrix}
1       & 0 \\
0       & -1
\end{pmatrix}\, .
\end{equation}
In matrix form 
\be\label{Stokes2matrix}
H(f,\hat{\bf{n}})=
\begin{pmatrix}
I (f,\hat{\bf{n}}) +Q (f,\hat{\bf{n}})\hspace*{3mm}    & U (f,\hat{\bf{n}}) -iV (f,\hat{\bf{n}})\\
U (f,\hat{\bf{n}})  +iV (f,\hat{\bf{n}}) \hspace*{3mm}    & I (f,\hat{\bf{n}}) -Q (f,\hat{\bf{n}})\,
\end{pmatrix}.
\ee
The coefficients of this decompositions define the Stokes parameters of the GW stochastic background, and are real functions of $f$ and $\hat{\bf{n}}$ describing intensity ($I$), linear polarization ($U$ and $Q$) and circular polarization ($V$).

For  the stochastic background generated by the superposition of astrophysical sources in an observation time $T$, we can  write
\be\label{tildehdeltan}
\tilde{h}_A(f,\hatn)=\sum_{i=1}^{\nev} \tilde{h}_{A,i}(f)\, \delta(\hatn,\hatn_i)\, ,
\ee
where $\hatn_i$ is the propagation direction of the $i$-th signal, and $\nev$ is the number of events reaching the detector  in the observation time $T$. Then, \eq{snrhab} becomes
\be\label{snrhabdiscrete}
h_{ab}(t,\vx)
=\sum_{i=1}^{\nev} \int_{-\infty}^{\infty} df \,\sum_{A=+,\times}  \tilde{h}_{A,i}(f) e^A_{ab}(\hatn_i) \, \,   e^{-2\pi i f(t-\hatn_i\cdot\vx /c )}
\, .
\ee

\subsection{Averages over extrinsic and intrinsic parameters}\label{sect:def_averages_ext_int}

We now discuss how to perform the average $\langle\ldots \rangle_U$ for an astrophysical background. We will  use a language appropriate to CBCs, although some aspects of the discussion below  hold for more general astrophysical backgrounds, e.g. due to supernovae. An  essential step is to separate the parameters of the waveform  of a CBC into two groups,  extrinsic and intrinsic parameters, and perform first the average over the extrinsic parameters.

In this work, we define  the extrinsic parameters of a CBC (labeled by an index $i$), to be: its arrival time $t_i$  (defined, e.g., as the time of arrival of the peak of the waveform, or the time of entry of the signal in the detector bandwidth); the polarization angle $\psi_i$; the propagation direction  $+\hatn_i$  of the GW (so that
the direction of the source in the sky is $\hat{\vO}_i=-\hatn_i$); and the inclination of the orbit, defined by the angle $\iota_i$ between the normal to the orbit and the line-of-sight (we will actually rather use $\cos\iota_i$).\footnote{
In the case of precessing spins the orbital plane is not fixed, and the notion of $\iota$ becomes time-dependent. Anyway, all computations below will remain valid replacing $\iota$ by $\theta_{JN}$, defined as the angle between the line-of-sight and the total angular momentum.}
So, for a set of $\nev$  events, the extrinsic parameters are
\be
\{ t_i, \psi_i, \hatn_i, \cos\iota_i\}_{i=1,\ldots ,\nev}\, .
\ee
The intrinsic parameters are all the other parameters of the binary, such as the component masses, spins, luminosity distance, orbit eccentricity, tidal deformability (for neutron stars), etc. 

A first rationale for this distinction is that events characterized by different  extrinsic parameters (but the same intrinsic parameters) can be intuitively seen as different realizations of the same event; e.g.,  signals all corresponding to a BBH  with given masses, spin, etc., but with different arrival times (within the time span of the observation), or arriving from different directions in the sky, can be considered as different realizations of the same event. In contrast, once we start sampling e.g. the mass distribution, there is
no longer much sense in which we could say that an ensemble of signals
always refers to the same BBH, just extracted with different component masses.

More technically, the averages over extrinsic parameters are made with a flat measure, that reflects some underlying symmetry principle. For instance, the average over the time of arrival $t_i$ of a generic quantity (denoted by the dots)  is just performed with a flat measure 
\be\label{ave_ti}
\langle\ldots\rangle_{t_i}= \frac{1}{T}  \int_{-T/2}^{T/2} dt_i\, (\ldots)\, ,
\ee
(where $T$ is the observation time), reflecting the invariance  under time translations of the probability of having a BBH or BNS merger  during the observation time $T$, at least on the timescale $T$ of the observation, which is typically at most months or years, and therefore is extremely small compared to the timescales over which the populations of BBHs and BNSs evolve.\footnote{If the  process were not stationary on the timescale $T$, we would rather have
$\int dt_i\, \mu(t_i)$ for some measure $\mu(t_i)$. This  happens on cosmological timescales, in which case the merger distribution has an important dependence on redshift.} Similarly, the polarization angle parameterizes the freedom of the observer to choose the axes, in the  plane transverse  to the propagation direction of the $i$-th signal, with respect to which the plus and cross polarizations of the $i$-th signal are defined, so the
average over the polarization angle $\psi_i$ is naturally performed with the flat measure 
\be\label{ave_psii}
\langle\ldots\rangle_{\psi_i}=\int_0^{2\pi}\, \frac{d\psi_i}{2\pi}\, (\ldots)\, .
\ee
The inclination angle of the orbit with respect to the line-of-sight  must also be distributed uniformly on the sphere, since the binary ``knows nothing'' about the location of the observer, and there is no reason why it should have preferential orientations  with respect to the observer's line of sight,\footnote{Of course, here we are referring to the intrinsic distribution of the source parameters. The detection probability depends on $\cos\iota$, and in particular  the amplitude of the waveform in the inspiral phase is maximized for $\cos^2\iota=1$, i.e. for face-on/face-off systems.}
so it is averaged as
\be\label{ave_cosi}
\langle\ldots\rangle_{\cos\iota_i}=\frac{1}{2}\int_{-1}^{1}d\cos\iota_i\, (\ldots)\, .
\ee
In many situations it is also natural to assume that, at least in a first approximation, extragalactic sources such as BBHs and BNSs are  uniformly distributed in the sky, in which case the average over this extrinsic parameter is performed as  
\be\label{ave_ni}
\langle\ldots\rangle_{\hatn_i}=\int_{S^2} \frac{d^2\hatn_i}{4\pi}\, (\ldots)\, ,
\ee
where $S^2$ denotes the 2-sphere. However, the appropriateness of this assumption really depends on the context; in particular, if one is interested in the anisotropies of the astrophysical background and in its correlation with the galaxy field~\cite{Contaldi:2016koz,Cusin:2017fwz,Cusin:2017mjm,Jenkins:2018lvb,Cusin:2018rsq,Jenkins:2018uac,Jenkins:2018kxc,Jenkins:2019uzp,Cusin:2019jhg,Cusin:2019jpv,Jenkins:2019nks,Bertacca:2019fnt,Pitrou:2019rjz,Bellomo:2021mer,Capurri:2021zli}, this average would select the ``monopole'' term of the distribution, while in this case one would be interested exactly in the small deviations from the uniform distributions, described by  the higher multipoles; so, more generally, as long as we assume that the integrations over arrival directions of different signals factorize, for an anisotropic background we will write
\be\label{aved2n_mu}
\langle \ldots \rangle_{\hatn_i}=
\int_{S^2}\frac{d^2\hatn_i}{4\pi}\, \mu(\hatn_i) \, (\ldots )\, ,
\ee
for some function $\mu(\hatn_i)$, normalized as
\be\label{normmu}
\int_{S^2}\frac{d^2\hatn}{4\pi}\, \mu(\hatn)=1\, ,
\ee
and that could reflect for instance a prior given by a galaxy catalog. Observe that, in general, the function $\mu(\hatn)$ could also depend  on some intrinsic parameters of the CBCs; for instance, the anisotropy of the CBC distribution could be correlated with the luminosity distance, or with the (source-frame) masses. It could also depend on further indices  labeling the type of CBC populations, reflecting the evolutionary channel. E.g.,  if anisotropies are present, one could expect a difference between, say, the anisotropies of CBCs that formed from  primordial black holes,  those that formed  from  population~III stars, and those originating in more recent epochs by ``field'' binaries. In the following, we are interested in  performing explicitly the integration over the extrinsic parameters, and all the steps below will still be valid even if $\mu(\hatn)$ should have a dependence on intrinsic parameters, or labels such as those identifying the formation channel, which we will not write explicitly. 

Physically, 
it is natural to assume that the integration over the extrinsic parameters  are independent. At the mathematical level, however, one must be careful about the order of the integration of $\hatn_i$,  $\psi_i$ and $\cos\iota_i$. Indeed, the angles $\iota_i$ and $\psi_i$ are defined only after the direction $\hatn_i$ has been specified. Given that for us $\hatn_i$ is the propagation direction of the $i$-th signal, the position of the $i$-th source in the sky is given by the unit  vector $\hat{\vO}_i= -\hatn_i$. The angles $\iota_i$ and $\psi_i$ define the orientation of the plane of the orbit of the binary, with respect to this direction:  if ${\bf \hat{L}}_i$ is the unit vector orthogonal to the plane of the orbit (i.e. the unit vector in the direction of the orbital angular momentum), 
$\iota_i$ and $\psi_i$ are the two polar angles that identify the direction of ${\bf \hat{L}}_i$, in a frame where the polar axis is given by $\hat{\vO}_i$. In particular,    $\cos\iota_i={\bf \hat{L}}_i\bdot \hat{\vO}_i$ while 
$\psi_i$ is the other polar angle, whose definition is completed specifying the orientation, in the plane orthogonal to $\hatn_i$, of two unit vectors $\hatu_i,\hatv_i$ orthogonal among them, such that
the  system of axis  $\{\hatu_i,\hatv_i,\hatn_i\}$ forms  an orthogonal right-handed oriented frame and $\psi_i$ is measured, e.g., counterclockwise from the $\hatu_i$ axis.\footnote{We follow the common convention (as in \cite{{Maggiore:2007ulw}}) that the frame with respect to which polarization is defined is right-handed with respect to the GW propagation direction $\hatn$; the same convention is used e.g. in \cite{Allen:2022dzg}, while \cite{Romano:2016dpx} rather define it with respect to the source direction $\hat{\vO}=-\hatn$.}
This means that the integrations over $\psi_i$ and $\cos\iota_i$  must always be performed before integrating over $\hatn_i$, since these quantities are no longer defined if we first integrate over $\hatn_i$. We therefore define 
the average over the extrinsic parameters  as
\be\label{allaverages}
\langle \ldots \rangle_{\rm ext} =\prod_{i=1}^{\nev}
\int_{-T/2}^{T/2} \frac{dt_i}{T}
\int_{S^2}\frac{d^2\hatn_i}{4\pi}\, \mu(\hatn_i) \,
\int_{-1}^{1}\frac{d\cos\iota_i}{2}
\int_0^{2\pi}\, \frac{d\psi_i}{2\pi}\, 
(\ldots)\, ,
\ee
where the integrations over $\psi_i$ and $\cos\iota_i$ are performed before the integration over $\hatn_i$.  Observe that, in order to define unambiguously the polarization angles of an ensemble of sources,  we must set up a global reference frame $\{\hatu,\hatv,\hatn\}$ on the sphere. 
Given a generic propagation direction $\hatn$, with coordinates
\be\label{n_Om}
\hatn=\hatx\sin\theta\cos\phi  + \haty\sin\theta\sin\phi  +\hatz \cos\theta \, ,
\ee 
one defines
\bees
\hatu(\hatn)&=&\hatx\cos\theta\cos\phi+\haty\cos\theta\sin\phi- \hatz\sin\theta \equiv \hat{\vtheta}\, ,\label{u_Om}\\
\hatv(\hatn)&=&-\hatx\sin\phi+\haty\cos\phi \equiv \hat{\vphi} \, .\label{v_Om}
\ees
It is worth noticing that \eqs{u_Om}{v_Om} can be equivalently written just in terms of $\hatn$ and the polar axis $\hatz$ chosen, as\footnote{Checking the equivalence is straightforward. As a quick strategy to obtain these expressions, one first observes that $\hatv(\hatn)$ in eq.~(\ref{v_Om}) is orthogonal to $\hatz$. Since $\hatv(\hatn)$ is also orthogonal to $\hatn$, it must be proportional to the cross product $\hatz\times\hatn$. By examining the orientation and using the identity $||\hatz\times\hatn||=[1-(\hatz\cdot\hatn)^2]^{1/2}$ for the normalization, one gets eq.~(\ref{v_Om_eq}). Finally eq.~(\ref{u_Om_eq}) follows from $\hatu(\hatn)=\hatv(\hatn)\times\hatn$ and the identity $(\hatz\times\hatn)\times\hatn=(\hatz\cdot\hatn)\hatn-\hatz$.}
\bees
\hatu(\hatn)&=&\frac{(\hatz\cdot\hatn)\hatn-\hatz}{[1-(\hatz\cdot\hatn)^2]^{1/2}}\,,\label{u_Om_eq}\\
\hatv(\hatn)&=&\frac{\hatz\times\hatn}{[1-(\hatz\cdot\hatn)^2]^{1/2}}\,\label{v_Om_eq}.
\ees
The fact that \eqs{u_Om_eq}{v_Om_eq} are undetermined at the two poles of the sphere (where $\hatn=\pm \hatz$) reflects the fact that the unit vectors $\hat{\vtheta}$, $\hat{\vphi}$ in \eqs{u_Om}{v_Om} are not unique at the two poles (which have $\theta=0$ and $\theta=\pi$, respectively, but $\phi$ undetermined).

With respect to these vectors, the plus and cross polarization tensors are defined globally on the sphere as
\bees
e^{+}_{ab}(\hatn)&=&\hatu_a(\hatn)\hatu_b(\hatn)-\hatv_a(\hatn)\hatv_b(\hatn)\, ,\label{eplusuv}\\
e^{\times}_{ab}(\hatn)&=&\hatu_a(\hatn)\hatv_b(\hatn)+\hatv_a(\hatn)\hatu_b(\hatn)\, .\label{ecrossuv}
\ees
Observe that this constructions leaves an overall  freedom, which consists in replacing
$\hatu(\hatn)$ and $\hatv(\hatn)$ by a rotated combination
\bees
\hatu'(\hatn)=\hatu(\hatn)\cos[\psi_0(\hatn)]-\hatv(\hatn)\sin[\psi_0(\hatn)]\, ,\label{hatup}\\
\hatv'(\hatn)=\hatu(\hatn)\sin[\psi_0(\hatn)]+\hatv(\hatn)\cos[\psi_0(\hatn)]\, ,\label{hatvp}
\ees
where  the  angle $\psi_0$ can  a priori be an arbitrary function of $\hatn$. As we will see in section~\ref{sect:deppsi}, the Stokes parameters describing the linear polarization of a discrete set of sources depend on the choice of the global reference frame $\{\hatu,\hatv,\hatn\}$ on the sphere (while circular polarization is independent of this choice). There is nothing wrong with it. Also the polarizations $h_{+}$ and $h_{\times}$ of a single GW are physical, observable quantities, despite the fact that they depend on the choice of the 
$\hatu,\hatv$ axes perpendicular to its propagation direction, and  in particular they transform  as a helicity-2 field when this frame is rotated in the transverse plane. Similarly, we will see that the Stokes parameters describing  the linear polarization of an ensemble of GW signals depend  on the choice of the global reference frame $\{\hatu,\hatv,\hatn\}$ on the sphere. For definiteness, one can use the choice defined by \eqst{n_Om}{v_Om}. The values of the linear polarizations with respect to any other choice of global frame, i.e. any non-vanishing function $\psi_0(\hatn)$ in \eqs{hatup}{hatvp}, can be obtained transforming  these Stokes parameters appropriately (as we will see, as helicity-4 fields).

As we will see below, the average over these extrinsic parameters allow us to extract, from the correlator  $\langle \tilde{h}_A^*(f,\hatn)
\tilde{h}_{A'}(f',\hatn')\rangle_U$, Dirac delta functions or Kronecker's deltas, or other simple mathematical structures.
In contrast, intrinsic parameters, such as the masses and spins of the component stars, the distance to the binary, etc., must be sampled according to non-trivial distributions that reflect specific astrophysical properties of the population, rather than uniform distributions reflecting  symmetry principles.\footnote{In general, in the literature, the distinction between extrinsic and intrinsic parameters is  rather performed stating that the extrinsic parameters are those  also have a dependence on the observer, while the intrinsic do not. For instance, the arrival time and the arrival direction depend on the observer's position on Earth, the polarization angle on the axes  in the transverse plane chosen by the observer to define the polarization, and $\cos\iota$ depends on the line of sight to the observer; in contrast, e.g. the (source frame) masses or the spins of the component stars are independent of the observer. 
We find that the distinction based on whether a parameter has a flat integration measure determined by symmetry principles, or a non-trivial integration measure due to astrophysical effects, 
is more pertinent. Indeed, the computation below will make it clear that 
the real reason to separate the averages into two groups is that the averages over extrinsic parameters produce Dirac or Kronecker delta's, or other simple structures.
Note   that, with our definition,   also the luminosity distance $d_L$ of the binary is included among the intrinsic parameters (despite the fact that, in principle, depends on the observer's location), because
any   change  in the luminosity distance of the source that produces observable effects 
corresponds to placing the event  at an appreciably different  redshift, and the   probability of having a coalescence at a given redshift depends on non-trivial astrophysics, and not just on symmetries.
Within the same logic, not only the source-frame masses are obviously included among the intrinsic parameters, but also the detector-frame masses, despite the factor $(1+z)$ that connects them to the source-frame masses. Another way to appreciate the difference between time of arrival (which we include among the extrinsic parameters) and  luminosity distance  $d_L$ (which instead we treat as an intrinsic parameter) is that the former  enters  in the phase of the waveform. Since phases are defined only modulo $2\pi$, an absolute change $\Delta t_i$ matters as long as $f\Delta t_i$ is non-negligible, even if $\Delta t_i$ is of course extremely small compared to the travel time of the signal from the source to the observer. In contrast, $d_L$ enters only in the amplitude, and changes in $d_L$ are appreciable only if $\Delta d_L/d_L$ is non-negligible. For instance, changing the observer's position on Earth by, say, 1000~km, produces a $\Delta t_i$ whose effect in the phase is significant, but the corresponding change $\Delta d_L$ only affects the waveform  through  $\Delta d_L/d_L$, which is utterly negligible, and in this sense $d_L$ does not depend on the observer's position, while $t_i$ does. 
Concerning the sky position parameters,
if anisotropy effects, depending on complicated astrophysics, were dominant, then, according to the distinction between extrinsic and intrinsic parameters  based on whether  the integration measure reflects underlying symmetry principles or not,  they should rather be considered among the intrinsic parameters. However, we expect that, in a first approximation, the distribution of CBCs will be isotropic, with only small corrections due to intrinsic anisotropies. In this case, it is convenient to include sky positions among the extrinsic parameters, as we do here.
\label{foot:defextint}}

Let us denote by $\langle\ldots \rangle_{\rm ext}$ the average over the extrinsic parameters, and  by $\langle\ldots \rangle_{\rm int}$ that over the intrinsic ones. These operations define what one means by  ``average over different realization of the Universe'' at the level of  the parameters characterizing an individual CBC. On top of it,  even the number $\nev$ of binaries coalescing in a given observation time $T$ is a stochastic variable, and the average over different realizations of the ``Universe'' should include also an average over a Poisson distribution for the number of events, with mean $\bnev$.\footnote{Observe that this is not equivalent to considering the uncertainty in the observed binary merger rate. Even if the merger rate were exactly determined, with no error,  still  $\nev$ would be a random quantity  which fluctuates, and takes different value in different observing runs of the same duration $T$.} Then, overall,
\be\label{aveextint}
\langle\ldots \rangle_U =  \langle\,\, \langle\ldots \rangle_{\rm ext} \,\, \rangle_{{\rm int}, \nev}\, ,
\ee
where we have denoted by $\langle \ldots  \rangle_{{\rm int}, \nev}$ the average over the intrinsic parameters as well as on $\nev$.

\section{Computation of  the spectral density of the astrophysical background}\label{sect:compSastro}

We now   compute 
the spectral density for an astrophysical stochastic background, by   evaluating explicitly the correlator $\langle\tilde{h}_A^*(f,\hatn)\tilde{h}_{A'}^*(f',\hatn')\rangle_U$. We compute first the average over the extrinsic parameters of all signals,
\be
\{  {\rm ext} \} =
\{t_k,\hatn_k,\cos\iota_k,\psi_k
\}_{k=1,\dots,\nev}\, .
\ee
Until now, when writing $\tilde{h}_A(f,\hatn)$ or $\tilde{h}_{A,i}(f)$,
we have left implicit the dependence of the GW signal on the extrinsic parameters, as well as that on the intrinsic parameters.  
If we write explicitly at least the dependence on the extrinsic parameters, then
\be\label{hAifull}
\tilde{h}_{A,i}=\tilde{h}_{A,i}(f;t_i,\cos\iota_i,\psi_i)\, ,
\ee
while $\hatn_i$ enters through the Dirac delta in \eq{tildehdeltan};
correspondingly, from \eq{tildehdeltan}, $\tilde{h}_{A}(f,\hatn)$ should in principle  be written as
\be\label{hAfull}
\tilde{h}_{A}=\tilde{h}_{A}(f,\hatn; \{t_k, \hatn_k, \cos\iota_k,  \psi_k\})\, ,
\ee
where the brackets $\{....\}$ denotes the collection of all $t_k, \hatn_k, \cos\iota_k,  \psi_k$ with $k=1,\ldots,\nev$. However, in order not to burden too much the notation, we will often either still keep implicit the dependence on the extrinsic parameters, or else we will just write in the argument those relevant to the specific steps of the computation that we are performing. In any case, it is useful to keep in mind that the full dependencies are those given in \eqs{hAifull}{hAfull}.

The averages over the extrinsic parameters are performed
while keeping  the intrinsic parameters, and the number of events $\nev$, fixed. 
We use the notation $\langle \ldots \rangle_{\{\psi_k\}}$ to denote the average over  all $\psi_k$ with  $k=1,\dots,\nev$, $\langle \ldots \rangle_{ \{\cos\iota_k\} }$ for the average over  all 
$\cos\iota_k$, $\langle \ldots \rangle_{ \{\hatn_k\} }$ for the average over  all $\hatn_k$, and
$\langle \ldots \rangle_{ \{t_k\} }$ for the average over  all $t_k$. A label such as
$\langle\ldots\rangle_{ \{t_k,\cos\iota_k,\psi_k,\} }$ means that we average over all the corresponding parameters.
We assume that the extrinsic parameters of different events are independent. Then, from \eq{tildehdeltan},
\begin{eqnarray}
\hspace*{1mm}\langle\tilde{h}_A^*(f,\hatn)\tilde{h}_{A'}(f',\hatn')
\rangle_{ \{t_k,\hatn_k,\cos\iota_k, \psi_k\} } &=& \sum_{i,j=1}^{\nev}\,
\langle\tilde{h}_{A,i}^*(f)\tilde{h}_{A',j}(f')\delta(\hatn,\hatn_i)\delta(\hatn',\hatn_j)\rangle_{ \{t_k,\hatn_k,\cos\iota_k, \psi_k\} }\,\nonumber\\
&&\hspace*{-50mm}=\sum_{i,j=1}^{\nev}\langle \[  \langle\tilde{h}_{A,i}^*(f)\tilde{h}_{A',j}(f')\rangle_{ \{t_k,\cos\iota_k, \psi_k\} }\delta(\hatn,\hatn_i)\delta(\hatn',\hatn_j)\]\rangle_{ \{\hatn_k\} }\,,\label{eq: avg ext sum i j}
\end{eqnarray}
where  we used the fact that the directions of propagation $\hatn_k$, with $k=1 \ldots ,\nev$, do not enter explicitly into $\tilde{h}_{A,i}(f)$ and $\tilde{h}_{A',j}(f)$ but, as we explained above, the average over $\hatn_k$ must be left as the external one, with respect to the averages over 
$\cos\iota_k$ and $\psi_k$.
Once again, 
in \eq{eq: avg ext sum i j} the full dependencies of $\tilde{h}_{A,i}$ and $\tilde{h}_{A}$ on the extrinsic parameters are given in \eqs{hAifull}{hAfull}, but we will usually write explicitly only the arguments on which we are averaging; so, e.g., when performing the averages over the polarization angles we write the correlator 
between $\tilde{h}_{A,i}$ and $\tilde{h}_{A',j}$ as
\be
\langle\tilde{h}^*_{A,i}(f;\psi_i)\tilde{h}_{A',j}(f';\psi_j)\rangle_{\{\psi_k\}}\, .
\ee

\subsection{Average over polarization angles}\label{averagesoverpsi}

It is convenient to start from the average over polarization angles (which, as discussed above \eq{allaverages}, must in any case be performed before the integration over the arrival directions).
Using \eq{allaverages}, 
\be\label{aveallpsi}
\langle\tilde{h}^*_{A,i}(f;\psi_i)\tilde{h}_{A',j}(f';\psi_j)\rangle_{\{\psi_k\}}=\prod_{k=1}^{\nev}\int_0^{2\pi}\frac{d\psi_k}{2\pi}\,
\tilde{h}^*_{A,i}(f;\psi_i)\tilde{h}_{A',j}(f';\psi_j)\, .
\ee
Since $\tilde{h}_{A,i}(f)$ 
only depends on $\psi_i$ and 
$\tilde{h}_{A,j}(f)$ only depends on $\psi_j$, in \eq{aveallpsi}
the integrations over the  parameters $\psi_k$ with $k\neq i, j$ play no role as they trivially reduce to unity. Thus,  if $i\neq j$,
\be\label{dpsi_i_dpsi_j}
\langle\tilde{h}^*_{A,i}(f;\psi_i)\tilde{h}_{A',j}(f';\psi_j)\rangle_{\{\psi_k\}}=
\int_0^{2\pi}\frac{d\psi_i}{2\pi}\int_0^{2\pi}\frac{d\psi_j}{2\pi} \tilde{h}^*_{A,i}(f;\psi_i)\tilde{h}_{A',j}(f';\psi_j)\, ,
\ee
while, for $i=j$,
\be\label{dpsi_i_dpsi_i}
\langle \tilde{h}_{A,i}^*(f;\psi_i)\tilde{h}_{A',i}(f';\psi_i)\rangle_{\{\psi_k\}}\nn\\
=
\int_0^{2\pi}\frac{d\psi_i}{2\pi} \tilde{h}_{A,i}^*(f;\psi_i)\tilde{h}_{A',i}(f';\psi_i)\, ,
\ee
(no sum over $i$).
To compute  the integrals in  \eqs{dpsi_i_dpsi_j}{dpsi_i_dpsi_i}
we need to extract explicitly the dependence of the integrands on the polarization angles.  With respect to a choice of axes in the plane transverse to $\hatn_i$ which defines the origin for $\psi_i$ [see the discussion above \eq{allaverages}], the polarization angle of the $i$-the event enters through a rotation by an angle $2\psi_i$ in the waveform, which  can be written as
\be
\label{eq: psi rot}
\tilde{h}_{A,i}(f;\psi_i)=
\sum_{B=+,\times}R_{AB}(2\psi_i)\,\tilde{h}_{B,i}(f;\psi_i=0)\,,
\ee
where $R_{AB}(2\psi_i)$ is the rotation matrix with entries
\be
R_{AB}(2\psi_i)=
\begin{pmatrix}
\cos{2\psi_i}       & -\sin{2\psi_i} \\
\sin{2\psi_i}       & \cos{2\psi_i}\,
\end{pmatrix}
_{AB}\,,
\label{eq: rot matrix}
\ee
and $\psi_i=0$ is the fixed reference angle. It is convenient to decompose
the rotation matrix as
\be\label{RABPauli}
R_{AB}(2\psi_i)=\delta_{AB}\cos 2\psi_i - i\sigma^2_{AB} \sin 2\psi_i\, ,
\ee
where $\sigma^{2}$ is the second Pauli matrix.
Then, the integrations in \eqs{dpsi_i_dpsi_j}{dpsi_i_dpsi_i} are straightforward, and in particular the integral in \eq{dpsi_i_dpsi_j}, with $i\neq j$, vanishes. 
We then obtain
\begin{eqnarray}
\hspace{-10mm}\langle\tilde{h}^*_{A,i}(f;\psi_i)\tilde{h}_{A',j}(f';\psi_j)\rangle_{\{\psi_k\}}\nn\\
&&\hspace{-45mm}=\frac12 \delta_{ij}\sum_{B,B'=+,\times}\left(\delta_{A A'}\delta_{B B'}-\sigma^{2}_{A A'}\sigma^{2}_{B B'}\right)\,\tilde{h}_{B,i}^*(f;\psi_i=0)\,\tilde{h}_{B',i}(f';\psi_i=0)\,\nn\\
&&\hspace{-45mm} =\frac{1}{2} \delta_{ij} \delta_{A A'} 
\sum_{B=+,\times} \tilde{h}_{B,i}^*(f;\psi_i=0) \,\tilde{h}_{B,i}(f';\psi_i=0) \nn\\
&&\hspace{-40mm}-\frac{i}{2}\delta_{ij}\sigma^{2}_{A A'}
\big[ \tilde{h}_{\times,i}^*(f;\psi_i=0)\,\tilde{h}_{+,i}(f';\psi_i=0) - \tilde{h}_{+,i}^*(f;\psi_i=0)\,\tilde{h}_{\times,i}(f';\psi_i=0)
\big]\, ,
\label{eq: avg pol 2}
\end{eqnarray}
where, to get the second equality, we  made  use of the identity\footnote{This identity can be proven observing that the left-hand side is a $2\times 2$ matrix (with complex matrix elements)  in the indices $(A,A')$, and can therefore be expanded in the basis of $\delta_{AA'}$ and $\sigma^a_{AA'}$, with (complex) coefficients that carry the $B,B'$ indices, i.e. it can be written in the form
\be\label{idAB1}
\delta_{AB}\delta_{A' B'}-\sigma^{2}_{AB}\sigma^{2}_{A' B'}=\delta_{A A'}q_{B B'}
+\sigma^a_{A A'}p^a_{B B'}\, ,
\ee
for some coefficients $q_{B B'}$ and $p^a_{B B'}$. Contracting both sides with $\delta_{AA'}$ fixes 
$q_{B B'}=\delta_{BB'}$,
while   contracting both side with $\sigma^b_{AA'}$ gives
$p^a_{B B'}=-\delta^{a2}\sigma^2_{BB'}$.
Proceeding analogously, one also gets the identities
\begin{eqnarray}
    \delta_{AB}\delta_{A' B'}-\sigma^{1}_{AB}\sigma^{1}_{A' B'} &&= -\sigma^{2}_{AA'}\sigma^{2}_{BB'} +\sigma^{3}_{AA'}\sigma^{3}_{B B'}\, , \label{idAB2} \\
    \delta_{AB}\delta_{A' B'}-\sigma^{3}_{AB}\sigma^{3}_{A' B'} &&= \sigma^{1}_{AA'}\sigma^{1}_{BB'} -\sigma^{2}_{AA'}\sigma^{2}_{BB'}\, . \label{idAB3}
\end{eqnarray}
}
\be
\label{eq: identity swap app}
\delta_{AB}\delta_{A' B'}-\sigma^{2}_{AB}\sigma^{2}_{A' B'}=\delta_{A A'}\delta_{B B'}-\sigma^{2}_{A A'}\sigma^{2}_{B B'}\,.
\ee
We observe that the average over the polarizations has produced a factor $\delta_{ij}$, which means that, after this average, different events are uncorrelated. This will also simplify the computation of the subsequent averages. We also notice that it has  produced only two structures in the $(A,A')$ indices, i.e.  $\delta_{AA'}$ and $\sigma^{2}_{A A'}$, out of the four possible combinations coming from the fact that the identity matrix and the three Pauli matrices are a complete basis for a $2\times 2$ matrix; i.e. the terms proportional to $\sigma^{1}_{A A'}$ and $\sigma^{3}_{A A'}$, and therefore the Stokes parameters
$U$ and $Q$ describing linear polarization, see \eq{Stokes2},
have been averaged to zero while, at this stage, we still have a non-vanishing circular polarization $V$.

In terms of $\tilde{h}_A(f,\hatn)$ given in \eq{tildehdeltan},
\eq{eq: avg pol 2} gives
\bees
&&\hspace*{-2mm}\langle \tilde{h}_A^*(f,\hatn)
\tilde{h}_{A'}(f',\hatn')\rangle_{ \{\psi_k\} }=\frac12  \delta_{A A'} 
\sum_{i=1}^{\nev}\sum_{B=+,\times}\hspace*{-1mm} \tilde{h}_{B,i}^*(f;\psi_i=0)\,\tilde{h}_{B,i}(f';\psi_i=0) \nn\\
&&\hspace*{70mm}\times\delta(\hatn,\hatn_i)\delta(\hatn',\hatn_i)
\nn\\
&&-\frac{i}{2}\sigma^{2}_{A A'}
\sum_{i=1}^{\nev}
\big[ \tilde{h}_{\times,i}^*(f;\psi_i=0)\,\tilde{h}_{+,i}(f';\psi_i=0) - \tilde{h}_{+,i}^*(f;\psi_i=0)\,\tilde{h}_{\times,i}(f';\psi_i=0)\big]\nn\\
&&\hspace*{22mm}\times \delta(\hatn,\hatn_i)\delta(\hatn',\hatn_i)
\, ,
\ees
where the presence of   $\delta_{ij}$ in \eq{eq: avg pol 2} allowed us to reduce the double sum over $i,j$ to a single sum. We further observe that 
\be\label{deltadelta}
\delta(\hatn,\hatn_i)\delta(\hatn',\hatn_i)=\delta(\hatn,\hatn')\delta(\hatn,\hatn_i)\, .
\ee
Therefore, we can also extract a term $\delta(\hatn,\hatn')$ from the sum over $i$, and we get
\bees
&&\hspace*{-2mm}\langle \tilde{h}_A^*(f,\hatn)
\tilde{h}_{A'}(f',\hatn')\rangle_{ \{\psi_k\} }=\frac{\delta(\hatn,\hatn')}{2} \Big[  \delta_{A A'}\sum_{i=1}^{\nev}\hspace*{-0.2mm}\sum_{B=+,\times}\hspace*{-1.2mm} \tilde{h}_{B,i}^*(f;\psi_i=0)\,\tilde{h}_{B,i}(f';\psi_i=0) \delta(\hatn,\hatn_i)
\nn\\
&&-i\sigma^{2}_{A A'}\sum_{i=1}^{\nev}
\big[ \tilde{h}_{\times,i}^*(f;\psi_i=0)\,\tilde{h}_{+,i}(f';\psi_i=0) - \tilde{h}_{+,i}^*(f;\psi_i=0)\,\tilde{h}_{\times,i}(f';\psi_i=0)\big] \delta(\hatn,\hatn_i)\Big]
\, .\nn\\
\label{corrhahadeltannp}
\ees

\subsection{Average over arrival times}\label{sect:arrivals}

We find technically convenient to compute next the average  over the times of arrival of the GW events (which can be done in any order with respect to the other integrations). Again, we need to make explicit the dependence on the times of arrival, that until now we kept  implicit in $\tilde{h}_{A,i}(f)$. For the $i$-th event, $t_i$ enters through a phase factor as
\be
\label{eq: arr time}
\tilde{h}_{A,i}(f;t_i)=e^{2\pi i f t_i}\, \tilde{h}_{A,i}(f,t_i=0)\ .
\ee
Since \eq{eq: avg pol 2} is already proportional to $\delta_{ij}$, we only need to compute the time average for $i=j$. This is given by
\bees
&&\hspace*{-6mm}\langle\tilde{h}^*_{A,i}(f;\psi_i,t_i)\tilde{h}_{A',i}(f';\psi_i,t_i)\rangle_{ \{\psi_k,t_k\} }=\prod_{k=1}^{\nev} \int_{-T/2}^{T/2}\frac{dt_k}{T}\, 
\langle\tilde{h}^*_{A,i}(f;\psi_i,t_i)\tilde{h}_{A',i}(f';\psi_i,t_i)\rangle_{ \{\psi_k\} }\nn\\
&&\hspace*{-2mm}=\int_{-T/2}^{T/2}\frac{dt_i}{T}\, 
\langle e^{-2\pi i f t_i}\tilde{h}^*_{A,i}(f;\psi_i,t_i=0) \, 
e^{2\pi i f' t_i} \tilde{h}_{A',i}(f';\psi_i,t_i=0)\rangle_{ \{\psi_k\} }\nn\\
&&\hspace*{-2mm}=\frac{1}{T}\delta_T(f-f')\, \langle\tilde{h}^*_{A,i}(f;\psi_i,t_i=0) \tilde{h}_{A',i}(f';\psi_i,t_i=0)\rangle_{ \{\psi_k\} }\, ,\label{ave_t_and_psi}
\ees
where $\delta_T(f-f')$ is given by 
\bees\label{defdeltaT}
\delta_T(f-f')&=&\int_{-T/2}^{T/2} dt\, e^{-2\pi i (f-f') t}\nn\\
&=&\frac{\sin[\pi (f-f') T]}{\pi (f-f') }\, .
\ees
Note  that, for $f=f'$, $\delta_T(f-f')$ takes the value  
\be\label{deltazero}
\delta_T(0) =T\, ,
\ee
and in the limit $|f-f'|T\ra\infty$ we have  $\delta_T(f-f')\ra \delta(f-f')$.
We stress that this specific expression for $\delta_T(f-f')$ emerges from the explicit computation, and is not just an  arbitrary choice of regularization of the Dirac delta. 

We now observe that, in the regime 
\be\label{condf1}
|f-f'|T\gg 1\, ,
\ee 
where $\delta_T(f-f')$ approaches $\delta(f-f')$,
we can use  the approximation
\be\label{deltaTG}
\delta_T(f-f') G(f,f')\simeq \delta_T(f-f') G(f,f)
\ee
for $G(f,f')$ a generic  function (regular at $f'=f$),
and therefore in \eq{ave_t_and_psi} we can replace the correlator $
\langle\tilde{h}^*_{A,i}(f) \tilde{h}_{A',i}(f')\rangle$ on the right-hand side [where we omit here for simplicity the other arguments in $\tilde{h}_{A,i}(f)$] by
$\langle\tilde{h}^*_{A,i}(f) \tilde{h}_{A',i}(f)\rangle$.\footnote{However, at least in the intermediate steps of the computations, even for $|f-f'|T\gg 1$ it can be necessary  to keep  $\delta_T(f-f') $ rather than replacing it immediately with $\delta(f-f')$, since this allows us to regularize  the divergence of $\delta(0)$ through \eq{deltazero}. For instance, the factor $(1/T)\delta_T(f-f')$ on the right-hand side of \eq{ave_t_and_psi} is equal to 1 for $f=f'$, while it would be undefined  for $f=f'$ if one would replace there 
$\delta_T(f-f') $  by $\delta(f-f') $.} 

Another regime where, in the correlator $ \langle\tilde{h}^*_{A,i}(f) \tilde{h}_{A',i}(f')\rangle$ that appears on the right-hand side of \eq{ave_t_and_psi}, we can approximately replace $f'$ by $f$ is when $|f-f'|$ is much smaller than the typical frequency scale $f_0$ on which 
$\langle\tilde{h}^*_{A,i}(f) \tilde{h}_{A',i}(f')\rangle$ differs significantly from
$\langle\tilde{h}^*_{A,i}(f) \tilde{h}_{A',i}(f)\rangle$. We can define for instance $f_0$ from
\be
\left| \frac{1}{ \langle\tilde{h}^*_{A,i}(f) \tilde{h}_{A',i}(f')\rangle}
\frac{\pa}{\pa f'}  \langle\tilde{h}^*_{A,i}(f) \tilde{h}_{A',i}(f')\rangle \right|_{f'=f}=\frac{1}{f_0}\, .
\ee
Then, we can approximately replace $ \langle\tilde{h}^*_{A,i}(f) \tilde{h}_{A',i}(f')\rangle$ by $ \langle\tilde{h}^*_{A,i}(f) \tilde{h}_{A',i}(f)\rangle$ if
\be\label{condf2}
|f-f'|\ll f_0\, .
\ee
The two conditions (\ref{condf1}) and (\ref{condf2}) are in ``OR'', i.e. it is enough that one of them be satisfied to replace $f'$ by $f$ in the correlator.  In particular, if the correlator has a power-law behavior  
\be\label{hhfalphafpalpha}
\langle\tilde{h}^*_{A,i}(f) \tilde{h}_{A',i}(f')\rangle\propto f^{\alpha}{f'}^{\alpha}\, ,
\ee 
 then $f_0=f/|\alpha|$. This is the case  for binary systems in the inspiral phase and with negligible eccentricity, in which case, for both polarizations, the frequency dependence of the GW amplitude factorizes as
$\tilde{h}_{A}(f)\propto f^{-7/6}$ (see e.g. eqs.~(4.34, 4.35) of \cite{Maggiore:2007ulw})
and therefore the correlator $\langle\tilde{h}^*_{A,i}(f) \tilde{h}_{A',i}(f')\rangle$ has the form
(\ref{hhfalphafpalpha}) with 
$\alpha= -7/6$.\footnote{From this it then follows that $\langle\tilde{h}^*_{A,i}(f) \tilde{h}_{A',i}(f)\rangle\propto f^{-7/3} $ and  then, for an ensemble of binaries in the inspiral phase,  $d\rho_{\rm gw}/d\log f\propto f^3 \langle\tilde{h}^*_{A,i}(f) \tilde{h}_{A',i}(f)\rangle\propto f^{2/3}$.}

 In this case
we can simply write $|\alpha|= {\cal O}(1)$ and $f_0\simeq f$. Then, we can replace the two-point correlator $\langle\tilde{h}^*_{A,i}(f) \tilde{h}_{A',i}(f')\rangle$ by a correlator that depends only on a single frequency $f$ if
\be\label{cond12}
|f'-f|\ll f\, ,\qquad  {\rm or}\qquad  |f'-f|\gg 1/T\, .
\ee
For ground-based detectors, and realistic values of $T$, one of the two conditions in \eq{cond12} is always satisfied. Indeed, if  $|f'-f|\ll f$ the first condition is satisfied, while if $|f'-f|\,\gsim\, f$, the second condition is satisfied when $fT\gg 1$, which, as we already mentioned in section~\ref{sect:basic}, for ground based detectors operating at
$f\, >\, (5-10)$~Hz, is  well satisfied already for stretches of data of about $60$~s, and extremely well satisfied if we consider a data stretch of days or weeks. This conclusion can also be reached observing  that a sufficient (but not necessary) condition for \eq{cond12} to hold is that
\be\label{cond12bis}
fT\gg 1 \, ,\qquad  {\rm or}\qquad  f'T\gg 1\, ,
\ee
which is satisfied for ground-based detectors as long as $T$ is larger than, say, 60~s.\footnote{The fact that \eq{cond12bis} implies \eq{cond12}
can be proved  by contradiction, assuming that \eq{cond12bis} holds and that  \eq{cond12} does not. If   \eq{cond12} does  not to hold, we must have $|f'-f|\,\gsim\, f$ {\em and} $ |f'-f|\,\lsim\, 1/T$, i.e. $f\, \lsim\, |f'-f|\, \lsim 1/T$, which is only possible if $fT\,\lsim\,1$. Since we assumed that   \eq{cond12bis} holds, we must then have $f'T\gg 1$. But if  $fT\,\lsim\,1$ and $f'T\gg 1$, then $  |f'-f| T\gg 1$, and therefore  \eq{cond12} holds, contradicting the hypothesis. Note however that \eq{cond12bis} is sufficient for \eq{cond12} to hold, but not necessary. E.g., taking $fT=1$ and $f'T=0.999$,  we have $|f'-f| T\ll fT$ and therefore $|f'-f| \ll f$, so \eq{cond12} holds, but still \eq{cond12bis} does not hold.}
However, for PTA, which operates  in the nHz region, the condition $fT\gg 1$ is only satisfied 
in a part of the frequency range explored. Indeed, the functioning principle of pulsar timing implies that, with an integration time $T$, PTA can access frequencies down to $1/T$; for such frequencies we therefore have by definition $fT={\cal O}(1)$, so near the lower limit  of the accessible bandwidth the condition $fT\gg 1$ is not satisfied (see also the discussion in footnote~\ref{foot:PTAenergy} below);  close to the lower limit of the frequency range explored by PTA we still have the possibility of satisfying \eq{cond12} by satisfying the first condition, $|f'-f|\ll f$, but this is only valid over a very small  range of frequencies. For PTA, it can therefore be useful to write the result in the full general form, rather than 
setting $f'=f$ in the correlator on the right-hand side of \eq{ave_t_and_psi}. In the following, we will therefore work with the most general expression, and then provide also the explicit results in the limit in which we can set $f'=f$ in the correlator.

With this understanding, we can now proceed with the computation and insert  into
\eq{ave_t_and_psi} the correlator averaged over polarizations obtained in \eq{eq: avg pol 2}. This gives 
\be
\langle\tilde{h}^*_{A,i}(f;\psi_i,t_i)\tilde{h}_{A',j}(f',\psi_j,t_j)\rangle_{ \{\psi_k,t_k\} }
=\frac{1}{2} \delta_{ij}\,\delta_T(f-f')\left[\mathcal{I}_i(f,f')\delta_{A A'}+\mathcal{V}_i(f,f')\sigma^{2}_{A A'}\right]\,,
\label{eq: avg pol t 2ffp}
\ee
where
\begin{eqnarray}
\mathcal{I}_i(f,f')&=&\frac{1}{T} \, \[\tilde{h}^*_{+,i}(f)\tilde{h}_{+,i}(f')+
\tilde{h}^*_{\times,i}(f)\tilde{h}_{\times,i}(f')\]\, ,\label{defcalIffp} \\
\mathcal{V}_i(f,f')&=&\frac{i}{T} \, \[\tilde{h}_{+,i}^*(f)\,\tilde{h}_{\times,i}(f')-\tilde{h}_{\times,i}^*(f)\,\tilde{h}_{+,i}(f') \]
\, ,\label{defcalVpp}
\end{eqnarray} 
and we have omitted for simplicity the dependence of $\tilde{h}_{A,i}(f)$ on the reference values $\psi_i=0$ and 
$t_i=0$.\footnote{Actually, we will see below that $\mathcal{I}_i(f,f')$ and $\mathcal{V}_i(f,f')$ are independent of $\psi_i$, while they have a residual dependence on $t_i$ that disappears when the condition (\ref{cond12}) holds.\label{footIVnopsi}}
When the condition (\ref{cond12}) is satisfied  we can simplify these expressions 
as
\be
\langle\tilde{h}^*_{A,i}(f;\psi_i,t_i)\tilde{h}_{A',j}(f',\psi_j,t_j)\rangle_{ \{\psi_k,t_k\} }
\simeq\frac{1}{2} \delta_{ij}\,\delta_T(f-f')\left[\mathcal{I}_i(f)\delta_{A A'}+\mathcal{V}_i(f)\sigma^{2}_{A A'}\right]\,,
\label{eq: avg pol t 2}
\ee
where
\begin{eqnarray}
\mathcal{I}_i(f)\equiv \mathcal{I}_i(f,f)
&=&\frac{|\tilde{h}_{+,i}(f)|^2+|\tilde{h}_{\times,i}(f)|^2}{T}\, ,\label{defcalI} \\
\mathcal{V}_i(f)\equiv \mathcal{V}_i(f,f) &=&\frac{2\operatorname{Im}\big[\tilde{h}_{\times,i}^*(f)\tilde{h}_{+,i}(f)\big]}{T}\, .\label{defcalV}
\end{eqnarray} 
Observe that the factors $1/T$ in these expressions have been inherited from \eq{ave_ti}. Note also that, if we set $f'=f$ in \eq{eq: avg pol t 2} [or in \eq{eq: avg pol t 2ffp}] and use \eq{deltazero},
they cancel with the factor $\delta_T(0)=T$.

Introducing the right-handed ($R$) and left-handed ($L$) GW fields
\begin{eqnarray}
\tilde{h}_{R,i}(f;\psi_i,t_i)=\frac{\tilde{h}_{+,i}(f;\psi_i,t_i)+i\,\tilde{h}_{\times,i}(f;\psi_i,t_i)}{\sqrt{2}}\,,\label{defhR}\\
\tilde{h}_{L,i}(f;\psi_i,t_i)=\frac{\tilde{h}_{+,i}(f;\psi_i,t_i)-i\,\tilde{h}_{\times,i}(f;\psi_i,t_i)}{\sqrt{2}}\,,\label{defhL}
\end{eqnarray}
\eqs{defcalIffp}{defcalVpp} can also be rewritten as
\begin{eqnarray}
\mathcal{I}_i(f,f')&=&\frac{\tilde{h}^*_{R,i}(f)\tilde{h}_{R,i}(f')+
\tilde{h}^*_{L,i}(f)\tilde{h}_{L,i}(f')}{T}\, ,\label{defcalILRpp} \\
\mathcal{V}_i(f,f')&=&\frac{\tilde{h}^*_{R,i}(f)\tilde{h}_{R,i}(f')-
\tilde{h}^*_{L,i}(f)\tilde{h}_{L,i}(f')}{T}\, ,\label{defcalVLRpp}
\end{eqnarray} 
and therefore 
\begin{eqnarray}
\mathcal{I}_i(f)&=&\frac{|\tilde{h}_{R,i}(f)|^2+|\tilde{h}_{L,i}(f)|^2}{T}\,,\label{defcalILR}\\
\mathcal{V}_i(f)
&=&\frac{|\tilde{h}_{R,i}(f)|^2-|\tilde{h}_{L,i}(f)|^2}{T}\label{defcalVLR}\, .
\end{eqnarray}
Note that $\mathcal{I}_i(f,f')$ and $\mathcal{V}_i(f,f')$ are invariant under rotations of the form (\ref{eq: psi rot}), i.e. 
\be\label{rotationofh}
\tilde{h}_{A,i}(f) \ra \sum_{B=+,\times}R_{AB}(2\psi_i)\,\tilde{h}_{B,i}(f)\, .
\ee
In terms of $\tilde{h}_{L,i}(f) $ and  $\tilde{h}_{R,i}(f) $, \eq{rotationofh} reads
\be\label{hLRhelicity2}
\tilde{h}_{L,i}(f)\ra e^{-2i\psi_i} \tilde{h}_{L,i}(f)\, ,\qquad
\tilde{h}_{R,i}(f)\ra e^{+2i\psi_i} \tilde{h}_{R,i}(f)\, ,
\ee
which  expresses the fact that  $\tilde{h}_{L,i}(f) $ and  $\tilde{h}_{R,i}(f) $ are eigenstates of helicity with eigenvalues $-2$ and $+2$, respectively.\footnote{Our sign convention on the definition of helicity is the same as in eq.~(2.197) of ref.~\cite{Maggiore:2007ulw}, that states that  an eigenstate of helicity with eigenvalue $h$   (not to be confused with a GW amplitude! both symbols are standard) transforms as $e^{ih\psi}$ under a rotation by an angle $\psi$. If one uses the opposite convention, so that  an eigenstate of helicity with eigenvalue $h$ transforms as $e^{-ih\psi}$, one must exchange the definitions of $h_R$ and $h_L$.}
Therefore $\mathcal{I}_i(f,f')$ and $\mathcal{V}_i(f,f')$ lose any dependence on the reference values $\psi_i=0$  introduced in \eq{eq: psi rot}.  Similarly,
$\mathcal{I}_i(f)$ and $\mathcal{V}_i(f)$ are  invariant under
\be\label{phaseshifth}
\tilde{h}_{A,i}(f) \ra e^{2\pi i f t_i}\, 
\tilde{h}_{A,i}(f) \, ,
\ee
and therefore $\mathcal{I}_i(f)$ and $\mathcal{V}_i(f)$ lose any dependence on the reference values  $t_i=0$ introduced in \eq{eq: arr time}.  This, however, is no longer true in general  for 
$\mathcal{I}_i(f,f')$ and $\mathcal{V}_i(f,f')$. The dependence on $t_i$ only disappears when the condition (\ref{cond12}) is satisfied.

We see that the averages over polarizations and over arrival times have produced a Kronecker delta $\delta_{ij}$ that decorrelates different events, and  a (finite-time) Dirac delta $\delta_T(f-f')$ that, in the limit $|f-f'|T\gg 1$,  decorrelates different frequencies. These two averages did not involve any stringent physical assumption (such as the isotropy of the source distribution), since we just assumed that the result does not depend on the system of axes set by the observer to measure $\psi$, and that the astrophysical population does not evolve appreciably over the observation time;  the corresponding result is therefore very general. Plugging \eq{eq: avg pol t 2ffp} into \eq{eq: avg ext sum i j}  
we see that
\bees
&&\hspace*{-6mm}\langle \tilde{h}_A^*(f,\hatn)
\tilde{h}_{A'}(f',\hatn')\rangle_{ \{t_k, \psi_k\} } \nn\\
&&\hspace*{-2mm}=\frac{1}{2} \,\delta_T(f-f')
\sum_{i=1}^{\nev} 
\[ \mathcal{I}_i(f,f')\delta_{A A'}+\mathcal{V}_i(f,f')\sigma^{2}_{A A'}\] \delta(\hatn,\hatn_i)\delta(\hatn',\hatn_i)  \,,
\ees
where the presence of $\delta_{ij}$ in \eq{eq: avg pol t 2ffp} allowed us to reduce \eq{eq: avg ext sum i j} to a sum over a single index $i$. We can now further use   \eq{deltadelta}, so that
\bees
&&\hspace*{-8mm}\langle \tilde{h}_A^*(f,\hatn)
\tilde{h}_{A'}(f',\hatn')\rangle_{ \{t_k,\psi_k\} } \nn \\
&&=\frac{1}{2} \,\delta_T(f-f') \delta(\hatn,\hatn')
\sum_{i=1}^{\nev}
\[ \mathcal{I}_i(f,f')\delta_{A A'}+\mathcal{V}_i(f,f')\sigma^{2}_{A A'}\]  \delta(\hatn,\hatn_i)  \, .\label{corrintffp}
\ees
We observe that, after averaging over polarization angles and times of arrival, the  correlator $\langle \tilde{h}_A^*(f,\hatn)
\tilde{h}_{A'}(f',\hatn')\rangle$  is proportional to both $\delta(\hatn,\hatn')$ and $\delta_T(f-f')$, and linear polarizations have been averaged to zero.

When the condition (\ref{cond12}) is satisfied, using \eq{deltaTG}, we have
\bees
&&\hspace*{-8mm}\langle \tilde{h}_A^*(f,\hatn)
\tilde{h}_{A'}(f',\hatn')\rangle_{ \{t_k,\psi_k\} } \nn \\
&&=\frac{1}{2} \,\delta_T(f-f') \delta(\hatn,\hatn')
\sum_{i=1}^{\nev}
\[ \mathcal{I}_i(f)\delta_{A A'}+\mathcal{V}_i(f)\sigma^{2}_{A A'}\]  \delta(\hatn,\hatn_i)  \, ,\label{corrint}
\ees
and eventually, as $(f-f')T\ra\infty$,  we can also replace $\delta_T(f-f')$ by  $\delta(f-f')$.

\subsection{Average over orbit inclinations}\label{avecosiota}

We next tackle the  computation of the average over $\{\cos\iota_k\}$.  Using \eq{allaverages} we see that, in each term of the sum over $i$ in \eq{corrint}, the average over $\cos\iota_k$ with $k\neq i$ gives one, and we only remain with the average over $\cos\iota_i$, so that 
\bees
&&\hspace*{-18mm}\langle \tilde{h}_A^*(f,\hatn)
\tilde{h}_{A'}(f',\hatn')\rangle_{ \{t_k, \psi_k, \cos\iota_k\} }=\frac{1}{2} \,\delta_T(f-f')  \delta(\hatn,\hatn') \nn\\
&&\hspace*{-0mm}\times
\sum_{i=1}^{\nev} 
\langle \mathcal{I}_i(f,f';\cos\iota_i)\delta_{A A'}+\mathcal{V}_i(f,f';\cos\iota_i)\sigma^{2}_{A A'}\rangle_{\cos\iota_i} \delta(\hatn,\hatn_i)  \,,\label{hhtpsicosiota}
\ees
where we have now written explicitly  the argument $\cos\iota_i$ in $\mathcal{I}_i$ and $\mathcal{V}_i$. However, since in \eq{hhtpsicosiota} the average of $\mathcal{I}_i(f,f')$  is just with respect to $\cos\iota_i$, with the same index $i$, when performing the average $\cos\iota_i$ becomes a dummy integration variable and we can simply denote it by it $\cos\iota$, omitting the index $i$. 
Similarly to the polarization angle and the time of arrival, the integration over $\cos\iota$ is also naturally carried out with a flat measure,  \eq{ave_cosi}, without the need of any non-trivial physical assumption: the binary knows nothing about the observer, and has no reason to be oriented preferentially in some direction with respect to the line of sight connecting it to the observer. However, the difference with the previous cases is that now the dependence on $\cos\iota$, for a full  waveform including the effect of higher modes,  is in general non-trivial.

In ref.~\cite{ValbusaDallArmi:2023ydl} the average over the orbit inclinations has been computed keeping only the quadrupolar mode and  neglecting all higher modes in the amplitude. In that case, the dependence on $\cos\iota$ becomes very simple, and is given by (see eqs.~(4.34) and (4.35) of ref.~\cite{Maggiore:2007ulw})
\bees
\tilde{h}_{+}(f;\cos\iota)&=&{\cal A}(f) e^{i\Psi(f)}\, \frac{1+\cos^2\iota}{2}\, ,\label{hplus}\\
\tilde{h}_{\times}(f;\cos\iota)&=&{\cal A}(f) i e^{i\Psi(f)}\, \cos\iota\, ,
\label{hcross}
\ees
where ${\cal A}(f)$ and $\Psi(f)$ are an amplitude and a phase, whose explicit expression we do not need here. Note that the dependence of the waveform on  $\cos\iota$ 
is a characteristic of the mode considered, independently of the post-Newtonian order at which the phase of the waveform is computed, and is  valid  even for a full inspiral-merger-ringdown waveform, again as far as only the lowest mode of the amplitude (i.e., the quadrupole) is retained.
Then, setting for instance $f'=f$ in the correlator,
\bees
|\tilde{h}_{+}(f)|^2 + |\tilde{h}_{\times}(f)|^2 &=& {\cal A}^2(f) Q_I(\cos\iota)\, ,\\
\operatorname{Im}\left[\tilde{h}_{+}(f)\tilde{h}_{\times}^*(f)\right]&=& -{\cal A}^2(f) Q_V(\cos\iota)\, ,
\ees
where
\bees
Q_I(\cos\iota) &=& \( \frac{1+\cos^2\iota}{2}\)^2+ \cos^2\iota\, ,\\
Q_V(\cos\iota) &=&\frac{1+\cos^2\iota}{2} \cos\iota\, .
\ees
Averaging over $\cos\iota$ with the flat measure (\ref{ave_cosi}), 
\bees
\langle Q_I(\cos\iota) \rangle_{\cos\iota}&=&\frac{4}{5}\, ,\label{QIave45}\\
\langle Q_V(\cos\iota) \rangle_{\cos\iota}&=&0\, .
\ees
Therefore, in the approximation in which one considers only the quadrupole mode in the waveform and neglects the  higher modes,  
the circular polarization averages to zero, as already found in ref.~\cite{ValbusaDallArmi:2023ydl}.

For the term related to intensity, using the fact that $Q_I(\cos\iota=1)=2$, so that $\langle Q_I(\cos\iota) \rangle_{\cos\iota}=(2/5)Q_I(\cos\iota=1)$, 
we could  write 
\be\label{Iave25Imax}
\langle \mathcal{I}_i(f;\cos\iota)\rangle_{\cos\iota} =\frac{2}{5}
\mathcal{I}_i(f;\cos\iota=1)\, .
\ee
The question of whether the circular polarization vanishes exactly or not is, however, quite important, because parity-violating mechanism in the early Universe could produce a cosmological stochastic background with a net circular polarization~\cite{Alexander:2004us}, and various techniques have been discussed to extract this effect from GW data~\cite{Seto:2008sr,Crowder:2012ik,Romano:2016dpx,Yagi:2017zhb,Domcke:2019zls,Martinovic:2021hzy,Callister:2023tws,Cruz:2024esk};  such  early-Universe signature might however be masked  by  
contamination from the astrophysical background, if even a small fraction of it should turn out to be  polarized. It is therefore interesting to see if the vanishing of the circular polarization after averaging over $\cos\iota$ also holds when we include higher modes in the waveform. At first sight one might doubt it, given the complicated dependence on $\sin\iota$ and $\cos\iota$ of the higher modes. However,  the vanishing of the circular polarization upon averaging over $\cos\iota$ is  in fact a general feature, which simply follows from parity arguments. Let us denote by $\hatn_{\rm orbit}$ the normal to the orbit. In polar coordinates, choosing as polar axis  the line-of-sight to the observer,
$\hatn_{\rm orbit}=(\sin\iota\cos\psi,\sin\iota\sin\psi,\cos\iota)$, where $\psi$ is the angle in the plane transverse to the polar axis [i.e. the polarization angle, see the discussion above \eq{allaverages}]. 
If we perform a parity transformation, the normal 
to the orbit changes sign, $\hatn_{\rm orbit}\ra -\hatn_{\rm orbit}$ which, recalling that $\iota \in [0,\pi]$, means 
$\{\iota,\psi\}\ra \{\pi-\iota,\psi+\pi\}$, so in particular 
\be\label{transcosiota}
\cos\iota\ra -\cos\iota\, , \qquad \sin\iota\ra \sin\iota\, .
\ee
At the same time, under a parity transformation, $h_+$ transform as  a scalar, $h_+\ra h_+$, while $h_{\times}$ as  a pseudoscalar, $h_{\times}\ra -h_{\times}$.\footnote{This follows from the fact that the plus and cross polarization of a GW with propagation direction $\hatn$ are defined introducing two unit vectors $\hatu$, $\hatv$ orthogonal to $\hatn$ and among them, chosen so that $\{\hatu,\hatv,\hatn\}$ form a right-handed oriented frame, i.e. $\hatu\times\hatv=\hatn$; e.g., when $\hatn=\hatz$, one could choose $\hatu=\hatx$ and $\hatv=\haty$. Then the polarization tensors associated with the plus and cross polarizations are, respectively, 
\be\label{eplusecross}
e_{ab}^+(\hatn)=\hat{\bf u}_a\hat{\bf u}_b
-\hat{\bf v}_a\hat{\bf v}_b\, ,
\qquad
e_{ab}^{\times}(\hatn)=\hat{\bf u}_a\hat{\bf v}_b+
\hat{\bf v}_a\hat{\bf u}_b\, ,
\ee
(see \eqs{eplusuv}{ecrossuv}, or  eq.~(1.54) of ref.~\cite{Maggiore:2007ulw}).
Under a parity transformation we have $\hatn\ra \hatn'=-\hatn$,
$\hatu\ra \hatu'=-\hatu$ and $\hatv\ra \hatv'=-\hatv$. However, 
$\{\hatu',\hatv',\hatn'\}$ no longer forms a right-handed frame, since
$\hatu'\times\hatv'=-\hatn'$. To recover a right-handed frame we must exchange   the sign of either $\hatu'$ or $\hatv'$. As a result, 
\be
e_{ab}^+(-\hatn)=e_{ab}^+(\hatn)\, ,\qquad
e_{ab}^{\times}(-\hatn)=-e_{ab}^{\times}(\hatn)\, .
\ee}
Therefore, the GWs produced by a CBC whose orbit has  a normal  $\hatn_{\rm orbit}$ are obtained from those produced by a CBC whose orbit has a normal $-\hatn_{\rm orbit}$, simply replacing $h_{+}\ra h_{+}$ and $h_{\times}\ra -h_{\times}$, or,  from \eqs{defhR}{defhL}, exchanging  $h_L$ with $h_R$. Since this follows from a general symmetry argument,  it  holds for the full waveform, including higher modes.\footnote{This can be checked explicitly for the first five higher-mode amplitudes $H_{+,\times}^{(a)}$ (with $a=1/2,\,1,\,3/2,\,2,\,5/2$), i.e., the amplitudes up to 2.5PN order, given explicitly in refs.~\cite{Blanchet:1996pi,Arun:2004ff}
where we see that, indeed, under the
transformation (\ref{transcosiota}), the  amplitudes $H_+^{(a)}$ are invariant while $H_{\times}^{(a)}$ change sign. It can also be verified by explicit numerical  calculation (which we performed up to $l=6$) that the $\cos\iota$ average of the combination ${\rm Im}[\tilde{h}_{+}\tilde{h}^*_{\times}]$  vanishes exactly for all higher modes due to the structure of the spin-weighted spherical harmonics commonly used to decompose the signal.}
We see from \eqs{defcalILRpp}{defcalVLRpp} that 
$\mathcal{I}_i(f,f')$ is even under $h_L\leftrightarrow h_R$, while 
$\mathcal{V}_i(f,f')$ is odd. As a result, in \eq{hhtpsicosiota}
the integral of $\mathcal{I}_i(f,f')$ over $\cos\iota$ between $-1$ and $1$ is twice the integral from 0 to 1, while that of $\mathcal{V}_i(f,f')$ vanishes. 

Therefore, after averaging over the inclination angle as in \eq{ave_cosi}, the circular polarization vanishes, even for the exact waveform including higher modes.
In contrast, the exact number in \eq{QIave45} is not protected by symmetries, and will be affected by higher modes. Then, for this term it can be better to leave the average over $\cos\iota$, rather than using \eq{Iave25Imax}. Depending on the accuracy required, one could perform the average numerically with the desired waveform including higher modes, or else use \eq{Iave25Imax}.

To sum up, after averaging  over the inclination of the orbits, we get 
\be\label{avefinocosiotapp}
\langle \tilde{h}_A^*(f,\hatn)
\tilde{h}_{A'}(f',\hatn')\rangle_{ \{t_k, \psi_k, \cos\iota_k\} } =\frac{1}{2} \,\delta_T(f-f')\delta(\hatn,\hatn') \delta_{A A'}
\sum_{i=1}^{\nev} 
\langle \mathcal{I}_i(f,f';\cos\iota)\rangle_{\cos\iota} \delta(\hatn,\hatn_i)  \,,
\ee
which, when the condition (\ref{cond12}) is satisfied, can be approximated as
\be\label{avefinocosiota}
\langle \tilde{h}_A^*(f,\hatn)
\tilde{h}_{A'}(f',\hatn')\rangle_{ \{t_k, \psi_k, \cos\iota_k\} } \simeq\frac{1}{2} \,\delta_T(f-f')\delta(\hatn,\hatn') \delta_{A A'}
\sum_{i=1}^{\nev} 
\langle \mathcal{I}_i(f;\cos\iota)\rangle_{\cos\iota} \delta(\hatn,\hatn_i)  \,.
\ee

%

\subsection{Average over arrival directions}

We next perform the average over the arrival directions $-\hatn_i$ or, equivalently, over the propagation directions $\hatn_i$.
Observe that, given a dependence of the waveform on the inclination angles, such as that in 
\eqs{hplus}{hcross} or the more complicated dependences involving higher modes, 
$\langle \mathcal{I}_i(f,f';\cos\iota)\rangle_{\cos\iota}$ in \eq{avefinocosiota}
is just a number, which depends on the $i$-th event through its waveform, but no longer carries a dependence on $\hatn_i$ (and, in this sense, can be written as an average over a generic dummy integration variable $\cos\iota$, rather than writing explicitly $\cos\iota_i$). Therefore, the subsequent average over arrival directions,
which in a generic non-isotropic case we perform using \eq{aved2n_mu}, gives
\bees
\hspace*{-12mm}\langle \tilde{h}_A^*(f,\hatn)
\tilde{h}_{A'}(f',\hatn')\rangle_{ \{t_k, \psi_k, \cos\iota_k, n_k\} } &=&\frac{1}{2} \,\delta_T(f-f')\delta(\hatn,\hatn') \delta_{A A'}\nn\\
&&\times 
\sum_{i=1}^{\nev} 
\langle \mathcal{I}_i(f,f';\cos\iota)\rangle_{\cos\iota} \langle \delta(\hatn,\hatn_i)  \rangle_{\hatn_i}\,,
\label{avefinocosiota2}
\ees
where, as usual, the average of $ \delta(\hatn,\hatn_i) $ over $\hatn_k$ with $k\neq i$ gives one. The remaining average is simply computed, 
\bees
\langle\delta(\hatn,\hatn_i)\rangle_{\hatn_i}&=&
\int_{S^2}\frac{d^2\hatn_i}{4\pi}\, \mu(\hatn_i)
\delta(\hatn,\hatn_i)\nn\\
&=&\frac{\mu(\hatn)}{4\pi}\, .\label{average_delta_mu}
\ees
Inserting this into \eq{corrintffp}, and including also
the averages over intrinsic parameters and over $\nev$  as in \eq{aveextint}, we get
\be
\langle \tilde{h}_A^*(f,\hatn)
\tilde{h}_{A'}(f',\hatn')\rangle_{ \{t_k, \psi_k, \cos\iota_k, n_k\} } =\frac{1}{2} \,\delta_T(f-f')\delta(\hatn,\hatn') \delta_{A A'}\frac{\mu(\hatn)}{4\pi}
\sum_{i=1}^{\nev} 
\langle \mathcal{I}_i(f,f';\cos\iota)\rangle_{\cos\iota}\, ,
\ee
and again, when the condition (\ref{cond12}) is satisfied, we can simplify this to
\be
\langle \tilde{h}_A^*(f,\hatn)
\tilde{h}_{A'}(f',\hatn')\rangle_{ \{t_k, \psi_k, \cos\iota_k, n_k\} } =\frac{1}{2} \,\delta_T(f-f')\delta(\hatn,\hatn') \delta_{A A'}\frac{\mu(\hatn)}{4\pi}
\sum_{i=1}^{\nev} 
\langle \mathcal{I}_i(f;\cos\iota)\rangle_{\cos\iota}\, .
\ee

\subsection{Average over intrinsic parameters}

We can finally add the average over the intrinsic parameters, as well as over the number of events $\nev$, obtaining what we have called the average over ``Universe realizations''. The result can be written as
\be
\langle \tilde{h}_A^*(f,\hatn)
\tilde{h}_{A'}(f',\hatn')\rangle_U=\delta_T(f-f')
\frac{\delta(\hatn,\hatn')}{4\pi} \frac{1}{2} \, [S_h^{\rm astro}(f,f'; \hatn)]_{AA'} \, ,\label{hhgeneralpp}
\ee
where
\be\label{defSastro_fhatnffp}
[S_h^{\rm astro}(f,f'; \hatn)]_{AA'}=\delta_{AA'} S_h^{\rm astro}(f,f'; \hatn)\, .
\ee
The function $S_h^{\rm astro}(f,f'; \hatn)$ factorizes as\footnote{We stress that this factorization holds also if $\mu(\hatn)$ depends on the intrinsic parameters of the CBC population, see the discussion below \eq{normmu}. All that we used, for this derivation, is that the integrations over the extrinsic parameters factorize as in \eq{allaverages}.
\label{foot:muintrinsic}} 
\be\label{defIastro_fhatn}
S_h^{\rm astro}(f,f'; \hatn)=\mu(\hatn)\, S_h^{\rm astro}(f,f')\, ,
\ee
where
\bees
S_h^{\rm astro}(f,f')&=&\big\langle\sum_{i=1}^{\nev}
\langle \mathcal{I}_i(f,f';\cos\iota)\rangle_{\cos\iota} \big\rangle_{ {\rm int}, \nev }\,\label{eq: Stokes Ipp}\\
&=&\frac{1}{T}\big\langle \Big[ \sum_{i=1}^{\nev}
\langle\, \(
\tilde{h}^*_{+,i}(f;\cos\iota)\tilde{h}_{+,i}(f';\cos\iota)+
\tilde{h}^*_{\times,i}(f;\cos\iota)\tilde{h}_{\times,i}(f';\cos\iota)
 \)
\,\rangle_{\cos\iota}
\Big]
\big\rangle_{ {\rm int}, \nev }\,\nonumber\\
&=&\frac{1}{T}\big\langle \Big[ \sum_{i=1}^{\nev}
\langle\, 
\( 
\tilde{h}^*_{R,i}(f;\cos\iota)\tilde{h}_{R,i}(f';\cos\iota)+
\tilde{h}^*_{L,i}(f;\cos\iota)\tilde{h}_{L,i}(f';\cos\iota)
 \)
\,\rangle_{\cos\iota}
\Big]
\big\rangle_{ {\rm int}, \nev }\,.\nonumber
\ees
In the limit where the condition (\ref{cond12}) is satisfied, appropriate in particular to ground-based GW detectors, 
\be
\langle \tilde{h}_A^*(f,\hatn)
\tilde{h}_{A'}(f',\hatn')\rangle_U=\delta_T(f-f')
\frac{\delta(\hatn,\hatn')}{4\pi} \delta_{AA'} \frac{1}{2} \, S_h^{\rm astro}(f; \hatn) \, ,\label{hhgeneral}
\ee
where
\be\label{defSastro_fhatn}
S_h^{\rm astro}(f; \hatn)=\mu(\hatn)\, S_h^{\rm astro}(f)\, ,
\ee
and
\bees
S_h^{\rm astro}(f)&=&\big\langle\sum_{i=1}^{\nev}
\langle \mathcal{I}_i(f;\cos\iota)\rangle_{\cos\iota} \big\rangle_{ {\rm int}, \nev }\nn\\
&=&\frac{1}{T}\big\langle \Big[ \sum_{i=1}^{\nev}
\langle\, \(
|\tilde{h}_{+,i}(f;\cos\iota)|^2+|\tilde{h}_{\times,i}(f;\cos\iota)|^2 \)
\,\rangle_{\cos\iota}
\Big]
\big\rangle_{ {\rm int}, \nev }\,\nonumber\\
&=&\frac{1}{T}\big\langle \Big[ \sum_{i=1}^{\nev}
\langle\, 
\( |\tilde{h}_{R,i}(f;\cos\iota)|^2+|\tilde{h}_{L,i}(f;\cos\iota)|^2 \)
\,\rangle_{\cos\iota}
\Big]
\big\rangle_{ {\rm int}, \nev }\,.\label{eq: Stokes I}
\ees
The isotropic case is recovered setting $\mu(\hatn)=1$.

\subsection{Summary}\label{sect:Summary}

To summarize this section, we have computed the spectral density due to an astrophysical background by performing explicitly the average of the extrinsic parameters. Going step-by-step, one can appreciate the effects of the different averages. 
The averages over the polarization angles decorrelate the different events, producing the factor $\delta_{ij}$ in \eq{eq: avg pol 2} or, equivalently, decorrelate the arrival directions, producing a Dirac delta 
$\delta(\hatn,\hatn')$ in \eq{corrhahadeltannp}. They also eliminate the linear polarizations, since in \eq{eq: avg pol 2}
the terms proportional to $\sigma^{1}_{A A'}$ and $\sigma^{3}_{A A'}$ have been averaged to zero. The averages over arrival times produce a regularized Dirac delta $\delta_T(f-f')$, which becomes a Dirac delta as the observation time $T\ra\infty$.\footnote{One could have also started from the integration over the times of arrival. In that case, for $i\neq j$, using \eq{eq: arr time} the correlator 
$\langle\tilde{h}^*_{A,i}(f;t_i)\tilde{h}_{A',j}(f',t_j)\rangle_{ \{t_k\} }$ becomes proportional to 
$$
\int_{-T/2}^{T/2} dt_i \int_{-T/2}^{T/2} dt_j\,  e^{-2\pi i ft_i+2\pi i f' t_j}\, ,
$$
which does not vanish but is rather proportional to 
$\delta_T(f)\delta_T(f')$. However, the product of these regularized Dirac delta goes to zero  in the limit in which $fT\gg 1$ or $f'T\gg 1$, so in this limit  the result vanishes for $i\neq j$ and  becomes proportional to $\delta_{ij}$.
Performing first the average over the polarization angles produces already a factor $\delta_{ij}$, so  we do not need to take here the limit $fT\gg 1$ or $f'T\gg 1$.  However, in \cref{sect:Stokesdiscrete}, when we consider shot noise,  we will need to perform the  average over time of arrivals while treating  the polarization angles  as a discrete sum, and in this case the result becomes proportional to $\delta_{ij}$ only in  the limit $fT\gg 1$ or $f'T\gg 1$.
\label{foot:decorij_ti}}
The averages over the inclination angles remove the remaining Stokes parameter, associated with circular polarization, so the result for  $\langle \tilde{h}_A^*(f,\hatn)
\tilde{h}_{A'}(f',\hatn')\rangle$  becomes proportional  to $\delta_{AA'}$. The average  over arrival directions is performed last (since $\hatn_i$ is necessary to define $\psi_i$ and $\cos\iota_i$). We have performed these computations for generic values of  the observation time $T$, i.e. without assuming from the start the condition (\ref{cond12}). In this general case, after performing these averages 
the two-point correlator takes the form given in \eq{hhgeneralpp}, with a spectral density 
$S_h^{\rm astro}(f,f'; \hatn)$ that depends on both frequencies $f$ and $f'$, and which is given explicitly in \eqs{defIastro_fhatn}{eq: Stokes Ipp}. In the limit (\ref{cond12}), which is  appropriate for ground-based interferometers (but not necessarily for PTA), we can approximate $S_h^{\rm astro}(f,f'; \hatn)$ with $S_h^{\rm astro}(f; \hatn)\equiv S_h^{\rm astro}(f,f; \hatn)$. Then
the two-point correlator takes the form given in \eq{hhgeneral}, with a spectral density 
$S_h^{\rm astro}(f; \hatn)$  given explicitly in \eqst{defSastro_fhatn}{eq: Stokes I}.

Finally, we should perform  the averages over the intrinsic parameters of the binary, and over the number of events that arrive during the observation time $T$.
When we have a large ensemble of CBCs, whether coming from actual detections or generated synthetically according to a given distribution of parameters, 
the  averages over these parameters  will  be automatically performed (modulo a dependence on the realization, i.e. shot noise effects, that we will discuss in \cref{sect:shot}) by the sum over the events.
Of course, this could have been done for all parameters, including times of arrival, polarization angles and arrival directions. However, in those cases we have been able to perform the averages analytically, and this allowed us to extract Dirac or Kronecker delta functions, that simplify considerably the structure of the result. Note that, for the average over $\cos\iota$, we are taking a mixed approach, where we use the information that the average sets to zero the circular polarization, while we compute  the average of the intensity parameter by summing over the events; this is because, as we have seen, the vanishing of the circular polarization is an exact result related to a parity symmetry (again, modulo shot noise effects that we will discuss in \cref{sect:shot}) while a numerical coefficient such as that in \eq{Iave25Imax} is not protected by any symmetry argument, and will be affected by the inclusion of higher modes in the waveform.

With this understanding, 
we can rewrite the result for  $S_h^{\rm astro}(f,f')$ as
\be\label{Shfinal1pp}
S_h^{\rm astro}(f,f')=
\frac{1}{T}\big\langle  \sum_{i=1}^{\nev}
 \[
\tilde{h}^*_{+,i}(f)\tilde{h}_{+,i}(f')+
\tilde{h}^*_{\times,i}(f)\tilde{h}_{\times,i}(f')
 \]
\big\rangle_{\nev }\, ,
\ee
keeping only the average over the Poisson distribution  of the number of events detected in the observation time $T$ (we do not write explicitly the dependence of $\tilde{h}_{A,i}(f)$ on $\cos\iota$ nor on the intrinsic parameters). As a further simplification, we can eliminate even the latter average, and just replace $\nev$ with its average value $\bnev$ over the observation time $T$,\footnote{Note that $\bnev$ should be written, more precisely, as $\bnev(T)$. We will leave this  dependence implicit, for notational simplicity.} so 
in this approximation we can simply write
\be\label{Shfinal2pp}
S_h^{\rm astro}(f,f')=
\frac{1}{T}  \sum_{i=1}^{\bnev}
 \[
\tilde{h}^*_{+,i}(f)\tilde{h}_{+,i}(f')+
\tilde{h}^*_{\times,i}(f)\tilde{h}_{\times,i}(f')
 \]\, .
\ee
In the limit (\ref{cond12}), the correlator  $\langle \tilde{h}_A^*(f,\hatn)
\tilde{h}_{A'}(f',\hatn')\rangle_U$ further simplifies and is given by \eqs{hhgeneral}{defSastro_fhatn}, with  
\be\label{Shastroavenev}
S_h^{\rm astro}(f)=\frac{1}{T}\big\langle  \sum_{i=1}^{\nev} \[
|\tilde{h}_{+,i}(f)|^2+|\tilde{h}_{\times,i}(f)|^2 \]
\big\rangle_{ \nev }\, .
\ee
Eliminating also the average over the number of events, and just replacing $\nev$ with its average value $\bnev$ over the observation time $T$, we finally get
\be\label{eq:Shsum}
S_h^{\rm astro}(f)= \frac{1}{T}
\sum_{i=1}^{\bnev}
\[  |\tilde{h}_{+,i}(f)|^2 + |\tilde{h}_{\times,i}(f)|^2 \] \, ,
\ee
which provides a simple expression for the spectral density generated by  an astrophysical background in an observation time $T$, when  the  condition (\ref{cond12})  is satisfied.

Let us now recall that the energy associated with a single GW signal with typical frequency $f$ is given by (see e.g. eq.~(1.135) of ref.~\cite{Maggiore:2007ulw})
\be
\rho_{\rm gw}=\frac{c^2}{32\pi G}\, \langle \dot{h}_{ab}\dot{h}_{ab}\rangle\, ,
\ee
where  the sum over the repeated indices $a,b$ is understood, and here $\langle \ldots \rangle$ denotes a temporal average over a time interval  much longer than the period of the GW. For a stochastic background, observed for a time $T$, the energy density is therefore  well defined only if $T$ is much larger than the period of  the Fourier modes that contribute significantly.\footnote{This is a point that can be relevant when interpreting physically the results of PTA. Indeed, 
while a correlator such as $\langle \tilde{h}_A^*(f,\hatn)
\tilde{h}_{A'}(f,\hatn')\rangle_U$ is well defined for all $T$, 
the  energy density of the stochastic background that produces this correlation is  well defined  only for $fT\gg 1$.  As the integration time of PTA increases, the lower limit of the  frequency range explored decreases as $1/T$, but the ``freshly conquered'' frequency range can be interpreted in terms  of  energy density of a stochastic GW background only after further integration time. In particular, for the current  stretch of data of duration $T=15$~yr, the condition $fT\gg 1$ implies  $f\gg 2\, {\rm nHz}$. 
\label{foot:PTAenergy}
}
We will therefore restrict now to this limit, so that  the background can be considered stationary and its energy density is well defined. 
For an astrophysical stochastic background we will also add an average over the extrinsic and intrinsic parameters (the ``Universe realizations''), writing
\be\label{rhogwaveU}
\rho_{\rm gw}=\frac{c^2}{32\pi G}\, \langle \dot{h}_{ab}\dot{h}_{ab}\rangle_U\, ,
\ee
For a  stochastic background  that is also  isotropic and unpolarized, and whose two-point correlator  is then given by \eq{ave}, a standard computation (see,  e.g., eq.~(7.201) of ref.~\cite{Maggiore:2007ulw}) then shows that
\be\label{rho3}
\frac{d\rho_{\rm gw}}{d\log f}=\frac{\pi c^2}{2G}f^3S_h(f)\, .
\ee
It is convenient to define the dimensionless quantity
\be \label{eq: Omegagw general def}
\omgw(f) \equiv \frac{1}{\rho_c}\,\frac{ d\rho_{\rm gw}}{d\log f} \, ,
\ee
where  $\rho_c = 3 c^2 H_0^2/\left(8\pi G\right)$ is the critical energy density for closing the Universe. 
In terms of $S_h(f)$, we therefore have
\be\label{eq:OmegaSh}
\omgw(f) = \frac{4\pi^2}{3H_0^2}\, f^3S_h(f)\, .
\ee
When the background is due to the superposition of a discrete number of astrophysical events, the spectral density (in the limit $fT\gg 1$ that we are considering here) is given by \eq{Shastroavenev} which, for large $\nev$, can also be approximated by \eq{eq:Shsum}. Using for simplicity the latter expression, the energy density per logarithmic interval of frequency, normalized again to $\rho_c$, is  then given by
\be\label{eq:Omegagw_tot_def}
\omgwa(f)=\dfrac{4\pi^2}{3H_0^2} \dfrac{f^3}{T}\sum_{i=1}^{\bnev} \[ \, |\tilde{h}_{+,i}(f)|^2 + |\tilde{h}_{\times,i}(f)|^2\]\, .
\ee
Observe that, in the limit of large observation time, this sum converges to a finite quantity. This can be better seen rewriting \eq{eq:Omegagw_tot_def} as
\be\label{eq:Omegagw_tot_rate}
\omgwa(f)=\dfrac{4\pi^2}{3H_0^2}\, f^3 \dfrac{\bnev}{T}\frac{1}{\bnev}\sum_{i=1}^{\bnev} \[ \, |\tilde{h}_{+,i}(f)|^2 + |\tilde{h}_{\times,i}(f)|^2\]\, .
\ee
In the large $T$ limit, $\bnev/T$ converges to the merger rate, while the term written after it is the average of 
$|\tilde{h}_{+,i}(f)|^2 + |\tilde{h}_{\times,i}(f)|^2$ over the population of events.\footnote{Despite the fact that \eq{eq:Shsum}, or \eq{eq:Omegagw_tot_def},
have  been often used in the literature (e.g.~\cite{Regimbau:2011rp,Regimbau:2012ir,Meacher:2014aca,Meacher:2015iua,Pan:2023naq}),  we have not been able to find  a  convincing derivation of it; typically, either the result is just stated, or is attributed to references that, in reality, do not contain any derivation  but just restate it, or a chain of citations eventually leads to Phinney's paper~\cite{Phinney:2001di}; this paper is the closest reference for this result but, in fact,  even this reference  does not  contain (and does not claim) a derivation of \eqs{eq:Shsum}{eq:Omegagw_tot_def}. The final result for $\omgw(f)$
in~\cite{Phinney:2001di} [his eq.~(32) combined with his eq.~(26)] is 
\be\label{result:Phinney}
\omgw(f)=\frac{4\pi^2}{3 H_0^2} c f^3 \int_0^{\infty}dz\, N(z)4\pi d_{\rm com}^2(z)\,
\langle  |\tilde{h}_{+}(f)|^2 + |\tilde{h}_{\times}(f)|^2\rangle_{\Omega_s}\, ,
\ee
where $d_{\rm com}(z)$ is the comoving distance (denoted $d_M(z)$ in \cite{Phinney:2001di}), $N(z)dz$ is the number of events per comoving volume which occur between redshift $z$ and $z+dz$, and $\langle  ... \rangle_{\Omega_s}$ denotes an angular average over the sources at redshift $z$. This expression does not even involve the observation time $T$ (and is derived without any reference to it), so its relation to \eq{eq:Omegagw_tot_def} is still quite implicit. Actually, from \eq{eq:Omegagw_tot_def} one could derive \eq{result:Phinney} by taking the limit $T\ra\infty$ (see \cite{Renzini:2024pxt} for a recent derivation). However, the inverse process, of deriving an expression for $\omgw(f)$ (or for the spectral density), for a finite observation time $T$ from its expression in the limit $T\ra\infty$, can only be heuristic, as many possible finite-$T$ expressions could have the same limit as $T\ra\infty$. The actual derivation, for finite observation time $T$, is the one that we have presented here, where $T$ enters from the very beginning in the form of the averages (\ref{allaverages}).
As one can appreciate from the steps that we have performed, this derivation involved several non-trivial conceptual and technical points,  that we attempted to make as explicit as possible. Observe that our formalism is applicable also when \eq{cond12} is not satisfied, and shows that in this general case the spectral density $S_h(f)$ must be replaced by a function  $S_h(f,f')$ whose  exact expression is given by \eq{Shfinal1pp}; as we  see from \eq{hhgeneralpp},
the other  modification for finite $T$ is that, in  \eq{ave}, the Dirac delta $\delta(f-f')$  is replaced by $\delta_T(f-f')$, which is given by \eq{defdeltaT}. This is not just an ``ad hoc'' regularization of the Dirac delta, but, as we have seen,  is the specific form of a regularized Dirac delta that emerges  from computing the integrals over the arrival times in \eq{ave_t_and_psi}.\label{foot:comparisonPhinney}}

\section{Energy density of anisotropic and polarized  astrophysical backgrounds}\label{sect:energy}

It is interesting to generalize the above computation of the energy density to a generic   anisotropic and polarized  background. This can be relevant both to cosmological backgrounds and to the astrophysical background, where  anisotropies, at some level, will be present; furthermore, even if in the previous  section we have shown that the average over extrinsic parameter sets to zero the polarization of the CBC background,  we will see in \cref{sect:shot} that some amount of polarization  can be generated by shot noise.

We work directly in the limit $T\ra\infty$ where the notion of GW energy density is rigorously defined.
The computation can be then performed writing the correlator as  in \eq{Stokes1}, with
a generic function 
$H_{AA'}(f,\hatn)$ (without even assuming a factorized form). 
Inserting \eq{snrhab} into \eq{rhogwaveU} we get
\bees
&&\hspace*{-12mm}\rho_{\rm gw}=\frac{c^2}{32\pi G}\,
\int_{-\infty}^{\infty} dfdf' \int_{S^2} d^2\hatn d^2\hatn'\,\sum_{A,A'=+,\times}
\nn\\
&&\hspace*{-2mm}\times  (2\pi i f) (-2\pi i f')\,
\langle \hti^*_A(f,\hatn)\hti_{A'}(f',\hatn')\rangle_U\,  e^A_{ab}(\hatn)e^{A'}_{ab}(\hatn')\, 
e^{2\pi i (f-f')t } e^{-2\pi i (f\hatn-f'\hatn')\cdot\vx /c }\nn\\
&&\hspace*{-5.8mm}= \frac{\pi c^2}{8 G}\,
\int_{0}^{\infty} df f^2 \int_{S^2} \frac{d^2\hatn}{4\pi}\,\sum_{A,A'=+,\times} H_{AA'}(f,\hatn)e^A_{ab}(\hatn)e^{A'}_{ab}(\hatn)
\, ,\label{derivazene1}
\ees
where we have written the integral from $f=-\infty$ to $f=\infty$
as twice an integral from $f=0$ to $f=\infty$
using  $\hti^*_A(f,\hatn)=\hti_A(-f,\hatn)$ [which follows from the reality of $h_{ab}(t)$ and of the polarization tensors $e^+_{ab}$, $e^{\times}_{ab}$, see \eq{eplusecross}].

The crucial point now is that, thanks to the Dirac delta $\delta(\hatn,\hatn')$ in \eq{Stokes1}, we recovered the structure $e^A_{ab}(\hatn)e^{A'}_{ab}(\hatn)$, in which the arguments in the two polarization tensors are the same. We can then use
\eq{normaee} (recall that the sum over the repeated spatial indices $a,b$ is understood), and then 
\bees
H_{AA'}(f,\hatn)e^A_{ab}(\hatn)e^{A'}_{ab}(\hatn)&=&
2\delta_{AA'}H_{AA'}(f,\hatn)\nn\\
&=&4 I(f,\hatn)\, ,
\ees
where, in the last line, we wrote $H_{AA'}(f,\hatn)$ as in 
\eq{Stokes2}, and used the fact that the Pauli matrices  are traceless.   

Therefore the Stokes parameters associated with linear and with circular polarization do not contribute to the energy density of a stationary stochastic background (independently of whether it is isotropic or anisotropic),
 and
\be\label{rhogwIfhatn}
\rho_{\rm gw}=\frac{\pi c^2}{2 G}\,
\int_{0}^{\infty} d\log f\, f^3 \int_{S^2} \frac{d^2\hatn}{4\pi}\,  I(f,\hatn)\, ,
\ee
so that
\bees
\frac{d\rho_{\rm gw}}{d\log fd^2\hatn}&=&\frac{\pi c^2}{2G}\, 
\frac{f^3 I(f,\hatn)}{4\pi}\nn\\
&=&\frac{c^2}{8G}\, f^3 I(f,\hatn)\label{drhodlofdn}
\, .
\ees
We can also define
\be\label{Shd2nI}
S_h(f)\equiv I(f)\equiv  \int_{S^2} \frac{d^2\hatn}{4\pi}\,I(f,\hatn)\, ,
\ee
so that  we recover the standard expressions 
\be
\rho_{\rm gw}=\frac{\pi c^2}{2 G}\,
\int_{0}^{\infty} d\log f\,  f^3 S_h(f)\, ,
\ee
and
\bees
\frac{d\rho_{\rm gw}}{d\log f}&=&\frac{\pi c^2}{2 G}\,f^3 S_h(f)\nn\\
&=&\frac{\pi c^2}{2 G}\,f^3  \int_{S^2} \frac{d^2\hatn}{4\pi}\,I(f,\hatn)
\, .\label{drhogwintI}
\ees
Therefore, we can write the result in any of the equivalent forms
\bees
\omgw(f) &=& \frac{4\pi^2}{3H_0^2}\, f^3S_h(f)\nn\\
&=& \frac{4\pi^2}{3H_0^2}\, f^3\, I(f)\nn\\
&=& \frac{4\pi^2}{3H_0^2}\, f^3\, \int_{S^2} \frac{d^2\hatn}{4\pi}\,I(f,\hatn)
\, .\label{OgwintI}
\ees
In the literature, quantities  analogous to 
$\omgw(f)$ for the Stokes parameters associated with polarization have  sometimes been defined. For instance, refs.~\cite{ValbusaDallArmi:2023ydl,Callister:2023tws}, together with the definition $\Omega^I=[(4\pi^2f^3)/(3H_0^2) ] I(f)$, also define a quantity $\Omega^V=[(4\pi^2f^3)/(3H_0^2) ] V(f)$ associated with circular polarization. One must be careful, however, with the interpretation of the latter quantity. As we showed here, there is no energy density associated with the $V$, $U$, and $Q$ Stokes parameters. Only the spectral density associated with $I$ contributes to the energy density. 
Of course, one can always multiply any spectral density, or any Stokes parameter, by $[(4\pi^2f^3)/(3H_0^2)] $, in analogy to 
\eq{eq:OmegaSh}. This can sometimes be useful because, while any spectral density $S_h(f)$ has dimensions of $1/f$, and in the GW context typically has numerical values which are not very natural, such as $10^{-44}\, {\rm Hz}^{-1}$,  the  quantity
$[(4\pi^2f^3)/(3H_0^2)]S_h(f)$ is dimensionless and is naturally compared with the actual physical energy density per unit logarithmic frequency (normalized to $\rho_c$) of a GW background. 
The same can be done for the spectral density of the noise $S_n(f)$ (defined from the noise-noise correlator),  from which it can be useful to form the dimensionless combination $[(4\pi^2f^3)/(3H_0^2)]S_n(f)$.  However, again, there is no physical energy density associated with this definition. One could simply work at the level of spectral densities, comparing e.g.  the spectral density  of the signal, $[S_h^{\rm astro}(f)]_{AA'}$, with the  spectral density of the noise, $S_n(f)$, or  comparing among them the contributions to $[S_h^{\rm astro}(f)]_{AA'}$ associated with the different Stokes parameters. Again, one can, of course,  multiply all these spectral densities by 
$[(4\pi^2f^3)/(3H_0^2)] $ to deal with dimensionless quantities, but one must then  be aware that only the spectral density associated with the $I$ Stokes parameter has an interpretation in terms of an energy density, and there is no energy density associated with the other Stokes parameters, or to noise.

\section{Effect of shot noise}\label{sect:shot}

In the previous section we have computed explicitly the averages over extrinsic parameters. However,  when we deal with a finite ensemble of $\nev$ events collected in an observation time $T$, corresponding to a given realization of the underlying stochastic process, these formal averages will not represent the actually observed quantities. For instance, we will not have an infinite number of sources with polarization angle $\psi$ distributed uniformly in $[0,2\pi]$, but rather a specific sample of sources, with polarization angles $\psi_1,\ldots ,\psi_{\nev}$; then, quantities that would vanish in a formal average over $\psi$ computed as in \eq{ave_psii}, will  in general be non-vanishing because the cancellations will only be partial, reflecting the discreteness of the underlying stochastic process, i.e. shot noise.

In this section, therefore, we no longer perform analytically the averages over the extrinsic parameters,  and we instead work directly with the original sums over discrete events, with the aim of extracting, in the large $\nev$ limit, the difference between the discrete and continuous average. We will then eventually compute the results numerically, on a given specific realization of events.
The formulas below will be written for a generic number of events  $\nev$. Typically, we will be interested in the situation where $\nev$ is the number of events, in the given observation time $T$, that were below the detection threshold, since the actual astrophysical stochastic background is made by the superposition of the unresolved signals. However, the results below are general, and could also be applied to the sum of resolved and unresolved events.

For the astrophysical stochastic background, there are two situations where the effect of shot noise can be especially important. One is the computation of the polarization of the background: we have seen that linear polarizations are set to zero by the average over the polarization angles $\psi_i$, while circular polarization survives this average but is set to zero by the average over  $\cos\iota_i$.
The second situation concerns the angular distribution of an astrophysical background. Even  if the background is assumed to be  isotropic, anisotropies will be generated by shot noise. In both cases, these shot noise contributions could mask very interesting effects, such as a polarized cosmological background that might be generated by parity-violation mechanism in the early Universe~\cite{Alexander:2004us,Seto:2008sr,Crowder:2012ik,Romano:2016dpx,Yagi:2017zhb,Domcke:2019zls,Martinovic:2021hzy,Callister:2023tws,Cruz:2024esk}, or actual anisotropies of the astrophysical background that  provide  information on the clustering of the  matter distribution~\cite{Contaldi:2016koz,Cusin:2017fwz,Cusin:2017mjm,Jenkins:2018lvb,Cusin:2018rsq,Jenkins:2018uac,Jenkins:2018kxc,Jenkins:2019uzp,Cusin:2019jhg,Cusin:2019jpv,Jenkins:2019nks,Bertacca:2019fnt,Pitrou:2019rjz,Bellomo:2021mer,Capurri:2021zli}. It is therefore important to evaluate these shot-noise effects. The shot noise contribution to the polarization  has recently been investigated in ref.~\cite{ValbusaDallArmi:2023ydl} (restricting, however, to circular polarization), while
the anisotropies of the astrophysical background induced by shot noise, and their effect on GW observations, have been discussed 
in ref.~\cite{Jenkins:2019uzp,Kouvatsos:2023bgd} (see also \cite{Meacher:2014aca}).

The formalism that we have developed in \cref{sect:compSastro} is well-suited to evaluate the shot noise contribution to the spectral density of the astrophysical background.  Below, we will  show how to evaluate these effects with our formalism. In this paper, we focus on the methodological aspects, and we will illustrate our 
results by summing over the contribution of all the sources that merge in a given time span (we will assume one year). However, at a specific detector network, some sources will be resolved and some will not; the resolved sources can then be subtracted, and the actual residual contribution will be due only to the unresolved sources, plus the accumulation of errors on the reconstruction of resolved sources. We defer a detailed study at specific 3G detector networks to subsequent work.

\subsection{Stokes parameters as discrete sum over sources}\label{sect:Stokesdiscrete}

Our starting point is given by \eq{tildehdeltan}, from which it follows that
\be\tilde{h}_A^*(f,\hatn)\tilde{h}_{A'}(f',\hatn')=\sum_{i,j=1}^{\nev} \tilde{h}_{A,i}^*(f;\psi_i,t_i,\cos\iota_i)\tilde{h}_{A',j}(f';\psi_j,t_j,\cos\iota_j) \delta(\hatn,\hatn_i)\delta(\hatn',\hatn_j) \,,\label{starting_shot}
\ee
which is \eq{eq: avg ext sum i j} before performing on it any average. Actually, the full dependence of $\tilde{h}_A(f,\hatn)$ and $\tilde{h}_{A'}(f',\hatn')$ on the extrinsic parameters is as in \eq{hAfull}, but on the left-hand side we do not write it explicitly these dependence, in order not to burden excessively the notation.

We are not interested in the shot noise effect on the statistics of the  times of arrival:  it is difficult to imagine physical mechanism that, on the timescale $T$ of the observation, would correlate the times of arrival of different CBCs, so the statistics of the times of arrival is just that of independent events, and does not carry specific astrophysical information. 
We therefore still perform the corresponding average as in \eq{ave_ti}. Then, we write 
\bees
\langle\tilde{h}_A^*(f,\hatn)\tilde{h}_{A'}(f',\hatn')\rangle_{ \{t_k\} }&=&\sum_{i,j=1}^{\nev} \langle \tilde{h}_{A,i}^*(f;\psi_i,t_i,\cos\iota_i)\tilde{h}_{A',j}(f';\psi_j,t_j,\cos\iota_j)\rangle_{ \{ t_k\} } \nn\\
&&\hspace*{5mm} \times  \delta(\hatn,\hatn_i)\delta(\hatn',\hatn_j) \,.
\label{eqshot1}
\ees
In the following we will restrict to the situation in which the condition (\ref{cond12bis}) is satisfied.\footnote{The details of the computation when  the condition (\ref{cond12bis}) is  satisfied will be presented in app.~\ref{app:extractpsi}, while
the most general results are discussed in app.~\ref{app:shotgeneral}.}
Then,  as discussed in  \cref{foot:decorij_ti},
the average over arrival times
produces a factor $(1/T)\delta_T(f-f')$, as well as a factor $\delta_{ij}$ 
and, from the discussion in \cref{sect:arrivals}, we can set $f'=f$ in the term multiplied by  $\delta_T(f-f')$. Then, 
\be\langle\tilde{h}_A^*(f,\hatn)\tilde{h}_{A'}(f',\hatn')
\rangle_{ \{t_k\} }=\delta(\hatn,\hatn')\delta_T(f-f')
\frac{1}{T}\sum_{i=1}^{\nev} \tilde{h}_{A,i}^*(f;\psi_i,\cos\iota_i)\tilde{h}_{A',i}(f;\psi_i,\cos\iota_i)\delta(\hatn,\hatn_i)\,.
\label{eqshot2}
\ee
This expression will be our starting point for evaluating the effect of shot noise, either on the polarization or on the anisotropies.

\subsubsection{Dependence on the polarization angles}\label{sect:deppsi}

We begin by extracting explicitly the dependence on the polarization angles $\psi_i$. We give here the main steps,  referring to app.~\ref{app:extractpsi} for full details.
Using \eq{eq: psi rot}, we have
\be\label{hhMAABB}
\tilde{h}_{A,i}^*(f;\psi_i,\cos\iota_i)\tilde{h}_{A',i}(f;\psi_i,\cos\iota_i)=M_{AA',BB'}(\psi_i)
 \tilde{h}_{B,i}^*(f;\psi_i=0,\cos\iota_i)\tilde{h}_{B',i}(f;\psi_i=0,\cos\iota_i)
\, ,
\ee
where
$M_{AA',BB'}(\psi_i)=R_{AB}(2\psi_i)R_{A'B'}(2\psi_i)$. This matrix is computed explicitly in terms of Pauli matrices in  app.~\ref{app:extractpsi}, and inserting it into \eq{hhMAABB} and then in \eq{eqshot2} we get
\bees
&&\hspace*{-5mm}\langle \tilde{h}_A^*(f,\hatn)
\tilde{h}_{A'}(f',\hatn')\rangle_{ \{t_k\} }  =\delta(\hatn,\hatn')\delta_T(f-f') \frac{1}{2}
\sum_{i=1}^{\nev}\delta(\hatn,\hatn_i)\label{eqshot4}
\\
&&\hspace*{-0mm}\times  
 \Big[
\mathcal{I}_i(f;\cos\iota_i)\delta_{A A'}+\mathcal{U}_i(f;\cos\iota_i,\psi_i)\sigma^{1}_{AA'}+\mathcal{V}_i(f;\cos\iota_i)\sigma^{2}_{AA'}
+\mathcal{Q}_i(f;\cos\iota_i,\psi_i)\sigma^{3}_{AA'}
 \Big] \, ,\nn
\ees
where
\bees
&&\mathcal{I}_i(f;\cos\iota_i) =\frac{1}{T}  \sum_{B=+,\times} |\tilde{h}_{B,i}(f;\psi_i=0,\cos\iota_i)|^2\, ,\label{expressionI}\\
&&\mathcal{V}_i(f;\cos\iota_i) =\frac{1}{T}\sum_{B,B'=+,\times} \label{expressionV} \[- \tilde{h}^*_{B,i}(f;\psi_i=0,\cos\iota_i) \sigma^2_{BB'}\tilde{h}_{B',i}(f;\psi_i=0,\cos\iota_i)\]\, ,\\
&&\mathcal{U}_i(f;\cos\iota_i,\psi_i) \label{expressionU}\\
&&\hspace{3mm}=\frac{1}{T}\hspace*{-1mm}\sum_{B,B'=+,\times} \hspace*{-1mm} \tilde{h}^*_{B,i}(f;\psi_i=0,\cos\iota_i)\[\sigma^1_{BB'}\cos 4\psi_i+\sigma^3_{BB'}\sin 4\psi_i\] \tilde{h}_{B',i}(f;\psi_i=0,\cos\iota_i)\, ,\nn \\
&&\mathcal{Q}_i(f;\cos\iota_i,\psi_i)\label{expressionQ}\\
&&\hspace{3mm}=\frac{1}{T}\hspace*{-1mm}\sum_{B,B'=+,\times} \hspace*{-1mm} \tilde{h}^*_{B,i}(f;\psi_i=0,\cos\iota_i) 
\[-\sigma^1_{BB'}\sin 4\psi_i+\sigma^3_{BB'}\cos 4\psi_i\]\tilde{h}_{B',i}(f;\psi_i=0,\cos\iota_i)\, .\nn 
\ees
As we  discussed below \eq{hLRhelicity2},
$\mathcal{I}_i$ and $\mathcal{V}_i$ are independent of the reference value $\psi_i=0$, so we can replace the argument $\psi_i=0$ in \eqs{expressionI}{expressionV} by an arbitrary value $\psi_i$, or simply suppress the argument $\psi_i$, as we will do below.
In contrast, we see from \eqs{expressionU}{expressionQ} that
$\mathcal{U}_i$ and $\mathcal{Q}_i$ depend on $\psi_i$ and, under a transformation $\psi_i\ra\psi_i+\psi_0$,  transform linearly among them, as
\bees
\mathcal{U}_i &\ra& \phantom{+}\cos(4\psi_0) \mathcal{U}_i+\sin(4\psi_0) \mathcal{Q}_i\, ,
\label{transfUi}\\
\mathcal{Q}_i  &\ra& -\sin(4\psi_0) \mathcal{U}_i+\cos(4\psi_0) \mathcal{Q}_i\, ,
\label{transfQi}
\ees
so that 
\bees
\mathcal{Q}_i + i\mathcal{U}_i &\ra& e^{i4\psi_0} (\mathcal{Q}_i +i\mathcal{U}_i)\label{transfUQ1}\, , \\
\mathcal{Q}_i - i\mathcal{U}_i &\ra& e^{-i4\psi_0} (\mathcal{Q}_i -i\mathcal{U}_i)\, .\label{transfUQ2}
\ees
Therefore, while $\mathcal{I}_i $ and $\mathcal{V}_i $ transform as helicity-0 fields under rotations in the plane transverse to $\hatn_i$, $\mathcal{U}_i$ and $\mathcal{Q}_i $ transform as helicity-4 fields~\cite{Seto:2008sr,Romano:2016dpx,Conneely:2018wis}. 

So, in conclusion, \eq{eqshot4} can be written as 
\be
\langle
\tilde{h}^*_{A}(f,\hatn)\ \tilde{h}_{A'}(f',\hatn')\rangle_{ \{t_k\} }=\frac{\delta(\hatn,\hatn')}{4\pi}\delta_T(f-f')\frac{1}{2}
[S_h^{\rm astro}(f; \hatn)]_{AA'}\, ,
\ee
where 
\be\label{ShastroAAp}
[S_h^{\rm astro}(f; \hatn)]_{AA'}= I(f; \hatn)\delta_{AA'}+V(f; \hatn)\sigma^{2}_{A A'}+U(f; \hatn)\sigma^{1}_{A A'}+Q(f; \hatn)\sigma^{3}_{A A'}\, ,
\ee
and, writing explicitly  also the dependence on the set of inclination angles $\cos\iota_i$, propagation directions $\hatn_i$ and polarization angles $\psi_i$ of the events (with 
$\{\cos\iota_k, \hatn_k,\psi_k\}$ denoting the ensemble of all $\cos\iota_k, \hatn_k,\psi_k$ for $k=1,\ldots ,\nev$),
\bees
I(f,\hatn;\{\cos\iota_k, \hatn_k\})&=&4\pi\sum_{i=1}^{\nev}\mathcal{I}_i(f;\cos\iota_i)\,  \delta(\hatn,\hatn_i)\, ,
\label{Iastro_shot}\\
V(f,\hatn;\{\cos\iota_k, \hatn_k\})&=&4\pi \sum_{i=1}^{\nev}\mathcal{V}_i(f;\cos\iota_i)\, \delta(\hatn,\hatn_i) \, ,\label{Vastro_shot}\\
U(f,\hatn;\{\cos\iota_k, \hatn_k,\psi_k\})&=&4\pi \sum_{i=1}^{\nev}\mathcal{U}_i(f;\cos\iota_i,\psi_i)\, \delta(\hatn,\hatn_i)\, ,\label{Uastro_shot}\\
Q(f,\hatn;\{\cos\iota_k, \hatn_k,\psi_k\})&=&4\pi \sum_{i=1}^{\nev}\mathcal{Q}_i(f;\cos\iota_i,\psi_i)\, \delta(\hatn,\hatn_i)\, ,\label{Qastro_shot}
\ees
with $\mathcal{I}_i(f;\cos\iota_i), \ldots , \mathcal{Q}_i(f;\cos\iota_i)$  given in 
\eqst{expressionI}{expressionQ} or , in an equivalent form in terms of the left-handed and right-handed fields $h_L$ and $h_R$, in \eqst{expressionILRapp}{expressionQLRapp}

From \eqs{expressionU}{expressionQ} we see that $\mathcal{U}_i$ and $\mathcal{Q}_i$ are given by  linear combinations of $ \cos 4\psi_i$ and $ \sin 4\psi_i$.
Therefore, if we now perform an analytic average over polarization angles, $\mathcal{U}_i$ and $\mathcal{Q}_i$ vanish, since 
\be\label{avecos4psi}
\langle\cos 4\psi_i \rangle_{\psi_i}=\int_0^{2\pi}\frac{d\psi_i}{2\pi} \cos 4\psi_i=0\, ,
\ee
and similarly for the average of $\sin 4\psi_i$, and we therefore recover the result of section~\ref{averagesoverpsi} that linear polarizations vanish upon averaging over the angles $\psi_i$. However, in the form of a discrete sum over sources, they are in general non-vanishing, i.e. shot noise generates a linear polarization.

Let us now recall, from the discussion below \eq{allaverages}, that in order to define the $\psi_i$ of an ensemble of sources, we must set up a given global reference frame on the sphere. The values $\psi_i$ are then defined with respect to this frame, and change as 
\be
\psi_i\ra \psi_i+\psi_0(\hatn_i)
\ee
when the axes in the transverse plane are rotated as in \eqs{hatup}{hatvp}. The numerical values of $U$ and $Q$ therefore depends on this choice. There is nothing wrong with it; simply, when making predictions for these quantities, one must make a choice for this global reference frame on the sphere (with that in \eqs{u_Om}{v_Om} being a natural one), and the same  choice must be used consistently when comparing  to observations. However, to somehow reduce the dependence on this choice of frame, one can observe that
under the ``global'' version of the transformation \eqs{hatup}{hatvp}, where  $\psi_0$ is independent of $\hatn$, 
\be\label{psiipsiipsi0}
\psi_i\ra\psi_i+\psi_0\, ,
\ee
all $\mathcal{U}_i$ and  $\mathcal{Q}_i $ transform among them as in 
\eqs{transfUi}{transfQi} and, since $\psi_0$ is the same for all the events, also $U$ and $Q$ transform in the same way, i.e.
\bees
U&\ra& \phantom{+}\cos(4\psi_0) U+\sin(4\psi_0) Q\, ,
\label{transfUtot}\\
Q&\ra& -\sin(4\psi_0) U+\cos(4\psi_0) Q\, .
\label{transfQtot}
\ees
It is therefore convenient to introduce
\be\label{defmathcalP}
P= \[ U^2+Q^2\]^{1/2}\, ,
\ee
which is invariant under the transformation (\ref{psiipsiipsi0}).

\subsubsection{Expansion in spherical harmonics}\label{expaspherharm}

We next discuss the angular dependence of these discrete distributions.
As with any  function of $\hatn$, scalar under spatial rotations, we can expand  $\delta(\hatn,\hatn_i)$  in spherical harmonics,
\be\label{deltaYlm}
\delta(\hatn,\hatn_i)=\sum_{l=0}^{\infty}\sum_{m=-l}^l \delta_{lm}(\hatn_i) Y_{lm}(\hatn)\, ,
\ee
with some coefficients $\delta_{lm}(\hatn_i)$. These are determined inverting \eq{deltaYlm},
\bees\label{invertdlm}
\delta_{lm}(\hatn_i)&=&\int_{S^2} d^2\hatn\, \delta(\hatn,\hatn_i)Y^*_{lm}(\hatn)\nn\\
&=& Y_{lm}^*(\hatn_i)\,  ,
\ees
so that
\be
\delta(\hatn,\hatn_i)=\sum_{l=0}^{\infty}\sum_{m=-l}^l  Y_{lm}^*(\hatn_i)Y_{lm}(\hatn)\, ,
\ee
which is just the completeness relation of spherical harmonics. Using the addition theorem for spherical harmonics,
\be\label{YYaddi}
\sum_{m=-l}^l\, Y_{lm}(\hatn)Y^*_{lm}(\hatn')=\frac{2l+1}{4\pi} P_l(\hatn\bdot\hatn')\, ,
\ee
where  $P_l(\cos\theta)$ are the  Legendre polynomials, we can also write
\be\label{deltaPl}
\delta(\hatn,\hatn_i)=\frac{1}{4\pi}\sum_{l=0}^{\infty}(2l+1)P_l(\hatn\bdot\hatn_i)\, .
\ee
Then, from \eq{Iastro_shot},
\be\label{expaIlm}
I(f,\hatn;\{\cos\iota_k, \hatn_k\})=
 \sum_{l=0}^{\infty}\sum_{m=-l}^l I_{lm}(f;\{\cos\iota_k, \hatn_k\})Y_{lm}(\hatn)\, ,
 \ee
where
\be\label{invertIlm}
I_{lm}(f;\{\cos\iota_k, \hatn_k\})=4\pi \sum_{i=1}^{\nev} 
\mathcal{I}_i(f;\cos\iota_i) Y_{lm}^*(\hatn_i)\,  ,
\ee
or, equivalently,
\be\label{Iexpalm2}
I(f,\hatn;\{\cos\iota_k, \hatn_k\})=\sum_{l=0}^{\infty} (2l+1) \sum_{i=1}^{\nev} \mathcal{I}_i(f;\cos\iota_i) 
 P_l(\hatn_i\bdot\hatn)
\, .
\ee
Similarly, from \eq{Vastro_shot},
\be\label{expaVlm}
V(f,\hatn;\{\cos\iota_k, \hatn_k\})=
 \sum_{l=0}^{\infty}\sum_{m=-l}^l V_{lm}(f;\{\cos\iota_k, \hatn_k\})Y_{lm}(\hatn)\, ,
 \ee
with 
\be\label{invertVlm}
V_{lm}(f;\{\cos\iota_k, \hatn_k\})=4\pi
\sum_{i=1}^{\nev} 
\mathcal{V}_i(f;\cos\iota_i) Y_{lm}^*(\hatn_i)\,  ,
\ee
or, equivalently,
\be\label{Vexpalm2}
V(f,\hatn;\{\cos\iota_k, \hatn_k\})=\sum_{l=0}^{\infty} (2l+1) \sum_{i=1}^{\nev} \mathcal{V}_i(f;\cos\iota_i) 
 P_l(\hatn_i\bdot\hatn)
\, .
\ee
The same holds for $Q$ and $U$, using \eqs{expressionU}{expressionQ}.
 Because of \eqs{transfUQ1}{transfUQ2}, it is  convenient to work in terms of 
\be\label{defQpm}
Q^{\pm}\equiv Q\pm iU\, .
\ee
Then
\be\label{expaQpm}
Q^{\pm}(f,\hatn;\{\cos\iota_k, \hatn_k,\psi_k\})=
 \sum_{l=0}^{\infty}\sum_{m=-l}^l Q^{\pm}_{lm}(f;\{\cos\iota_k, \hatn_k,\psi_k\})Y_{lm}(\hatn)\, ,
\ee
which is inverted as
\be\label{invertQpm}
Q^{\pm}_{lm}(f;\{\cos\iota_k, \hatn_k\,\psi_k\})=4\pi
\sum_{i=1}^{\nev} 
\mathcal{Q}^{\pm}_i(f;\cos\iota_i,\psi_i) Y_{lm}^*(\hatn_i)\,  ,
\ee
where, from \eqs{expressionU}{expressionQ},
\be\label{expressionQpm}
\mathcal{Q}^{\pm}_i(f;\cos\iota_i,\psi_i)
=
\frac{e^{\pm i4\psi_i}}{T}\sum_{B,B'=+,\times} \hspace*{-1mm} \tilde{h}^*_{B,i}(f;\psi_i=0,\cos\iota_i) 
(\sigma^3\pm i\sigma^1)_{BB'}\tilde{h}_{B',i}(f;\psi_i=0,\cos\iota_i)\, . 
\ee
Similarly to \eqs{Iexpalm2}{Vexpalm2}, we then have
\be\label{Qpmexpalm2}
Q^{\pm}(f,\hatn;\{\cos\iota_k, \hatn_k,\psi_k\})=\sum_{l=0}^{\infty} (2l+1) \sum_{i=1}^{\nev} \mathcal{Q}^{\pm}_i(f;\cos\iota_i,\psi_i) 
 P_l(\hatn_i\bdot\hatn)
\, .
\ee
We conclude that, in general, shot noise can generate both circular and linear polarizations. 
For circular polarization, the most general shot-noise result is given by \eq{Vastro_shot} or, equivalently, by \eq{Vexpalm2}, with 
$\mathcal{V}_i(f;\cos\iota_i) $ given in \eq{expressionV}.
For linear polarizations, we have seen that there is a further dependence  on the polarization angles $\psi_i$. The result obtained from a full discrete sum over polarizations, arrival directions and orbit orientations is given by \eqs{Uastro_shot}{Qastro_shot}, or equivalently by
\eq{invertQpm}, 
with  $\mathcal{Q}^{\pm}_i(f;\cos\iota_i,\psi_i)$ given by \eq{expressionQpm}, and is again non-vanishing.
In the literature it has been argued that, for an isotropic source distribution,  linear polarization is strictly zero~\cite{Seto:2008sr,Romano:2016dpx}. The case of an isotropic source distribution corresponds to keeping only the term $l=0$ in \eq{Qpmexpalm2}, since the index $l$ in this expression is related to the  multipole expansion of the source distribution,\footnote{More precisely, we see from \eq{deltaPl} that retaining only the term $l=0$ is equivalent to replacing $\delta(\hatn,\hatn_i)$ by
$1/(4\pi)$, i.e. smearing uniformly over the sphere the probability of having a source in a given direction.} and we see from \eq{Qpmexpalm2} that the result becomes
\be\label{Qpmlzero}
Q^{\pm}(f;\{\cos\iota_i, \hatn_i,\psi_i\})_{l=0}=
\sum_{i=1}^{\nev} \mathcal{Q}^{\pm}_i(f;\cos\iota_i,\psi_i) \, ,
\ee
and is in general non-vanishing, as we will also check numerically in section~\ref{sect:numresStokes}. Observe that this expression could have been obtained more directly by averaging \eqs{Uastro_shot}{Qastro_shot} over $\hatn$,
\be
\int_{S^2} \frac{d^2\hatn}{4\pi}\, Q^{\pm}(f,\hatn;\{\cos\iota_k, \hatn_k,\psi_k\})=\sum_{i=1}^{\nev} \mathcal{Q}^{\pm}_i(f;\cos\iota_i,\psi_i) \, .
\ee
In app.~\ref{app:Linpol} we compare this result with the symmetry argument proposed in the literature.

\subsection{Numerical results}\label{sect:numresStokes}

The expressions  given by \eqst{Iastro_shot}{Qastro_shot},
together with \eqst{expressionI}{expressionQ}, or the corresponding expression given in section~\ref{expaspherharm} expanding in spherical harmonics, are a natural starting point for a numerical evaluation of   the Stokes parameters  generated by a given finite ensemble of sources, including the effect of shot noise. While the above formalism holds for a generic set of $\nev$ signals, in practical applications the sums will only run over the unresolved sources, since these are those that contribute to the stochastic background. Similarly to what has been discussed in the literature for the energy density (see refs.~\cite{Cutler:2005qq,Harms:2008xv,Regimbau:2016ike,Pan:2019uyn,Sachdev:2020bkk,Sharma:2020btq,Lewicki:2021kmu,Perigois:2021ovr,Zhou:2022otw,Zhou:2022nmt,Zhong:2022ylh,Pan:2023naq,Zhong:2024dss,Li:2024iua,Belgacem:2024ntv}), i.e. for the $I$ Stokes parameter, one will also have to add to this the effect due to the accumulation of errors from the reconstruction of the resolved sources.

All these aspects, however, will be very dependent on the specific detector network considered, since this determines which sources will be resolved (in a given noise realization), and the accuracy of their reconstruction. We leave a full analysis for specific 3G detector networks for future work. In this paper we are rather concerned with the methodological aspects, and we illustrate here our results computing the Stokes parameters on an ensemble of BBHs and BNSs corresponding to one year of data in our population model, independently of whether, at a given network, these sources  will be resolved or unresolved. In this sense, for a given number of events $\nev$, our results are really an upper bound on the actual contribution of the astrophysical background   to the Stokes parameters or, equivalently, corresponds to the limit in which the threshold for CBC detection is set very high, so that all $\nev$ events are deemed unresolved.

To compute the sums in \eqst{Iastro_shot}{Qastro_shot}
we use the same catalogs of BBHs and BNSs described in ref.~\cite{Branchesi:2023mws}, and from it we extract a random sample of events corresponding to one year of observation which, in our population model, corresponds to about $1\times 10^5$ BBH and $7\times 10^5$ BNS.\footnote{For the BNS case, we use the catalog featuring a Gaussian mass distribution of the sources, and adimensional tidal deformabilities computed assuming the APR equation of state~\cite{Akmal:1998cf}.}

We consider two separate cases, the generation of polarizations from shot noise, and the generation of anisotropies from shot noise.

\subsubsection{Polarization from shot noise}

To study how circular and linear polarizations can be generated by shot-noise effects we consider 
\eqss{Iexpalm2}{Vexpalm2}{Qpmexpalm2} and we restrict to the $l=0$ multipole, reflecting the fact that the underlying source distribution is expected to be isotropic. We  therefore write
\bees
I(f;\{\cos\iota_i\})&=& \sum_{i=1}^{\nev} \mathcal{I}_i(f;\cos\iota_i) \, ,\label{Iexpalm2lzero}\\
V(f;\{\cos\iota_i\})&=& \sum_{i=1}^{\nev} \mathcal{V}_i(f;\cos\iota_i) \, ,\label{Vexpalm2lzero}\\
Q^{\pm}(f;\{\cos\iota_i, \psi_i\})&=& \sum_{i=1}^{\nev} \mathcal{Q}^{\pm}_i(f;\cos\iota_i,\psi_i) \, .
\ees
A circular polarization $V$ is therefore generated by shot noise on the inclination angle, i.e. by the fact that, rather than integrating over $\cos\iota$ (which, as we have seen as in section~\ref{avecosiota}, sets $V$ to zero), we have a sum over a finite number of events with a discrete set of values $\cos\iota_i$. Linear polarizations, expressed by $Q^{\pm}$, or equivalently by $Q={\rm Re}\, Q^{+}$ and  $U={\rm Im}\, Q^{+}$, are instead generated by shot noise on both orbit inclination and polarization angle.

\begin{figure*}[t]
    \centering
    \includegraphics[width=\linewidth]{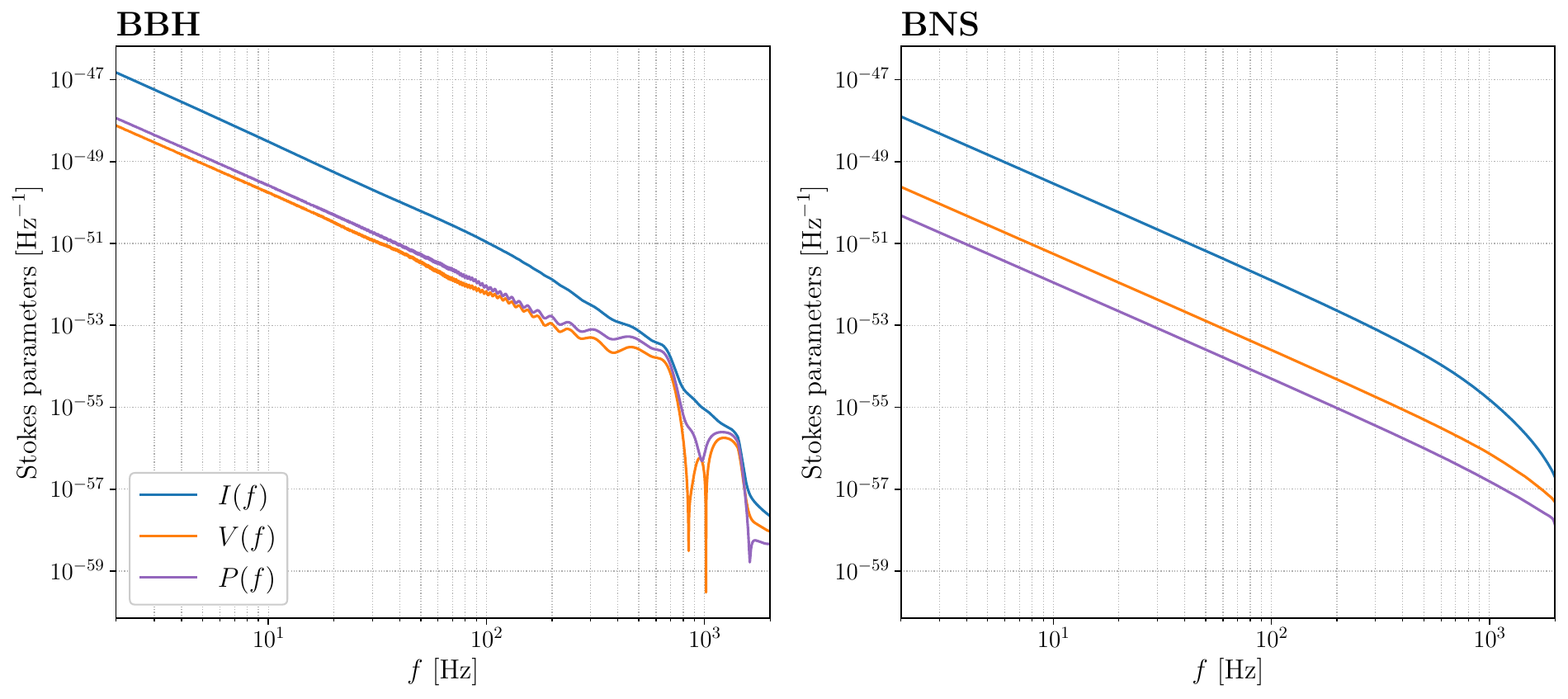}
    \caption{The quantities $I(f)$, $V(f)$  and 
    $P(f)=\sqrt{U^2(f)+Q^2(f)}$, for an ensemble of $\sim10^5$  BBHs (left panel) and $\sim7\times 10^5$  BNSs (right panel), corresponding to 1~yr of data in our population model.
    }
    \label{fig:StokesallParams}
\end{figure*}

The results are shown in 
\cref{fig:StokesallParams}, where we plot   $I(f)$, $V(f)$ and  $P(f)=\sqrt{U^2(f)+Q^2(f)}$ for our population model of BBHs (left panel) and  BNSs (right panel).\footnote{As discussed at the end of section~\ref{sect:deppsi}, the physical information about linear polarization is mostly contained in the  quantity 
$P$ defined in \eq{defmathcalP}, while the  contributions to it from $U$ and $Q$ can be rotated arbitrarily using \eqs{transfUtot}{transfQtot} so, in order not to clutter the plots,  we will show this quantity, rather than $Q$, $U$ and $P$ separately.}
We see that the polarization parameters, even if suppressed with respect to $I$, are indeed present, and follow roughly the same frequency dependence. For BBHs the linear polarization parameter $P$ is actually of the same order of magnitude as the circular polarization parameter $V$ and in fact even slightly larger. For the BNS case, the polarization terms are more suppressed with respect to $I$, which can be attributed to the larger sample of sources, and the polarization parameter $V$ is dominant with respect to the other (suggesting a potentially different scaling with the number of events). The frequency dependence of the  Stokes parameters in 
\cref{fig:StokesallParams} can be understood from the fact that, in the inspiral phase, for binary systems with negligible eccentricity,
$\tilde{h}_{A}(f)\propto f^{-7/6}$ for both polarizations, and therefore all Stokes parameters scale as $f^{-7/3}$, as indeed can be checked from the plots. As the frequency increases, more and more binaries  enter the merger phase and eventually no longer contribute to the signal, so all Stokes parameters drop. Since BNSs are lighter than BBHs, they stay longer in the inspiral phase, and correspondingly the smooth power-law behavior extends to higher frequencies.

\subsubsection{Anisotropies from shot noise}\label{sect:anishot}

\begin{figure*}[t]
    \centering
    \includegraphics[width=\linewidth]{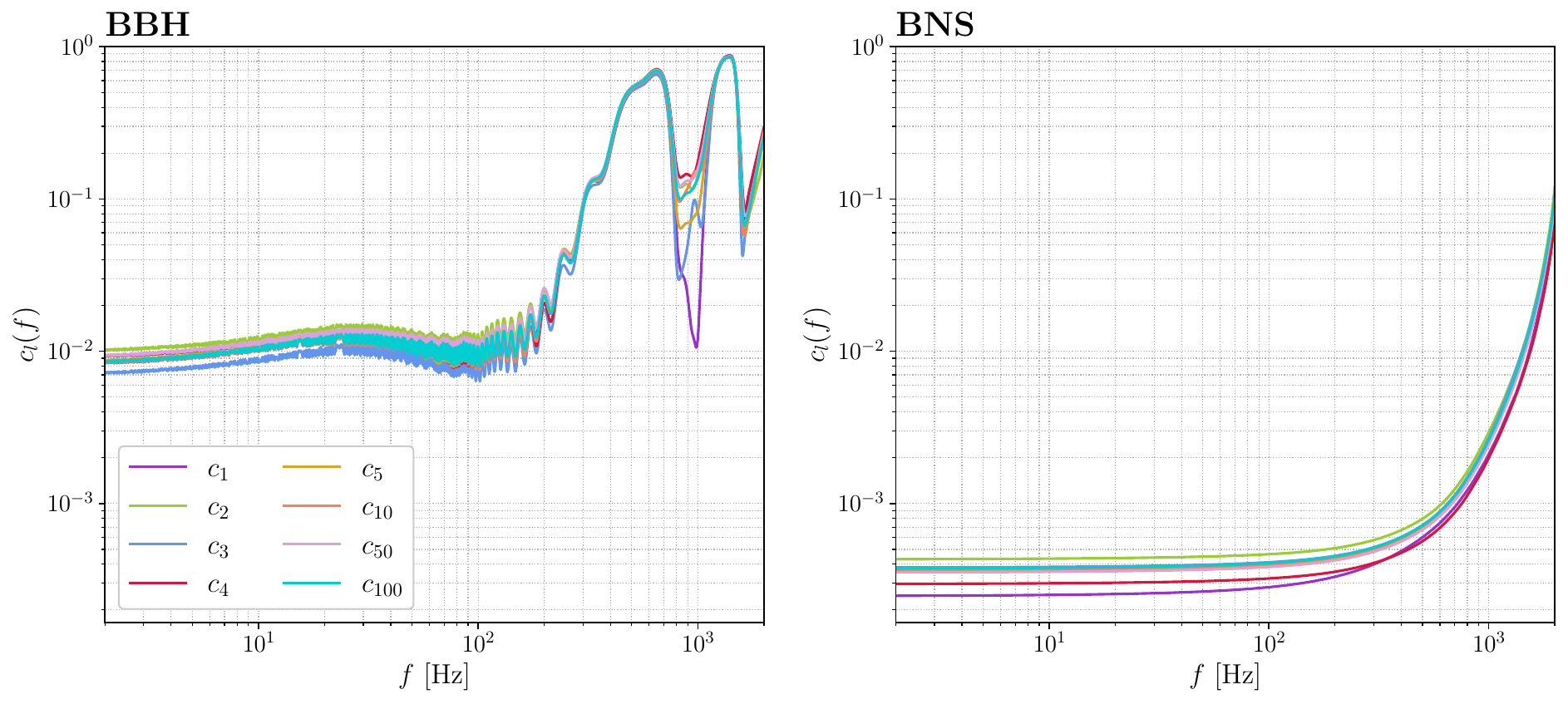}
    \caption{The  functions  $c_l(f)$ for a selection of values $l=[1,\,2,\,3,\,4,\,5,\,10,\,50,\,100]$, for an ensemble of $\sim 10^5$  BBHs (left panel) and $\sim7\times 10^5$  BNSs (right panel), corresponding to 1~yr of data in our population model.
    }
    \label{fig:Anisotropy_cls_fullPop_allParams}
\end{figure*}

We next address the effect of shot noise on the anisotropies of the angular distribution. 
These will be a foreground to the (more physically interesting)  anisotropies of the actual distribution of astrophysical sources, that can carry important information on the clustering of the underlying matter distribution. To this purpose we focus on the intensity $I$ given by \eq{invertIlm} (a similar treatment can be performed for the other Stokes parameters).
We consider the case in which the underlying distribution of sources is isotropic and we study the deviations from isotropy induced by shot noise.
In this case we can follow the standard approach used for the CMB multipoles (see, e.g., section~20.1 of \cite{Maggiore:2018sht}):
for a given value of $l$, the coefficients $I_{lm}$ with $m=-l, \ldots , l$ are statistically equivalent and all information is contained in the quantities
\be
C_l(f;\{\cos\iota_k, \hatn_k\})=\frac{1}{2l+1}\sum_{m=-l}^l |I_{lm}(f;\{\cos\iota_k, \hatn_k\})|^2\, ,
\ee
where, as in \eq{hAfull},
the brackets $\{....\}$ denotes the collection of all $ \cos\iota_k,  \hatn_k,$ with $k=1,\ldots,\nev$.
Then, using \eqs{invertIlm}{YYaddi}, we get
\be\label{Deltalnevnev}
C_l(f;\{\cos\iota_k, \hatn_k\})=4\pi
\sum_{i,j=1}^{\nev} 
\mathcal{I}_i(f;\cos\iota_i)\mathcal{I}_j(f;\cos\iota_j) P_l(\hatn_i\bdot\hatn_j)\, .
\ee
Since $P_0(\cos\theta)=1$,  we  have
\be
C_0(f;\{\cos\iota_k, \hatn_k\})=4\pi I^2(f;\{\cos\iota_k, \hatn_k\})\, .
\ee
We are interested in computing the functions $C_l(f;\{\cos\iota_k, \hatn_k\})$ for $l\neq 0$ which,  compared to 
$C_0(f;\{\cos\iota_k, \hatn_k\})$, will give a measure of the anisotropies of the astrophysical background generated by shot noise.
It is therefore convenient to define 
\bees
c_l(f;\{\cos\iota_k, \hatn_k\})&=&\frac{C_l(f;\{\cos\iota_k, \hatn_k\})}{C_0(f;\{\cos\iota_k, \hatn_k\})}\label{defclsmall}\\
&=& \frac{1}{I^2(f;\{\cos\iota_k, \hatn_k\})}\sum_{i,j=1}^{\nev} 
\mathcal{I}_i(f;\cos\iota_i)\mathcal{I}_j(f;\cos\iota_j)  P_l(\hatn_i\bdot\hatn_j)\, .\nn
\ees
We have computed numerically these quantities for our distribution of BBHs and of BNSs, using again one year of data and taking all sources as unresolved. As already discussed at the beginning of this section, this only has an illustrative purpose: at any given detector network, one will have to separate the resolved from unresolved sources, sum over the unresolved ones, and take into account the error in the reconstruction of the resolved sources.
\Cref{fig:Anisotropy_cls_fullPop_allParams} shows the functions $c_l(f)$, for the  sample of events that we extracted,  and a selection of values of $l$. 

For BBHs we see that, for all multipoles shown, below about 100~Hz the anisotropies are at the percent level, and then they grow, reaching values of order one in the kHz region. This can be understood from the fact that most BBHs merge at lower frequencies; therefore, the effective number of BBHs contributing to the result in the kHz region is just a small fraction of the total, and the relative effect of shot noise increases when the effective number of sources decreases. 

For BNS we observe that the anisotropies are considerably smaller as compared to the BBH case, with values of ${\cal O}(10^{-4})$ up to the $100$~Hz region. This can be traced on the one hand to the larger amount of sources present in our  BNS catalog, and on the other hand to the morphology of their signals in band. Indeed, most of the BNS systems merge at ${\cal O}({\rm kHz})$ frequencies, and sweep all through the frequency band of ground-based detectors. Also in this case, for $f\gtrsim300~{\rm Hz}$ we observe a raise in the spectrum up to values of $\sim10^{-1}$: this is again because part of the sources, especially at high redshifts, start to merge.

\section{Conclusions}\label{sect:concl}

In this paper we have shown  how the spectral density of an ensemble of compact binary coalescences emerges from the average over the extrinsic parameters of the population: the times of arrival, the polarization angles, the orbit inclinations and  the arrival directions. We have seen the role of each of these averages in arriving at the final result
(\ref{eq:Shsum}), which is often used in the literature although, to our knowledge, it lacked a real derivation (except for the heuristic argument that, in the limit where the observation time $T\ra\infty$, it reduces to the expression given in \cite{Phinney:2001di}, see \cref{foot:comparisonPhinney}). The step-by-step procedure that we have described allows us to understand the role of the various averages: the averages over the polarization angles decorrelate the different events or, equivalently, different  arrival directions, producing a Dirac delta 
$\delta(\hatn,\hatn')$ in \eq{corrhahadeltannp}, and furthermore  eliminates the linear polarizations; the averages over arrival times produce a (regularized) Dirac delta $\delta_T(f-f')$, which in the limit $|f-f'|T\ra\infty$ decorrelate different frequencies; and the averages over the inclination angles remove the remaining Stokes parameter, associated with circular polarization. 

We have  computed the  exact result for a finite observation time $T$, and not just in the limit of large $T$.
We have seen that, in this case, the correlator
$\langle \tilde{h}_A^*(f,\hatn)
\tilde{h}_{A'}(f',\hatn')\rangle_U$ is given by \eq{hhgeneralpp}. Note that it depends on the function  
$\delta_T(f-f')$ given by \eq{defdeltaT}, whose specific form emerges explicitly from the computation and is not just an ``ad hoc'' regularization of the Dirac delta, and
the spectral density  that appears in \eq{hhgeneralpp}
is a function $S_h(f,f')$ of two variables $f$ and $f'$, given by \eq{Shfinal1pp}. Only when the condition (\ref{cond12}) is satisfied (which is normally the case for ground-based detectors, but  not necessarily for PTA) it reduces to the function $S_h(f)$ given by \eq{Shastroavenev} or (after the further simplification of neglecting the fact that also the number of events in the observation time $T$ is a stochastic variable) reduces to \eq{eq:Shsum}, which is the expression commonly used in the literature.

For a finite sample of events, however, all these averages must be replaced with the sum over the actual events of the specific realizations, and the above cancellations are only approximate. Shot noise therefore generates a certain amount of polarization even in the CBC background, and a certain amount of anisotropy. In particular, we have seen that shot noise generates a linear polarization  even when the underlying distribution of sources is isotropic. We have computed numerically  these shot noise contributions with a realistic catalog of BBHs and of BNSs (treating all sources as undetected). This formalism can be useful to disentangle possible  polarized backgrounds produced by mechanisms in the early Universe,  or anisotropies in the stochastic background that reflect the underlying matter distribution, from contamination effects due to shot noise of the astrophysical background.

\normalsize

\let\oldaddcontentsline\addcontentsline
\renewcommand{\addcontentsline}[3]{}
\begin{acknowledgments}

We thank the referee for a very useful report.
The research of  F.I., 
M.~Mag. and N.M. is supported by  the  Swiss National Science Foundation, grant 200020$\_$191957, and  by the SwissMap National Center for Competence in Research. E.B. and M.~Mag. are supported by the SNSF
grant CRSII5$\_$213497. 
The work of M.~Man. received support from the French government under the France 2030 investment plan, as part of the Initiative d'Excellence d'Aix-Marseille Universit\'e -- A*MIDEX AMX-22-CEI-02. 
Computations made use of the Baobab cluster at the University of Geneva.
\end{acknowledgments}
\let\addcontentsline\oldaddcontentsline

\appendix

\section{Computation of the dependence on \texorpdfstring{$\psi_i$}{psi}}\label{app:extractpsi}

In this appendix we perform in details the computation discussed in section~\ref{sect:deppsi}. We start from \eq{eqshot2}.
Using \eq{eq: psi rot}, we have
\be
\tilde{h}_{A,i}^*(f;\psi_i,\cos\iota_i)\tilde{h}_{A',i}(f;\psi_i,\cos\iota_i)=\sum_{BB'}M_{AA',BB'}
 \tilde{h}_{B,i}^*(f;\psi_i=0,\cos\iota_i)\tilde{h}_{B',i}(f;\psi_i=0,\cos\iota_i)
\, ,
\ee
where
\be
\label{MAABBdef}
M_{AA',BB'}=R_{AB}(2\psi_i)R_{A'B'}(2\psi_i)\, .
\ee
Treating $M_{AA',BB'}$ as a $2\times 2$ complex matrix in the $(A,A')$ indices, we can decompose it in the basis of the identity matrix 
$\delta_{AA'}$ and the Pauli matrices $\sigma^a_{AA'}$ ($a=1,2,3$), with complex coefficients that are matrices in the $(B,B')$ indices, 
\be\label{MAABBPauli}
M_{AA',BB'}=\delta_{AA'}\alpha_{BB'}+\sum_{a=1}^3\sigma^a_{AA'}\beta^a_{BB'}\, .
\ee
The matrix $\alpha_{BB'}$ can be obtained contracting \eq{MAABBPauli} with $\delta_{AA'}$. This gives
\be
2\alpha_{BB'}=\sum_{AA'}\delta_{AA'}M_{AA',BB'}= \[ R^T(2\psi_i)R(2\psi_i)\]_{BB'}\, ,
\ee
where $R^T(2\psi_i)$ is the transpose matrix. Since $R$ is a rotation matrix, $ [R^T(2\psi_i)R(2\psi_i)]_{BB'}=\delta_{BB'}$ 
and we get
\be
\label{2alphass}
2\alpha_{BB'}=\delta_{BB'}\, .
\ee
Similarly, contracting both sides of \eq{MAABBPauli} with  
$\sigma^b_{A'A}$ and using ${\rm Tr} (\sigma^a\sigma^b)=2\delta^{ab}$, we get
\be
2\beta^b_{BB'}=\sum_{AA'}\sigma^b_{A'A}M_{AA',BB'}= \[ R^T(2\psi_i)(\sigma^b)^T R(2\psi_i)\]_{BB'}.\label{RtsR}
\ee
Writing $R(2\psi_i)$ as in \eq{RABPauli} and recalling from \eq{Pauli} that $(\sigma^2)^T=-\sigma^2$ while $(\sigma^1)^T=\sigma^1$ and
$(\sigma^3)^T=\sigma^3$, we get
\bees
2\beta^1_{BB'}&=& \sigma^{1}_{BB'}\cos 4\psi_i 
+ \sigma^{3}_{BB'}\sin 4\psi_i \, ,\label{2beta1ss}\\
2\beta^2_{BB'}&=&-\sigma^{2}_{BB'}\, ,\label{2beta2ss}\\
2\beta^3_{BB'}&=&-\sigma^{1}_{BB'} \sin 4\psi_i 
+\sigma^{3}_{BB'}\cos 4\psi_i \, .\label{2beta3ss}
\ees
Inserting this expression for $M_{AA',BB'}$ into \eq{hhMAABB} and then in \eq{eqshot2} we get \eq{eqshot4}, with $\mathcal{I}_i, \ldots, \mathcal{Q}_i$ given by \eqst{expressionI}{expressionQ}.

As we  discussed below \eq{hLRhelicity2},
$\mathcal{I}_i$ and $\mathcal{V}_i$ are independent of the reference value $\psi_i=0$.
In contrast, as we will see in a moment, $\mathcal{Q}_i$ and $\mathcal{U}_i$ depend on $\psi_i$.
Using \eqs{RtsR}{2beta1ss} we see that
\be
\sigma^{1}\cos 4\psi_i + \sigma^{3}\sin 4\psi_i=
R^T(2\psi_i)\sigma^1 R(2\psi_i)\, ,
\ee 
and therefore \eq{expressionU} can be rewritten as 
\be\label{expressionUpsiapp}
\mathcal{U}_i(f;\cos\iota_i,\psi_i)=\frac{1}{T}\sum_{A,A'=+,\times} 
\tilde{h}^*_{A,i}(f;\psi_i,\cos\iota_i) \sigma^1_{AA'} \tilde{h}_{A',i}(f;\psi_i,\cos\iota_i)\, ,
\ee
and similarly
\be\label{expressionQpsiapp}
\mathcal{Q}_i(f;\cos\iota_i,\psi_i)=\frac{1}{T}\sum_{A,A'=+,\times} 
\tilde{h}^*_{A,i}(f;\psi_i,\cos\iota_i) \sigma^3_{AA'} \tilde{h}_{A',i}(f;\psi_i,\cos\iota_i)\, .
\ee
In terms of $h_{L,i}$ and $h_{R,i}$, we have 
\be\label{expressionILRapp}
\mathcal{I}_i(f;\cos\iota_i)=\frac{1}{T} \big[ |\tilde{h}_{R,i}(f;\cos\iota_i)|^2+|\tilde{h}_{L,i}(f;\cos\iota_i)|^2 \big]\,,
\ee
and 
\be\label{expressionVLRapp}
\mathcal{V}_i(f;\cos\iota_i)=\frac{1}{T} \big[ |\tilde{h}_{R,i}(f;\cos\iota_i)|^2-|\tilde{h}_{L,i}(f;\cos\iota_i)|^2 \big]\, ,
\ee
(we are now also writing explicitly the argument $\cos\iota_i$). Similarly, we get
\be\label{expressionULRapp}
\mathcal{U}_i(f;\cos\iota_i,\psi_i)=\frac{i}{T} 
\big[
\tilde{h}^*_{R,i}(f;\psi_i,\cos\iota_i)\tilde{h}_{L,i}(f;\psi_i,\cos\iota_i)-\tilde{h}^*_{L,i}(f;\psi_i,\cos\iota_i)\tilde{h}_{R,i}(f;\psi_i,\cos\iota_i) 
\big]\, ,
\ee
and
\be\label{expressionQLRapp}
\mathcal{Q}_i(f;\cos\iota_i,\psi_i)=\frac{1}{T} 
\big[
\tilde{h}^*_{R,i}(f;\psi_i,\cos\iota_i)\tilde{h}_{L,i}(f;\psi_i,\cos\iota_i)
+\tilde{h}^*_{L,i}(f;\psi_i,\cos\iota_i)\tilde{h}_{R,i}(f;\psi_i,\cos\iota_i) 
\big]\, .
\ee
Using \eq{hLRhelicity2} we then see that, under $\psi_i\ra\psi_i+\psi_0$,  $\mathcal{U}_i$ and $\mathcal{Q}_i$ transform linearly among them, as
\bees
\mathcal{U}_i &\ra& \phantom{+}\cos(4\psi_0) \mathcal{U}_i+\sin(4\psi_0) \mathcal{Q}_i\, ,
\label{transfUiapp}\\
\mathcal{Q}_i  &\ra& -\sin(4\psi_0) \mathcal{U}_i+\cos(4\psi_0) \mathcal{Q}_i\, ,
\label{transfQiapp}
\ees
so that 
\bees
\mathcal{Q}_i + i\mathcal{U}_i &\ra& e^{i4\psi_0} (\mathcal{Q}_i +i\mathcal{U}_i)\label{transfUQ1app}\, , \\
\mathcal{Q}_i - i\mathcal{U}_i &\ra& e^{-i4\psi_0} (\mathcal{Q}_i -i\mathcal{U}_i)\, .\label{transfUQ2app}
\ees

\section{Shot noise for finite observation time}\label{app:shotgeneral}

The expression in \eq{eqshot4} was derived from \eq{eqshot2}, which is valid only when the condition (\ref{cond12bis}) is satisfied. In the more general case where this assumption does not hold, we must restart from \eq{eqshot1}. The average over arrival times appearing there can be written as (see \cref{foot:decorij_ti})
\bees
&&\langle \tilde{h}_{A,i}^*(f;\psi_i,t_i,\cos\iota_i)\tilde{h}_{A',j}(f';\psi_j,t_j,\cos\iota_j)\rangle_{ \{ t_k\} }\,\\
&=&
\begin{dcases*}
\frac{1}{T}\delta_T(f-f')\tilde{h}_{A,i}^*(f;\psi_i,t_i=0,\cos\iota_i)\tilde{h}_{A',i}(f';\psi_i,t_i=0,\cos\iota_i)
& if $i=j$\,,\\
\frac{1}{T^2}\delta_T(f)\delta_T(f')\tilde{h}_{A,i}^*(f;\psi_i,t_i=0,\cos\iota_i)\tilde{h}_{A',j}(f';\psi_j,t_j=0,\cos\iota_j)
& if $i\neq j$\,.\nn
\end{dcases*}
\ees
Plugging it into \eq{eqshot1} gives
\bees
&&\langle\tilde{h}_A^*(f,\hatn)\tilde{h}_{A'}(f',\hatn')\rangle_{ \{t_k\} }\,\label{eqshotapp1}\\
&=&\delta(\hatn,\hatn')\delta_T(f-f')\frac{1}{T}\sum_{i=1}^{\nev} \delta(\hatn,\hatn_i)\tilde{h}_{A,i}^*(f;\psi_i,t_i=0,\cos\iota_i)\tilde{h}_{A',i}(f';\psi_i,t_i=0,\cos\iota_i)\nn\\
&+&\delta_T(f)\delta_T(f')\frac{1}{T^2}\sum_{i\neq j}^{\nev}\delta(\hatn,\hatn_i)\delta(\hatn',\hatn_j)\tilde{h}_{A,i}^*(f;\psi_i,t_i=0,\cos\iota_i)\tilde{h}_{A',j}(f';\psi_j,t_j=0,\cos\iota_j)\,.\nn
\ees
One can already observe that the right-hand side of \eq{eqshotapp1} contains two contributions: a sum with $i=j$ and a second one with $i\neq j$. While the first sum ($i=j$) comes with a Dirac delta $\delta(\hatn,\hatn')$ that completely decorrelates different directions of observation (or, equivalently, directions of propagation), the same statement does not hold for the second one. Thus, for a finite observation time $T$ (more precisely, when the condition (\ref{cond12bis}) is not satisfied), shot noise produces correlations between different directions in the astrophysical background. To avoid any misunderstanding, note that the sum with $i\neq j$ is not intended as a sum over $i<j$; for example, both the combinations $(i=1,j=2)$ and $(i=2,j=1)$ contribute to it.\footnote{These two combinations also carry unequal dependence as functions of $\hatn$ and $\hatn'$ because $\delta(\hatn,\hatn_1)\delta(\hatn',\hatn_2)\neq\delta(\hatn,\hatn_2)\delta(\hatn',\hatn_1)$.}

In the rest of this appendix we calculate more explicitly the two sums above. To do this, we manipulate $\tilde{h}_{A,i}^*(f;\psi_i,t_i=0,\cos\iota_i)\tilde{h}_{A',j}(f';\psi_j,t_j=0,\cos\iota_j)$ (for generic $i,j$, valid both for $i=j$ and $i\neq j$) via a generalization of the procedure outlined in app. \ref{app:extractpsi}. Using \eq{eq: psi rot}, we write
\bees
&&\tilde{h}_{A,i}^*(f;\psi_i,t_i=0,\cos\iota_i)\tilde{h}_{A',j}(f';\psi_j,t_j=0,\cos\iota_j)\,\\
&=&\sum_{BB'}\mathcal{M}_{AA',BB'}
 \tilde{h}_{B,i}^*(f;\psi_i=0,t_i=0,\cos\iota_i)\tilde{h}_{B',j}(f';\psi_j=0,t_j=0,\cos\iota_j)
\,\nn,
\ees
where
\be
\label{MAABBgendef}
\mathcal{M}_{AA',BB'}=R_{AB}(2\psi_i)R_{A'B'}(2\psi_j)\,.
\ee
Of course, $\mathcal{M}_{AA',BB'}$ reduces to $M_{AA',BB'}$ in \eq{MAABBdef} when $i=j$.
Decomposing $\mathcal{M}_{AA',BB'}$ into the basis made by the identity $\delta_{AA'}$ and the Pauli matrices $\sigma^a_{AA'}$ ($a=1,2,3$), we now get
\be\label{MAABBgenPauli}
\mathcal{M}_{AA',BB'}=\delta_{AA'}\mathcal{\kappa}_{BB'}+\sum_{a=1}^3\sigma^a_{AA'}\mathcal{\lambda}^a_{BB'}\,,
\ee
where
\bees
2\kappa_{BB'}&=&\sum_{AA'}\delta_{AA'}M_{AA',BB'}= \[ R^T(2\psi_i)R(2\psi_j)\]_{BB'}\,,\label{eqkappa}\\
2\lambda^b_{BB'}&=&\sum_{AA'}\sigma^b_{A'A}M_{AA',BB'}= \[ R^T(2\psi_i)(\sigma^b)^T R(2\psi_j)\]_{BB'}\,.\label{eqlambdab}
\ees
Using \eq{RABPauli} one obtains
\bees
2\kappa_{BB'}&=&\delta_{BB'}\cos[2(\psi_i-\psi_j)]+i\sigma^2_{BB'}\sin[2(\psi_i-\psi_j)]\,,\label{eqkappanew}\\
2\lambda^1_{BB'}&=&\sigma^1_{BB'}\cos[2(\psi_i+\psi_j)]+\sigma^3_{BB'}\sin[2(\psi_i+\psi_j)]\,,\label{eqlambda1}\\
2\lambda^2_{BB'}&=&-i\delta_{BB'}\sin[2(\psi_i-\psi_j)]-\sigma^2_{BB'}\cos[2(\psi_i-\psi_j)]\,,\label{eqlambda2}\\
2\lambda^3_{BB'}&=&-\sigma^1_{BB'}\sin[2(\psi_i+\psi_j)]+\sigma^3_{BB'}\cos[2(\psi_i+\psi_j)]\,,\label{eqlambda3}
\ees
which correctly reduce to \eq{2alphass} and \eqst{2beta1ss}{2beta3ss} when $i=j$.
Then we can express \eq{eqshotapp1} as
\bees
&&\langle \tilde{h}_A^*(f,\hatn)
\tilde{h}_{A'}(f',\hatn')\rangle_{ \{t_k\} }  =\delta(\hatn,\hatn')\delta_T(f-f') \frac{1}{2}
\sum_{i=1}^{\nev}\delta(\hatn,\hatn_i)\label{eqshotapp2}
\\
&&\hspace*{0mm}\times  
 \Big[
\mathcal{I}_i(f,f';\cos\iota_i)\delta_{A A'}+\mathcal{U}_i(f,f';\cos\iota_i,\psi_i)\sigma^{1}_{AA'}+\mathcal{V}_i(f,f';\cos\iota_i)\sigma^{2}_{AA'}
+\mathcal{Q}_i(f,f';\cos\iota_i,\psi_i)\sigma^{3}_{AA'}
 \Big] \,\nn\\
&&\hspace*{2mm}+\frac{\delta_T(f)\delta_T(f')}{T}\frac{1}{2}\sum_{i\neq j}^{\nev}\delta(\hatn,\hatn_i)\delta(\hatn',\hatn_j)\,\nn\\
&&\hspace*{4mm}\times\Big[
 \mathscr{I}_{ij}(f,f';\cos\iota_i,\cos\iota_j,\psi_i,\psi_j)\delta_{A A'}+\mathscr{U}_{ij}(f,f';\cos\iota_i,\cos\iota_j,\psi_i,\psi_j)\sigma^{1}_{AA'}\,\nn\\
 &&\hspace*{5mm}+\mathscr{V}_{ij}(f,f';\cos\iota_i,\cos\iota_j,\psi_i,\psi_j)\sigma^{2}_{AA'}
 +\mathscr{Q}_{ij}(f,f';\cos\iota_i,\cos\iota_j,\psi_i,\psi_j)\sigma^{3}_{AA'}
 \Big] \,\nn,
\ees
where, omitting for brevity the dependence on the reference time $t_i=0$,
\be\label{expressionIshot}
\mathcal{I}_i(f,f';\cos\iota_i) =\frac{1}{T}  \sum_{B=+,\times} \tilde{h}_{B,i}^*(f;\psi_i=0,\cos\iota_i)\tilde{h}_{B,i}(f';\psi_i=0,\cos\iota_i)\, ,\, ,
\ee
\be\label{expressionVshot}
\mathcal{V}_i(f,f';\cos\iota_i) =\frac{1}{T}\sum_{B,B'=+,\times}  \tilde{h}^*_{B,i}(f;\psi_i=0,\cos\iota_i) (-\sigma^2_{BB'})\tilde{h}_{B',i}(f';\psi_i=0,\cos\iota_i)\, ,
\ee
\bees
&&\mathcal{U}_i(f,f';\cos\iota_i,\psi_i) \label{expressionUshot}\\
&&\hspace{3mm}=\frac{1}{T}\hspace*{-1mm}\sum_{B,B'=+,\times} \hspace*{-1mm} \tilde{h}^*_{B,i}(f;\psi_i=0,\cos\iota_i)\[\sigma^1_{BB'}\cos 4\psi_i+\sigma^3_{BB'}\sin 4\psi_i\] \tilde{h}_{B',i}(f';\psi_i=0,\cos\iota_i)\, ,\nn 
\ees
\bees
&&\mathcal{Q}_i(f,f';\cos\iota_i,\psi_i)\label{expressionQshot}\\
&&\hspace{3mm}=\frac{1}{T}\hspace*{-1mm}\sum_{B,B'=+,\times} \hspace*{-1mm} \tilde{h}^*_{B,i}(f;\psi_i=0,\cos\iota_i) 
\[-\sigma^1_{BB'}\sin 4\psi_i+\sigma^3_{BB'}\cos 4\psi_i\]\tilde{h}_{B',i}(f';\psi_i=0,\cos\iota_i)\,,\nn 
\ees
and, again omitting for brevity the reference times $t_i=0$, $t_j=0$,
\bees
&&\mathscr{I}_{ij}(f,f';\cos\iota_i,\cos\iota_j,\psi_i,\psi_j)\,\label{expressionIextra}\\
&&\hspace{3mm}=\frac{1}{T}\sum_{B,B'=+,\times} \tilde{h}_{B,i}^*(f;\psi_i=0,\cos\iota_i)\{\delta_{BB'}\cos[2(\psi_i-\psi_j)]+i\sigma^2_{BB'}\sin[2(\psi_i-\psi_j)]\}\nn\\
&&\hspace{30mm}\times\tilde{h}_{B',j}(f';\psi_j=0,\cos\iota_j)\,,\nn
\ees
\bees
&&\mathscr{V}_{ij}(f,f';\cos\iota_i,\cos\iota_j,\psi_i,\psi_j)\,\label{expressionVextra}\\
&&\hspace{3mm}=\frac{1}{T}\sum_{B,B'=+,\times} \tilde{h}_{B,i}^*(f;\psi_i=0,\cos\iota_i)\{-i\delta_{BB'}\sin[2(\psi_i-\psi_j)]-\sigma^2_{BB'}\cos[2(\psi_i-\psi_j)]\}\nn\\
&&\hspace{30mm}\times\tilde{h}_{B',j}(f';\psi_j=0,\cos\iota_j)\,,\nn
\ees
\bees
&&\mathscr{U}_{ij}(f,f';\cos\iota_i,\cos\iota_j,\psi_i,\psi_j)\,\label{expressionUextra}\\
&&\hspace{3mm}=\frac{1}{T}\sum_{B,B'=+,\times} \tilde{h}_{B,i}^*(f;\psi_i=0,\cos\iota_i)\{\sigma^1_{BB'}\cos[2(\psi_i+\psi_j)]+\sigma^3_{BB'}\sin[2(\psi_i+\psi_j)]\}\nn\\
&&\hspace{30mm}\times\tilde{h}_{B',j}(f';\psi_j=0,\cos\iota_j)\,,\nn
\ees
\bees
&&\mathscr{Q}_{ij}(f,f';\cos\iota_i,\cos\iota_j,\psi_i,\psi_j)\,\label{expressionQextra}\\
&&\hspace{3mm}=\frac{1}{T}\sum_{B,B'=+,\times} \tilde{h}_{B,i}^*(f;\psi_i=0,\cos\iota_i)\{-\sigma^1_{BB'}\sin[2(\psi_i+\psi_j)]+\sigma^3_{BB'}\cos[2(\psi_i+\psi_j)]\}\nn\\
&&\hspace{30mm}\times\tilde{h}_{B',j}(f';\psi_j=0,\cos\iota_j)\,.\nn
\ees
Observe that when $i=j$ the quantities $\mathscr{I}_{ij}$, $\mathscr{V}_{ij}$, $\mathscr{U}_{ij}$, $\mathscr{Q}_{ij}$ are equal to $\mathcal{I}_i$, $\mathcal{V}_i$, $\mathcal{U}_i$, $\mathcal{Q}_i$, respectively. Furthermore, $\mathscr{I}_{ij}$ and $\mathscr{V}_{ij}$ in \eqs{expressionIextra}{expressionVextra} depend on $\psi_i$ and $\psi_j$ only through the difference $\psi_i-\psi_j$, while $\mathscr{U}_{ij}$ and $\mathscr{Q}_{ij}$ \eqs{expressionUextra}{expressionQextra} depend on $\psi_i$ and $\psi_j$ only through the sum $\psi_i+\psi_j$.
Note also that, just as in the discussion below \eq{expressionQ}, $\mathcal{I}_i$ and $\mathcal{V}_i$ in \eqs{expressionIshot}{expressionVshot} are independent of the reference value $\psi_i=0$, so we could replace this by any $\psi_i$ or just suppress it in the arguments without ambiguity. Instead, as in the same discussion, $\mathcal{U}_i$ and $\mathcal{Q}_i$ in \eqs{expressionUshot}{expressionQshot} depend on $\psi_i$ and transform linearly among them under a shift of it. We also remark that the reference time $t_i=0$ has been omitted in \eqst{expressionIshot}{expressionQshot} only for brevity but, unless $f=f'$, it actually matters in $\mathcal{I}_i$, $\mathcal{V}_i$, $\mathcal{U}_i$, $\mathcal{Q}_i$. Similarly, in \eqst{expressionIextra}{expressionQextra} we omitted the times $t_i=0, t_j=0$ for notational simplicity, but they are relevant in $\mathscr{I}_{ij}$, $\mathscr{V}_{ij}$, $\mathscr{U}_{ij}$, $\mathscr{Q}_{ij}$ (in this case even when $f=f'$, unless we shift $t_i=0$ and $t_j=0$ by the same amount).

Using eqs. (\ref{RtsR}), (\ref{2beta1ss}) and (\ref{2beta3ss}), we can also rewrite \eqs{expressionUshot}{expressionQshot} as
\bees
&&\mathcal{U}_i(f,f';\cos\iota_i,\psi_i)=\frac{1}{T}\hspace*{-1mm}\sum_{B,B'=+,\times} \hspace*{-1mm} \tilde{h}^*_{B,i}(f;\psi_i,\cos\iota_i)\sigma^1_{BB'}\tilde{h}_{B',i}(f';\psi_i,\cos\iota_i)\,,\label{expressionUshot2} \\
&&\mathcal{Q}_i(f,f';\cos\iota_i,\psi_i)=\frac{1}{T}\hspace*{-1mm}\sum_{B,B'=+,\times} \hspace*{-1mm} \tilde{h}^*_{B,i}(f;\psi_i,\cos\iota_i)\sigma^3_{BB'}\tilde{h}_{B',i}(f';\psi_i,\cos\iota_i)\,.\label{expressionQshot2}
\ees
Similarly, using \eqst{eqkappa}{eqlambda3}, the results in \eqst{expressionIextra}{expressionQextra} can be recast into
\bees
&&\mathscr{I}_{ij}(f,f';\cos\iota_i,\cos\iota_j,\psi_i,\psi_j)=\frac{1}{T}\sum_{B=+,\times} \tilde{h}_{B,i}^*(f;\psi_i,\cos\iota_i)\tilde{h}_{B,j}(f';\psi_j,\cos\iota_j)\,,\label{expressionsIVUQextra}\\
&&\mathscr{V}_{ij}(f,f';\cos\iota_i,\cos\iota_j,\psi_i,\psi_j)=\frac{1}{T}\sum_{B,B'=+,\times} \tilde{h}_{B,i}^*(f;\psi_i,\cos\iota_i)(-\sigma^2_{BB'})\tilde{h}_{B',j}(f';\psi_j,\cos\iota_j)\,,\nn\\
&&\mathscr{U}_{ij}(f,f';\cos\iota_i,\cos\iota_j,\psi_i,\psi_j)=\frac{1}{T}\sum_{B,B'=+,\times} \tilde{h}_{B,i}^*(f;\psi_i,\cos\iota_i)\sigma^1_{BB'}\tilde{h}_{B',j}(f';\psi_j,\cos\iota_j)\,,\nn\\
&&\mathscr{Q}_{ij}(f,f';\cos\iota_i,\cos\iota_j,\psi_i,\psi_j)=\frac{1}{T}\sum_{B,B'=+,\times} \tilde{h}_{B,i}^*(f;\psi_i,\cos\iota_i)\sigma^3_{BB'}\tilde{h}_{B',j}(f';\psi_j,\cos\iota_j)\,.\nn
\ees

\section{Linear polarizations and spatial isotropy}\label{app:Linpol}

From \eqst{defQpm}{expressionQpm} we see that, once we include the effect of shot noise,   linear polarization is in principle allowed, and we see from \eq{Qpmlzero} that we get a non-vanishing result even when we only keep the $l=0$ multipole, which corresponds to smearing uniformly over the sphere the probability of having a source in a given direction. The corresponding numerical results have been shown in \cref{fig:StokesallParams}.
This is apparently at odd with the argument that $Q^{\pm}(f;\hatn) $ have helicities $\pm 4$, and can then be 
expanded in $(\pm 4)$ spin-weighted  spherical harmonics,
${}_{\pm 4}\hspace*{-0.2mm}Y_{lm}({\hatn})$, as
\be\label{UQspin4}
Q^{\pm}(f;\hatn)  =\sum_{l=4}^{\infty}\sum_{m=-l}^l  C_{lm}^{\pm}(f)\,
{}_{\pm 4}\hspace*{-0.2mm}Y_{lm}({\hatn})\, ,
\ee 
with the expansion starting from $l=4$, since ${}_{\pm 4}Y_{lm}({\hatn})$ vanish for $l<4$. In ref.~\cite{Seto:2008sr} it is then argued that an isotropic distribution of sources, that only has the $l=0$ multipole, cannot produce a non-vanishing value for $Q^{\pm}(f;\hatn) $, that  start from $l=4$.

In fact, the above conclusion is not correct, as we next explain.
To this purpose, it is useful to realize  that the labels $(l,m)$ in \eq{UQspin4} are in fact an abuse of notation, which induces one to believe that these are the indices associated with the orbital angular momentum operator. This is not true, and these indices are associated with the total angular momentum operator, and should be more appropriately denoted as $(j,j_z)$, so \eq{UQspin4} should rather be written as
\be\label{UQspin4jjz}
Q^{\pm}(f;\hatn) =\sum_{j=4}^{\infty}\sum_{j_z=-j}^j  C_{jj_z}^{\pm}(f)\,
{}_{\pm 4}Y_{jj_z}({\hatn})\, .
\ee 
The issue was already discussed in section~3.5.2 of ref.~\cite{Maggiore:2007ulw} (see in particular Note~47 there) and follows from the fact that 
the tensor spherical harmonics  (from which the spin-weighted spherical harmonics can be  derived, see e.g.  App.~F and G of ref.~\cite{Romano:2016dpx} for the explicit relations for the helicity 1 and 2 cases) are obtained  by coupling
the scalar spherical harmonics $Y_{ll_z}(\hatn)$
to the spin functions $\chi_{s s_z}$, with the appropriate Clebsch--Gordan
coefficients, to give a state with total angular momentum $|j,j_z\rangle$.
 
The loophole in the above argument, then, is that the quantity 
$$
\sum_{B,B'=+,\times} \hspace*{-1mm} \tilde{h}^*_{B,i}(f;\psi_i,\cos\iota_i) 
(\sigma^3\pm i\sigma^1)_{BB'}\tilde{h}_{B',i}(f;\psi_i,\cos\iota_i)
$$ 
from which  $Q^{\pm}(f;\hatn)$ is constructed
is a spin-4 operator [as we also see from the fact that, extracting the $\psi_i$ dependence, we get a factor $e^{\pm i4\psi_i}$, see \eq{expressionQpm}], i.e. it has $s=4$, $s_z=\pm 4$. An isotropic distribution of sources can generate only  the 
$l=0,l_z=0$ multipole in the orbital part of $Q^{\pm}(f;\hatn)$ but this,  when combined  with the helicities $s=4$, $s_z=\pm 4$ of the $Q^{\pm}(f;\hatn)$ operators, results in an object which has $j=4$, $j_z=\pm 4$, and which can therefore appear as the lowest term of the expansion (\ref{UQspin4jjz}). 

Another way to see that \eq{UQspin4} [or, in more appropriate notation, \eq{UQspin4jjz}] is consistent with the existence of an $l=0$ term in \eq{Qpmexpalm2} is to observe that the integral of the spin-weighted spherical harmonics over the sphere is  in general non-vanishing. Note in fact that the spin-weighted spherical harmonics satisfy the orthogonality condition (see e.g. eq.~(E8) of \cite{Romano:2016dpx})
\be\label{sYortho}
\int_{S^2} d^2\hatn\, \,
{}_{s}Y_{jj_z}({\hatn}) \,
{}_{s}Y^*_{j'j'_z}({\hatn})=\delta_{jj'}\delta_{j_zj_z'}\, ,
\ee
but this only holds if the spin index $s$ is the same for the two spin-weighted spherical harmonics that appear in \eq{sYortho}. There is no similar orthogonality condition between ${}_{s}Y_{jj_z}$ and 
${}_{s'}Y^*_{j'j'_z}$ with $s\neq s'$. In particular, the spin-weighted spherical harmonics ${}_{s}Y_{jj_z}$ with $s\neq 0$ in general are not orthogonal to ${}_{0}Y_{00}$, which is just a constant, and so the integral of ${}_{s}Y_{jj_z}({\hatn})$ over the sphere is in general non-vanishing.\footnote{Note also that  eq.~(E13) of \cite{Romano:2016dpx} only holds if $s_1+s_2+s_3=0$, even if that condition is not written explicitly there; therefore,  one cannot show that the integral of ${}_{s_1}Y_{jj_z}({\hatn})$ vanishes by setting $s_1\neq 0$, $s_2=s_3=0$ there.} 
For instance, from the explicit expressions of the spin-weighted spherical harmonics, we find that
\be
\int_{S^2} \frac{d^2\hatn}{4\pi}\, \,
{}_{\pm1}Y_{10}({\hatn}) =\pm \frac{1}{8}\, \sqrt{\frac{3\pi}{2}}\, ,
\ee
\be
\int_{S^2} \frac{d^2\hatn}{4\pi}\, \,
{}_{\pm 2}Y_{20}({\hatn}) =
\frac{1}{2}\sqrt{\frac{5}{6\pi}}
\, ,
\ee
and
\be
\int_{S^2} \frac{d^2\hatn}{4\pi}\, \,
{}_{\pm 4}Y_{40}({\hatn}) =  \frac{1}{2}\sqrt{\frac{7}{10\pi}}\, .
\ee
Note that a necessary condition for obtaining a non-vanishing integral over the sphere is that $j_z=0$, because of the dependence $e^{ij_z\phi}$ of the spin-weighted spherical harmonics.

We conclude that (contrary to statements in the literature) an isotropic distribution of sources can in principle produce a GW background which  also has a linear polarization. 

\section{Alternative regularizations for finite $T$}\label{app:finiteT}

In this appendix we explore an alternative approach to the finite-$T$ expression for the correlator, and we compare it with the one that we have adopted in the text. To simplify the notation, in this appendix we just use the notation 
$x(t)$ for a generic GW signal, omitting the polarization labels and the dependence on $\hatn$ (actually, the discussion can be  applied to signals of any kind, as long as they are characterized by  a well defined arrival time and different signals are uncorrelated). We now assume that $x(t)$ includes the contribution of all signals arriving from $t=-\infty$ to $t=+\infty$, and therefore its spectral density has the form
\be\label{deltaTxxSh0}
\langle \tilde{x}^*(f)\tilde{x}(f')\rangle
=\delta(f-f') S_x(f)\, ,
\ee
for some function $S_x(f)$.\footnote{Again for notational simplicity, we include here into the definition of $S_x(f)$ the  factor $1/2$ that normally appears on the right-hand side.}
The fact that we only collect the signals that arrive over the time $t\in [-T/2,T/2]$ is then taken into account by defining
\be\label{defyt}
y(t)=\left\{
\begin{array}{ll}
x(t)\qquad\qquad &{\rm if} \, t\in [-T/2,T/2] \\
0\qquad\qquad     &{\rm otherwise}\, ,
\end{array}
\right.
\ee
i.e. setting $y(t)=0$ outside the observation window.  From \eq{defyt},
\be
\tilde{y}(f)=\int_{-\infty}^{\infty}df'\, \tilde{x}(f')\delta_T(f-f')\, ,
\ee
and therefore
\bees
\langle \tilde{y}^*(f)\tilde{y}(f')\rangle 
&=&
\int_{-\infty}^{\infty}df'' df'''  \delta_T(f''-f)\delta_T(f'''-f')
\langle \tilde{x}^*(f'')\tilde{x}(f''')\rangle\nn\\ 
&=&
\int_{-\infty}^{\infty}df''  \delta_T(f''-f)\delta_T(f''-f')
S_x(f'')
\, . \label{deltaTyySh}
\ees
Note that
\be\label{yyTinfty}
{\rm lim}_{T\ra\infty} \langle \tilde{y}^*(f)\tilde{y}(f')\rangle=\delta(f-f') S_x(f)\, .
\ee
However, for finite $T$ \eq{deltaTyySh} is quadratic in the regularized Dirac delta, while the expressions that we found in the main text are always linear in $\delta_T(f-f')$. So,
the question that we want to address  here is what is the relation between
this approach and the one that we have followed  in the main text
(at finite $T$; as shown by \eq{yyTinfty}, they become the same in the limit $T\ra\infty$).\footnote{We thank the referee for suggesting this comparison.}

To address this question, we first repeat the derivation of the spectral density in the main text using this simplified notation. We begin by 
specifying that the overall signal $x(t)$ is due to a superposition of  events having an arrival time $t_i$,
\be\label{xditsumi}
x(t)=\sum_i x_i(t;t_i,\Psi_i)\, ,
\ee
where we generically denote by $\Psi_i$ the set of all other (extrinsic and intrinsic) parameters of the $i$-th signal. Writing $x(t)$ as a sum over distinct signals 
is possible because we are dealing with  signals whose duration is short with respect to the observation time $T$ (i.e. CBCs, in contrast for isntance to periodic signals), and therefore there is a well-defined notion of whether the signal is inside or outside the observation window $ t\in [-T/2,T/2]$. 

In the approach that we have followed in the main text,  the sum over $i$ in \eq{xditsumi} runs only over the $\nev$ events collected in the observation time. 
Proceeding as in \cref{sect:arrivals} we now extract explicitly the dependence on $t_i$, as
\be
\tilde{x}_i(f;t_i,\Psi_i)=e^{2\pi i f t_i}\tilde{x}(f;t_i=0,\Psi_i)\, ,
\ee 
and we then write
\bees
&&\hspace*{-3mm}\langle \tilde{x}_i^*(f;t_i,\Psi_i)\tilde{x}_i(f';t_i,\Psi_i)\rangle_{ \{t_i,\Psi_i\} }
=\hspace*{-0.5mm}
\int_{-T/2}^{T/2}\frac{dt_i}{T}\,e^{-2\pi i (f-f') t_i}\, 
\langle \tilde{x}_i^*(f;t_i=0,\Psi_i)\tilde{x}_i(f';t_i=0,\Psi_i)\rangle_{ \{\Psi_i\} }\nn\\
&&=\frac{1}{T}\delta_T(f-f')\langle \tilde{x}_i^*(f;t_i=0,\Psi_i)\tilde{x}_i(f';t_i=0,\Psi_i)\rangle_{\{\Psi_i\}}
\, .
\ees
What is crucial is that, when computing the correlator,  we have  assumed that the $i$-th signal arrived during the observation window $ [-T/2,T/2]$, since we averaged its arrival time over such an interval, rather than from $t_i=-\infty$ to $t_i=+\infty$. 
Using also the fact that $\langle \tilde{x}_i^*(f;t_i,\Psi_i)\tilde{x}_j(f';t_j,\Psi_j)\rangle_{ \{t_k,\Psi_k\} }$ vanishes for $i\neq j$ because of the average over the $\Psi_k$ (in particular, over the polarization angles $\psi_k$), we then get
\be\label{deltaTxxSh}
\langle \tilde{x}^*(f)\tilde{x}(f')\rangle_{ \{t_k,\Psi_k\} }
=\delta_T(f-f') S_x(f,f')\, ,
\ee
where, more precisely,
$\tilde{x}(f)=\tilde{x}(f;\{ t_k,\Psi_k\})$, and 
\be\label{Shxxnev}
S_x(f,f') =\frac{1}{T}\sum_{i=1}^{\nev(T)}\, 
\langle \tilde{x}_i^*(f;t_i=0,\Psi_i)\tilde{x}_i(f';t_i=0,\Psi_i)\rangle_{ \{\Psi_k\} }\, .
\ee
Observe that $\nev(T)$  in \eq{Shxxnev}
is the number of events that arrive in the time interval $[-T/2,T/2]$, because for each of these events we have averaged its time of arrival over such interval, i.e. we have assumed that, for each $i$, $t_i\in [-T/2,T/2]$.
Of course, the number of events that arrive in the observation interval is itself a stochastic variable. In our approach, we can then use the specific value of $\nev$ obtained in a given realization (e.g. in a given observation run, or a in given simulation) or else 
we can perform at the end also an average over $\nev$ with a Poisson distribution, or (more simply), we just replace $\nev$ by its average value $\bnev$.
In the limit $T\ra\infty$, \eq{deltaTxxSh} becomes
\be\label{xstarxdelta}
\langle \tilde{x}^*(f)\tilde{x}(f')\rangle_{ \{t_k,\Psi_k\} }
=\delta(f-f') S_x(f)\, ,
\ee
where 
\bees
S_x(f) &=&\lim_{T\ra\infty} \frac{1}{T}\sum_{i=1}^{\nev(T)}\, 
\langle \tilde{x}_i^*(f;t_i=0,\Psi_i)\tilde{x}_i(f;t_i=0,\Psi_i)\rangle_{ \{\Psi_k\} }\nn\\
&=&\lim_{T\ra\infty} \frac{\nev(T)}{T}\, \frac{1}{\nev(T)}\sum_{i=1}^{\nev(T)}\, 
\langle \tilde{x}_i^*(f;t_i=0,\Psi_i)\tilde{x}_i(f;t_i=0,\Psi_i)\rangle_{ \{\Psi_k\} }\nn\\
&=& R_0\, \lim_{\nev\ra\infty}
\frac{1}{\nev}\sum_{i=1}^{\nev}\, 
\langle \tilde{x}_i^*(f;t_i=0,\Psi_i)\tilde{x}_i(f;t_i=0,\Psi_i)\rangle_{ \{\Psi_k\} }
\, ,\label{ShlimTinfty}
\ees
where $R_0=\lim_{T\ra\infty} \nev(T)/T$ is the event rate.

Now, let us compare this with  the approach  leading to \eq{deltaTyySh}. In this approach, $x(t)$ is considered as a stochastic variable whose spectral density is given by \eq{deltaTxxSh0}, which is independent of $T$, while the dependence on $T$ only enters through \eq{defyt}. Since the
quantity $S_x(f)$ that appears in  \eq{deltaTxxSh0}, or equivalently in \eq{xstarxdelta}, is obtained summing the contribution of all events from $t=-\infty$ to $t=+\infty$, it  can be written as 
in \eq{ShlimTinfty}, and depends on the rate $R_0$ and on a sort of  `typical' signal obtained averaging   
the correlator $\langle \tilde{x}_i^*(f;t_i=0,\Psi_i)\tilde{x}_i(f;t_i=0,\Psi_i)\rangle_{ \{\Psi_k\} }$ over an infinite number of events. Therefore, it is not related to any specific realization of events during the observation time $T$. The dependence on $T$ only enters at a later stage through the definition of $y(t)$, in a way that however can no longer make any reference to the specific sample of events observed in the given run. It therefore represent a sort of `average' realization. In this sense, the approach based on \eqst{deltaTxxSh0}{deltaTyySh} can be considered a formal mathematical regularization for finite $T$, which can be useful to provide another rigorous derivation of the spectral density in the limit $T\ra\infty$. However, its specific expression at finite $T$ does not have 
a relation with the specific  ensemble of events (their number, as well as their specific signals) 
arriving in this observation interval for a given realization of the stochastic process, but it rather represents a sort of `average' realization.

\bibliographystyle{utphys}
\bibliography{bibliography.bib}

\end{document}

%% file: mydefs.tex
\newcommand{\nn}{\nonumber}

\renewcommand\({\left(}
\renewcommand\){\right)}
\renewcommand\[{\left[}
\renewcommand\]{\right]}

\newcommand{\ra}{\rightarrow}

\def\lsim{\raise 0.4ex\hbox{$<$}\kern -0.8em\lower 0.62
ex\hbox{$\sim$}}

\def\gsim{\raise 0.4ex\hbox{$>$}\kern -0.8em\lower 0.62
ex\hbox{$\sim$}}

\def\lbar{{\hbox{$\lambda$}\kern -0.7em\raise 0.6ex
\hbox{$-$}}}

\newcommand\eq[1]{eq.~(\ref{#1})}
\newcommand\eqs[2]{eqs.~(\ref{#1}) and (\ref{#2})}

\newcommand\eqss[3]{eqs.~(\ref{#1}), (\ref{#2}) and (\ref{#3})}

\newcommand\eqst[2]{eqs.~(\ref{#1})--(\ref{#2})}

\newcommand\pa{\partial}
\newcommand\p{\partial}

\newcommand\ee{\end{equation}}
\newcommand\be{\begin{equation}}
\def\bea{\begin{array}}
\def\eea{\end{array}}\def\ea{\end{array}}
\newcommand\ees{\end{eqnarray}}
\newcommand\bees{\begin{eqnarray}}
\def\nn{\nonumber}

\def\v#1{\hbox{\boldmath$#1$}}

\def\vtheta{\v{\theta}}
\def\vphi{\v{\phi}}
\def\vO{\v{\Omega}}

\def\dslash{\hspace{-1mm}\not{\hbox{\kern-2pt $\partial$}}}
\def\Dslash{\not{\hbox{\kern-2pt $D$}}}
\def\pslash{\not{\hbox{\kern-2.1pt $p$}}}
\def\kslash{\not{\hbox{\kern-2.3pt $k$}}}
\def\qslash{\not{\hbox{\kern-2.3pt $q$}}}

\newcommand{\vx}{{\bf x}}

\newcommand{\bdot}{{\bf\cdot}} 

\def\p1{{\bf p}_1}
\def\p2{{\bf p}_2}
\def\k1{{\bf k}_1}
\def\k2{{\bf k}_2}

\newcommand{\hatn}{\hat{\bf n}}
\newcommand{\hatx}{\hat{\bf x}}
\newcommand{\haty}{\hat{\bf y}}
\newcommand{\hatz}{\hat{\bf z}}

\newcommand{\hatu}{\hat{\bf u}}
\newcommand{\hatv}{\hat{\bf v}}

\newcommand{\dddM}{\kern 0.2em \raise 1.9ex\hbox{$...$}\kern -1.0em \hbox{$M$}}
\newcommand{\dddQ}{\kern 0.2em \raise 1.9ex\hbox{$...$}\kern -1.0em \hbox{$Q$}}
\newcommand{\dddI}{\kern 0.2em \raise 1.9ex\hbox{$...$}\kern -1.0em\hbox{$I$}}
\newcommand{\dddJ}{\kern 0.2em \raise 1.9ex\hbox{$...$}\kern-1.0em
\hbox{$J$}}
\newcommand{\dddcalJ}{\kern 0.2em \raise 1.9ex\hbox{$...$}\kern-1.0em
\hbox{${\cal J}$}}

\newcommand{\dddO}{\kern 0.2em \raise 1.9ex\hbox{$...$}\kern -1.0em
\hbox{${\cal O}$}}
\def\dddz{\raise 1.5ex\hbox{$...$}\kern -0.8em \hbox{$z$}}
\def\dddd{\raise 1.8ex\hbox{$...$}\kern -0.8em \hbox{$d$}}
\def\dddbd{\raise 1.8ex\hbox{$...$}\kern -0.8em \hbox{${\bf d}$}}
\def\ddbd{\raise 1.8ex\hbox{$..$}\kern -0.8em \hbox{${\bf d}$}}
\def\dddx{\raise 1.6ex\hbox{$...$}\kern -0.8em \hbox{$x$}}

\newcommand{\hti}{\tilde{h}}

\newcommand{\nev}{N_{\rm ev}}
\newcommand{\bnev}{\bar{N}_{\rm ev}}

\newcommand{\omgwa}{\Omega_{\rm gw}^{\rm astro}}

\newcommand{\omgw}{\Omega_{\rm gw}}

%% file: spectral_density_arxiv_v3.bbl
\providecommand{\href}[2]{#2}\begingroup\raggedright\begin{thebibliography}{10}

\bibitem{Maggiore:1999vm}
M.~Maggiore, ``{Gravitational wave experiments and early universe cosmology},''
  \href{http://dx.doi.org/10.1016/S0370-1573(99)00102-7}{{\em Phys. Rept.}
  {\bfseries 331} (2000) 283--367},
\href{http://arxiv.org/abs/gr-qc/9909001}{{\ttfamily arXiv:gr-qc/9909001
  [gr-qc]}}.

\bibitem{Maggiore:2007ulw}
M.~Maggiore, {\em {Gravitational Waves. Vol. 1: Theory and Experiments}}.
\newblock Oxford University Press, 2007.

\bibitem{Regimbau:2011rp}
T.~Regimbau, ``{The astrophysical gravitational wave stochastic background},''
  \href{http://dx.doi.org/10.1088/1674-4527/11/4/001}{{\em Res. Astron.
  Astrophys.} {\bfseries 11} (2011) 369--390},
  \href{http://arxiv.org/abs/1101.2762}{{\ttfamily arXiv:1101.2762
  [astro-ph.CO]}}.

\bibitem{Romano:2016dpx}
J.~D. Romano and N.~J. Cornish, ``{Detection methods for stochastic
  gravitational-wave backgrounds: a unified treatment},''
  \href{http://dx.doi.org/10.1007/s41114-017-0004-1}{{\em Living Rev. Rel.}
  {\bfseries 20} no.~1, (2017) 2},
  \href{http://arxiv.org/abs/1608.06889}{{\ttfamily arXiv:1608.06889 [gr-qc]}}.

\bibitem{Maggiore:2018sht}
M.~Maggiore, {\em {Gravitational Waves. Vol. 2: Astrophysics and Cosmology}}.
\newblock Oxford University Press,
2018.
\newblock

\bibitem{Caprini:2018mtu}
C.~Caprini and D.~G. Figueroa, ``{Cosmological Backgrounds of Gravitational
  Waves},'' \href{http://dx.doi.org/10.1088/1361-6382/aac608}{{\em Class.
  Quant. Grav.} {\bfseries 35} (2018) 163001},
  \href{http://arxiv.org/abs/1801.04268}{{\ttfamily arXiv:1801.04268
  [astro-ph.CO]}}.

\bibitem{Christensen:2018iqi}
N.~Christensen, ``{Stochastic Gravitational Wave Backgrounds},''
  \href{http://dx.doi.org/10.1088/1361-6633/aae6b5}{{\em Rept. Prog. Phys.}
  {\bfseries 82} no.~1, (2019) 016903},
  \href{http://arxiv.org/abs/1811.08797}{{\ttfamily arXiv:1811.08797 [gr-qc]}}.

\bibitem{Lawrence:2023buo}
J.~Lawrence, K.~Turbang, A.~Matas, A.~I. Renzini, N.~van Remortel, and J.~D.
  Romano, ``{A stochastic search for intermittent gravitational-wave
  backgrounds},'' \href{http://dx.doi.org/10.1103/PhysRevD.107.103026}{{\em
  Phys. Rev. D} {\bfseries 107} no.~10, (2023) 103026},
  \href{http://arxiv.org/abs/2301.07675}{{\ttfamily arXiv:2301.07675 [gr-qc]}}.

\bibitem{NANOGrav:2023gor}
{\bfseries NANOGrav} Collaboration, G.~Agazie {\em et~al.}, ``{The NANOGrav 15
  yr Data Set: Evidence for a Gravitational-wave Background},''
  \href{http://dx.doi.org/10.3847/2041-8213/acdac6}{{\em Astrophys. J. Lett.}
  {\bfseries 951} no.~1, (2023) L8},
  \href{http://arxiv.org/abs/2306.16213}{{\ttfamily arXiv:2306.16213
  [astro-ph.HE]}}.

\bibitem{EPTA:2023fyk}
{\bfseries EPTA} Collaboration, J.~Antoniadis {\em et~al.}, ``{The second data
  release from the European Pulsar Timing Array III. Search for gravitational
  wave signals},'' \href{http://dx.doi.org/10.1051/0004-6361/202346844}{{\em
  Astron. Astrophys.} {\bfseries 678} (2023) A50},
  \href{http://arxiv.org/abs/2306.16214}{{\ttfamily arXiv:2306.16214
  [astro-ph.HE]}}.

\bibitem{Reardon:2023gzh}
D.~J. Reardon {\em et~al.}, ``{Search for an Isotropic Gravitational-wave
  Background with the Parkes Pulsar Timing Array},''
  \href{http://dx.doi.org/10.3847/2041-8213/acdd02}{{\em Astrophys. J. Lett.}
  {\bfseries 951} no.~1, (2023) L6},
  \href{http://arxiv.org/abs/2306.16215}{{\ttfamily arXiv:2306.16215
  [astro-ph.HE]}}.

\bibitem{Xu:2023wog}
H.~Xu {\em et~al.}, ``{Searching for the Nano-Hertz Stochastic Gravitational
  Wave Background with the Chinese Pulsar Timing Array Data Release I},''
  \href{http://dx.doi.org/10.1088/1674-4527/acdfa5}{{\em Res. Astron.
  Astrophys.} {\bfseries 23} no.~7, (2023) 075024},
  \href{http://arxiv.org/abs/2306.16216}{{\ttfamily arXiv:2306.16216
  [astro-ph.HE]}}.

\bibitem{LIGOScientific:2019vic}
{\bfseries LIGO Scientific, Virgo} Collaboration, B.~P. Abbott {\em et~al.},
  ``{Search for the isotropic stochastic background using data from Advanced
  LIGO\textquoteright{}s second observing run},''
  \href{http://dx.doi.org/10.1103/PhysRevD.100.061101}{{\em Phys. Rev. D}
  {\bfseries 100} no.~6, (2019) 061101},
  \href{http://arxiv.org/abs/1903.02886}{{\ttfamily arXiv:1903.02886 [gr-qc]}}.

\bibitem{Perigois:2020ymr}
C.~P\'erigois, C.~Belczynski, T.~Bulik, and T.~Regimbau, ``{StarTrack
  predictions of the stochastic gravitational-wave background from compact
  binary mergers},'' \href{http://dx.doi.org/10.1103/PhysRevD.103.043002}{{\em
  Phys. Rev. D} {\bfseries 103} no.~4, (2021) 043002},
  \href{http://arxiv.org/abs/2008.04890}{{\ttfamily arXiv:2008.04890
  [astro-ph.CO]}}.

\bibitem{Regimbau:2022mdu}
T.~Regimbau, ``{The Quest for the Astrophysical Gravitational-Wave Background
  with Terrestrial Detectors},''
  \href{http://dx.doi.org/10.3390/sym14020270}{{\em Symmetry} {\bfseries 14}
  no.~2, (2022) 270}.

\bibitem{Hild:2008ng}
S.~Hild, S.~Chelkowski, and A.~Freise, ``{Pushing towards the ET sensitivity
  using 'conventional' technology},''  (2008) ,
\href{http://arxiv.org/abs/0810.0604}{{\ttfamily arXiv:0810.0604 [gr-qc]}}.

\bibitem{Punturo:2010zz}
M.~Punturo {\em et~al.}, ``{The Einstein Telescope: A third-generation
  gravitational wave observatory},''
\href{http://dx.doi.org/10.1088/0264-9381/27/19/194002}{{\em Class. Quant.
  Grav.} {\bfseries 27} (2010) 194002}.

\bibitem{Hild:2010id}
S.~Hild {\em et~al.}, ``{Sensitivity Studies for Third-Generation Gravitational
  Wave Observatories},''
  \href{http://dx.doi.org/10.1088/0264-9381/28/9/094013}{{\em Class. Quant.
  Grav.} {\bfseries 28} (2011) 094013},
\href{http://arxiv.org/abs/1012.0908}{{\ttfamily arXiv:1012.0908 [gr-qc]}}.

\bibitem{Maggiore:2019uih}
M.~Maggiore {\em et~al.}, ``{Science Case for the Einstein Telescope},''
  \href{http://dx.doi.org/10.1088/1475-7516/2020/03/050}{{\em JCAP} {\bfseries
  2020} no.~03, (2020) 050}, \href{http://arxiv.org/abs/1912.02622}{{\ttfamily
  arXiv:1912.02622 [astro-ph.CO]}}.

\bibitem{Reitze:2019iox}
D.~Reitze {\em et~al.}, ``{Cosmic Explorer: The U.S. Contribution to
  Gravitational-Wave Astronomy beyond LIGO},'' {\em Bull. Am. Astron. Soc.}
  {\bfseries 51} no.~7, (2019) 035,
  \href{http://arxiv.org/abs/1907.04833}{{\ttfamily arXiv:1907.04833
  [astro-ph.IM]}}.

\bibitem{Evans:2021gyd}
M.~Evans {\em et~al.}, ``{A Horizon Study for Cosmic Explorer: Science,
  Observatories, and Community},''  (9, 2021) ,
  \href{http://arxiv.org/abs/2109.09882}{{\ttfamily arXiv:2109.09882
  [astro-ph.IM]}}.

\bibitem{Evans:2023euw}
M.~Evans {\em et~al.}, ``{Cosmic Explorer: A Submission to the NSF MPSAC ngGW
  Subcommittee},''  (6, 2023) ,
  \href{http://arxiv.org/abs/2306.13745}{{\ttfamily arXiv:2306.13745
  [astro-ph.IM]}}.

\bibitem{Gupta:2023lga}
I.~Gupta {\em et~al.}, ``{Characterizing Gravitational Wave Detector Networks:
  From A$^\sharp$ to Cosmic Explorer},''  (7, 2023) ,
  \href{http://arxiv.org/abs/2307.10421}{{\ttfamily arXiv:2307.10421 [gr-qc]}}.

\bibitem{Branchesi:2023mws}
M.~Branchesi {\em et~al.}, ``{Science with the Einstein Telescope: a comparison
  of different designs},''
  \href{http://dx.doi.org/10.1088/1475-7516/2023/07/068}{{\em JCAP} {\bfseries
  2023} no.~07, (2023) 068}, \href{http://arxiv.org/abs/2303.15923}{{\ttfamily
  arXiv:2303.15923 [gr-qc]}}.

\bibitem{Iacovelli:2024mjy}
F.~Iacovelli, E.~Belgacem, M.~Maggiore, M.~Mancarella, and N.~Muttoni,
  ``{Combining underground and on-surface third-generation gravitational-wave
  interferometers},''
  \href{http://dx.doi.org/10.1088/1475-7516/2024/10/085}{{\em JCAP} {\bfseries
  2024} no.~10, (2024) 085}, \href{http://arxiv.org/abs/2408.14946}{{\ttfamily
  arXiv:2408.14946 [gr-qc]}}.

\bibitem{Cutler:2005qq}
C.~Cutler and J.~Harms, ``{BBO and the neutron-star-binary subtraction
  problem},'' \href{http://dx.doi.org/10.1103/PhysRevD.73.042001}{{\em Phys.
  Rev. D} {\bfseries 73} (2006) 042001},
  \href{http://arxiv.org/abs/gr-qc/0511092}{{\ttfamily arXiv:gr-qc/0511092}}.

\bibitem{Harms:2008xv}
J.~Harms, C.~Mahrdt, M.~Otto, and M.~Priess, ``{Subtraction-noise projection in
  gravitational-wave detector networks},''
  \href{http://dx.doi.org/10.1103/PhysRevD.77.123010}{{\em Phys. Rev. D}
  {\bfseries 77} (2008) 123010},
  \href{http://arxiv.org/abs/0803.0226}{{\ttfamily arXiv:0803.0226 [gr-qc]}}.

\bibitem{Regimbau:2016ike}
T.~Regimbau, M.~Evans, N.~Christensen, E.~Katsavounidis, B.~Sathyaprakash, and
  S.~Vitale, ``{Digging deeper: Observing primordial gravitational waves below
  the binary black hole produced stochastic background},''
  \href{http://dx.doi.org/10.1103/PhysRevLett.118.151105}{{\em Phys. Rev.
  Lett.} {\bfseries 118} no.~15, (2017) 151105},
  \href{http://arxiv.org/abs/1611.08943}{{\ttfamily arXiv:1611.08943
  [astro-ph.CO]}}.

\bibitem{Pan:2019uyn}
Z.~Pan and H.~Yang, ``{Probing Primordial Stochastic Gravitational Wave
  Background with Multi-band Astrophysical Foreground Cleaning},''
  \href{http://dx.doi.org/10.1088/1361-6382/abb074}{{\em Class. Quant. Grav.}
  {\bfseries 37} no.~19, (2020) 195020},
  \href{http://arxiv.org/abs/1910.09637}{{\ttfamily arXiv:1910.09637
  [astro-ph.CO]}}.

\bibitem{Sachdev:2020bkk}
S.~Sachdev, T.~Regimbau, and B.~S. Sathyaprakash, ``{Subtracting compact binary
  foreground sources to reveal primordial gravitational-wave backgrounds},''
  \href{http://dx.doi.org/10.1103/PhysRevD.102.024051}{{\em Phys. Rev. D}
  {\bfseries 102} no.~2, (2020) 024051},
  \href{http://arxiv.org/abs/2002.05365}{{\ttfamily arXiv:2002.05365 [gr-qc]}}.

\bibitem{Sharma:2020btq}
A.~Sharma and J.~Harms, ``{Searching for cosmological gravitational-wave
  backgrounds with third-generation detectors in the presence of an
  astrophysical foreground},''
  \href{http://dx.doi.org/10.1103/PhysRevD.102.063009}{{\em Phys. Rev. D}
  {\bfseries 102} no.~6, (2020) 063009},
  \href{http://arxiv.org/abs/2006.16116}{{\ttfamily arXiv:2006.16116 [gr-qc]}}.

\bibitem{Lewicki:2021kmu}
M.~Lewicki and V.~Vaskonen, ``{Impact of LIGO-Virgo black hole binaries on
  gravitational wave background searches},''
  \href{http://dx.doi.org/10.1140/epjc/s10052-023-11323-2}{{\em Eur. Phys. J.
  C} {\bfseries 83} no.~2, (2023) 168},
  \href{http://arxiv.org/abs/2111.05847}{{\ttfamily arXiv:2111.05847
  [astro-ph.CO]}}.

\bibitem{Perigois:2021ovr}
C.~P\'erigois, F.~Santoliquido, Y.~Bouffanais, U.~N. Di~Carlo, N.~Giacobbo,
  S.~Rastello, M.~Mapelli, and T.~Regimbau, ``{Gravitational background from
  dynamical binaries and detectability with 2G detectors},''
  \href{http://dx.doi.org/10.1103/PhysRevD.105.103032}{{\em Phys. Rev. D}
  {\bfseries 105} no.~10, (2022) 103032},
  \href{http://arxiv.org/abs/2112.01119}{{\ttfamily arXiv:2112.01119
  [astro-ph.CO]}}.

\bibitem{Zhou:2022otw}
B.~Zhou, L.~Reali, E.~Berti, M.~\c{C}al\i{}\c{s}kan, C.~Creque-Sarbinowski,
  M.~Kamionkowski, and B.~S. Sathyaprakash, ``{Compact Binary Foreground
  Subtraction in Next-Generation Ground-Based Observatories},''  (9, 2022) ,
  \href{http://arxiv.org/abs/2209.01221}{{\ttfamily arXiv:2209.01221 [gr-qc]}}.

\bibitem{Zhou:2022nmt}
B.~Zhou, L.~Reali, E.~Berti, M.~\c{C}al\i{}\c{s}kan, C.~Creque-Sarbinowski,
  M.~Kamionkowski, and B.~S. Sathyaprakash, ``{Subtracting compact binary
  foregrounds to search for subdominant gravitational-wave backgrounds in
  next-generation ground-based observatories},''
  \href{http://dx.doi.org/10.1103/PhysRevD.108.064040}{{\em Phys. Rev. D}
  {\bfseries 108} no.~6, (2023) 064040},
  \href{http://arxiv.org/abs/2209.01310}{{\ttfamily arXiv:2209.01310 [gr-qc]}}.

\bibitem{Zhong:2022ylh}
H.~Zhong, R.~Ormiston, and V.~Mandic, ``{Detecting cosmological gravitational
  wave background after removal of compact binary coalescences in future
  gravitational wave detectors},''
  \href{http://dx.doi.org/10.1103/PhysRevD.107.064048}{{\em Phys. Rev. D}
  {\bfseries 107} no.~6, (2023) 064048},
  \href{http://arxiv.org/abs/2209.11877}{{\ttfamily arXiv:2209.11877 [gr-qc]}}.

\bibitem{Pan:2023naq}
Z.~Pan and H.~Yang, ``{Improving the detection sensitivity to primordial
  stochastic gravitational waves with reduced astrophysical foregrounds},''
  \href{http://dx.doi.org/10.1103/PhysRevD.107.123036}{{\em Phys. Rev. D}
  {\bfseries 107} no.~12, (2023) 123036},
  \href{http://arxiv.org/abs/2301.04529}{{\ttfamily arXiv:2301.04529 [gr-qc]}}.

\bibitem{Zhong:2024dss}
H.~Zhong, B.~Zhou, L.~Reali, E.~Berti, and V.~Mandic, ``{Searching for
  cosmological stochastic backgrounds by notching out resolvable compact binary
  foregrounds with next-generation gravitational-wave detectors},''
  \href{http://dx.doi.org/10.1103/PhysRevD.110.064047}{{\em Phys. Rev. D}
  {\bfseries 110} no.~6, (2024) 064047},
  \href{http://arxiv.org/abs/2406.10757}{{\ttfamily arXiv:2406.10757 [gr-qc]}}.

\bibitem{Li:2024iua}
M.~Li, J.~Yu, and Z.~Pan, ``{Improving the detection sensitivity to primordial
  stochastic gravitational waves with reduced astrophysical foregrounds -- II:
  subthreshold binary neutron stars},''  (3, 2024) ,
  \href{http://arxiv.org/abs/2403.01846}{{\ttfamily arXiv:2403.01846 [gr-qc]}}.

\bibitem{Belgacem:2024ntv}
E.~Belgacem, F.~Iacovelli, M.~Maggiore, M.~Mancarella, and N.~Muttoni,
  ``{Confusion noise from astrophysical backgrounds at third-generation
  gravitational-wave detector networks},''
  \href{http://arxiv.org/abs/2411.04029}{{\ttfamily arXiv:2411.04029 [gr-qc]}}.

\bibitem{Jenkins:2019uzp}
A.~C. Jenkins and M.~Sakellariadou, ``{Shot noise in the astrophysical
  gravitational-wave background},''
  \href{http://dx.doi.org/10.1103/PhysRevD.100.063508}{{\em Phys. Rev. D}
  {\bfseries 100} no.~6, (2019) 063508},
  \href{http://arxiv.org/abs/1902.07719}{{\ttfamily arXiv:1902.07719
  [astro-ph.CO]}}.

\bibitem{ValbusaDallArmi:2023ydl}
L.~Valbusa~Dall'Armi, A.~Nishizawa, A.~Ricciardone, and S.~Matarrese,
  ``{Circular Polarization of the Astrophysical Gravitational Wave
  Background},'' \href{http://dx.doi.org/10.1103/PhysRevLett.131.041401}{{\em
  Phys. Rev. Lett.} {\bfseries 131} no.~4, (2023) 041401},
  \href{http://arxiv.org/abs/2301.08205}{{\ttfamily arXiv:2301.08205
  [astro-ph.CO]}}.

\bibitem{KAGRA:2021duu}
{\bfseries KAGRA, VIRGO, LIGO Scientific} Collaboration, R.~Abbott {\em
  et~al.}, ``{Population of Merging Compact Binaries Inferred Using
  Gravitational Waves through GWTC-3},''
  \href{http://dx.doi.org/10.1103/PhysRevX.13.011048}{{\em Phys. Rev. X}
  {\bfseries 13} no.~1, (2023) 011048},
  \href{http://arxiv.org/abs/2111.03634}{{\ttfamily arXiv:2111.03634
  [astro-ph.HE]}}.

\bibitem{Seto:2008sr}
N.~Seto and A.~Taruya, ``{Polarization analysis of gravitational-wave
  backgrounds from the correlation signals of ground-based interferometers:
  Measuring a circular-polarization mode},''
  \href{http://dx.doi.org/10.1103/PhysRevD.77.103001}{{\em Phys. Rev. D}
  {\bfseries 77} (2008) 103001},
  \href{http://arxiv.org/abs/0801.4185}{{\ttfamily arXiv:0801.4185
  [astro-ph]}}.

\bibitem{Renzini:2020rjw}
A.~Renzini, \href{http://dx.doi.org/10.25560/83803}{{\em {Mapping the
  Gravitational-Wave Background}}}.
\newblock PhD thesis, Imperial Coll., London, 2020.

\bibitem{Renzini:2022alw}
A.~I. Renzini, B.~Goncharov, A.~C. Jenkins, and P.~M. Meyers, ``{Stochastic
  Gravitational-Wave Backgrounds: Current Detection Efforts and Future
  Prospects},'' \href{http://dx.doi.org/10.3390/galaxies10010034}{{\em
  Galaxies} {\bfseries 10} no.~1, (2022) 34},
  \href{http://arxiv.org/abs/2202.00178}{{\ttfamily arXiv:2202.00178 [gr-qc]}}.

\bibitem{Contaldi:2016koz}
C.~R. Contaldi, ``{Anisotropies of Gravitational Wave Backgrounds: A Line Of
  Sight Approach},''
  \href{http://dx.doi.org/10.1016/j.physletb.2017.05.020}{{\em Phys. Lett. B}
  {\bfseries 771} (2017) 9--12},
  \href{http://arxiv.org/abs/1609.08168}{{\ttfamily arXiv:1609.08168
  [astro-ph.CO]}}.

\bibitem{Cusin:2017fwz}
G.~Cusin, C.~Pitrou, and J.-P. Uzan, ``{Anisotropy of the astrophysical
  gravitational wave background: Analytic expression of the angular power
  spectrum and correlation with cosmological observations},''
  \href{http://dx.doi.org/10.1103/PhysRevD.96.103019}{{\em Phys. Rev.}
  {\bfseries D96} no.~10, (2017) 103019},
\href{http://arxiv.org/abs/1704.06184}{{\ttfamily arXiv:1704.06184
  [astro-ph.CO]}}.

\bibitem{Cusin:2017mjm}
G.~Cusin, C.~Pitrou, and J.-P. Uzan, ``{The signal of the gravitational wave
  background and the angular correlation of its energy density},''
  \href{http://dx.doi.org/10.1103/PhysRevD.97.123527}{{\em Phys. Rev.}
  {\bfseries D97} no.~12, (2018) 123527},
\href{http://arxiv.org/abs/1711.11345}{{\ttfamily arXiv:1711.11345
  [astro-ph.CO]}}.

\bibitem{Jenkins:2018lvb}
A.~C. Jenkins and M.~Sakellariadou, ``{Anisotropies in the stochastic
  gravitational-wave background: Formalism and the cosmic string case},''
  \href{http://dx.doi.org/10.1103/PhysRevD.98.063509}{{\em Phys. Rev.}
  {\bfseries D98} no.~6, (2018) 063509},
\href{http://arxiv.org/abs/1802.06046}{{\ttfamily arXiv:1802.06046
  [astro-ph.CO]}}.

\bibitem{Cusin:2018rsq}
G.~Cusin, I.~Dvorkin, C.~Pitrou, and J.-P. Uzan, ``{First predictions of the
  angular power spectrum of the astrophysical gravitational wave background},''
  \href{http://dx.doi.org/10.1103/PhysRevLett.120.231101}{{\em Phys. Rev.
  Lett.} {\bfseries 120} (2018) 231101},
\href{http://arxiv.org/abs/1803.03236}{{\ttfamily arXiv:1803.03236
  [astro-ph.CO]}}.

\bibitem{Jenkins:2018uac}
A.~C. Jenkins, M.~Sakellariadou, T.~Regimbau, and E.~Slezak, ``{Anisotropies in
  the astrophysical gravitational-wave background: Predictions for the
  detection of compact binaries by LIGO and Virgo},''
  \href{http://dx.doi.org/10.1103/PhysRevD.98.063501}{{\em Phys. Rev.}
  {\bfseries D98} no.~6, (2018) 063501},
\href{http://arxiv.org/abs/1806.01718}{{\ttfamily arXiv:1806.01718
  [astro-ph.CO]}}.

\bibitem{Jenkins:2018kxc}
A.~C. Jenkins, R.~O'Shaughnessy, M.~Sakellariadou, and D.~Wysocki,
  ``{Anisotropies in the astrophysical gravitational-wave background: The
  impact of black hole distributions},''
  \href{http://dx.doi.org/10.1103/PhysRevLett.122.111101}{{\em Phys. Rev.
  Lett.} {\bfseries 122} no.~11, (2019) 111101},
\href{http://arxiv.org/abs/1810.13435}{{\ttfamily arXiv:1810.13435
  [astro-ph.CO]}}.

\bibitem{Cusin:2019jhg}
G.~Cusin, I.~Dvorkin, C.~Pitrou, and J.-P. Uzan, ``{Stochastic gravitational
  wave background anisotropies in the mHz band: astrophysical dependencies},''
  \href{http://dx.doi.org/10.1093/mnrasl/slz182}{{\em Mon. Not. Roy. Astron.
  Soc.} {\bfseries 493} no.~1, (2020) L1--L5},
  \href{http://arxiv.org/abs/1904.07757}{{\ttfamily arXiv:1904.07757
  [astro-ph.CO]}}.

\bibitem{Cusin:2019jpv}
G.~Cusin, I.~Dvorkin, C.~Pitrou, and J.-P. Uzan, ``{Properties of the
  stochastic astrophysical gravitational wave background: astrophysical sources
  dependencies},'' \href{http://dx.doi.org/10.1103/PhysRevD.100.063004}{{\em
  Phys. Rev. D} {\bfseries 100} no.~6, (2019) 063004},
  \href{http://arxiv.org/abs/1904.07797}{{\ttfamily arXiv:1904.07797
  [astro-ph.CO]}}.

\bibitem{Jenkins:2019nks}
A.~C. Jenkins, J.~D. Romano, and M.~Sakellariadou, ``{Estimating the angular
  power spectrum of the gravitational-wave background in the presence of shot
  noise},'' \href{http://dx.doi.org/10.1103/PhysRevD.100.083501}{{\em Phys.
  Rev. D} {\bfseries 100} no.~8, (2019) 083501},
  \href{http://arxiv.org/abs/1907.06642}{{\ttfamily arXiv:1907.06642
  [astro-ph.CO]}}.

\bibitem{Bertacca:2019fnt}
D.~Bertacca, A.~Ricciardone, N.~Bellomo, A.~C. Jenkins, S.~Matarrese,
  A.~Raccanelli, T.~Regimbau, and M.~Sakellariadou, ``{Projection effects on
  the observed angular spectrum of the astrophysical stochastic gravitational
  wave background},'' \href{http://dx.doi.org/10.1103/PhysRevD.101.103513}{{\em
  Phys. Rev. D} {\bfseries 101} no.~10, (2020) 103513},
  \href{http://arxiv.org/abs/1909.11627}{{\ttfamily arXiv:1909.11627
  [astro-ph.CO]}}.

\bibitem{Pitrou:2019rjz}
C.~Pitrou, G.~Cusin, and J.-P. Uzan, ``{Unified view of anisotropies in the
  astrophysical gravitational-wave background},''
  \href{http://dx.doi.org/10.1103/PhysRevD.101.081301}{{\em Phys. Rev. D}
  {\bfseries 101} no.~8, (2020) 081301},
  \href{http://arxiv.org/abs/1910.04645}{{\ttfamily arXiv:1910.04645
  [astro-ph.CO]}}.

\bibitem{Bellomo:2021mer}
N.~Bellomo, D.~Bertacca, A.~C. Jenkins, S.~Matarrese, A.~Raccanelli,
  T.~Regimbau, A.~Ricciardone, and M.~Sakellariadou, ``{CLASS\_GWB: robust
  modeling of the astrophysical gravitational wave background anisotropies},''
  \href{http://dx.doi.org/10.1088/1475-7516/2022/06/030}{{\em JCAP} {\bfseries
  06} no.~06, (2022) 030}, \href{http://arxiv.org/abs/2110.15059}{{\ttfamily
  arXiv:2110.15059 [gr-qc]}}.

\bibitem{Capurri:2021zli}
G.~Capurri, A.~Lapi, C.~Baccigalupi, L.~Boco, G.~Scelfo, and T.~Ronconi,
  ``{Intensity and anisotropies of the stochastic gravitational wave background
  from merging compact binaries in galaxies},''
  \href{http://dx.doi.org/10.1088/1475-7516/2021/11/032}{{\em JCAP} {\bfseries
  2021} no.~11, (2021) 032}, \href{http://arxiv.org/abs/2103.12037}{{\ttfamily
  arXiv:2103.12037 [gr-qc]}}.

\bibitem{Allen:2022dzg}
B.~Allen, ``{Variance of the Hellings-Downs correlation},''
  \href{http://dx.doi.org/10.1103/PhysRevD.107.043018}{{\em Phys. Rev. D}
  {\bfseries 107} no.~4, (2023) 043018},
  \href{http://arxiv.org/abs/2205.05637}{{\ttfamily arXiv:2205.05637 [gr-qc]}}.

\bibitem{Alexander:2004us}
S.~H.-S. Alexander, M.~E. Peskin, and M.~M. Sheikh-Jabbari, ``{Leptogenesis
  from gravity waves in models of inflation},''
  \href{http://dx.doi.org/10.1103/PhysRevLett.96.081301}{{\em Phys. Rev. Lett.}
  {\bfseries 96} (2006) 081301},
  \href{http://arxiv.org/abs/hep-th/0403069}{{\ttfamily arXiv:hep-th/0403069}}.

\bibitem{Crowder:2012ik}
S.~G. Crowder, R.~Namba, V.~Mandic, S.~Mukohyama, and M.~Peloso, ``{Measurement
  of Parity Violation in the Early Universe using Gravitational-wave
  Detectors},'' \href{http://dx.doi.org/10.1016/j.physletb.2013.08.077}{{\em
  Phys. Lett. B} {\bfseries 726} (2013) 66--71},
  \href{http://arxiv.org/abs/1212.4165}{{\ttfamily arXiv:1212.4165
  [astro-ph.CO]}}.

\bibitem{Yagi:2017zhb}
K.~Yagi and H.~Yang, ``{Probing Gravitational Parity Violation with
  Gravitational Waves from Stellar-mass Black Hole Binaries},''
  \href{http://dx.doi.org/10.1103/PhysRevD.97.104018}{{\em Phys. Rev. D}
  {\bfseries 97} no.~10, (2018) 104018},
  \href{http://arxiv.org/abs/1712.00682}{{\ttfamily arXiv:1712.00682 [gr-qc]}}.

\bibitem{Domcke:2019zls}
V.~Domcke, J.~Garcia-Bellido, M.~Peloso, M.~Pieroni, A.~Ricciardone, L.~Sorbo,
  and G.~Tasinato, ``{Measuring the net circular polarization of the stochastic
  gravitational wave background with interferometers},''
  \href{http://dx.doi.org/10.1088/1475-7516/2020/05/028}{{\em JCAP} {\bfseries
  2020} no.~05, (2020) 028}, \href{http://arxiv.org/abs/1910.08052}{{\ttfamily
  arXiv:1910.08052 [astro-ph.CO]}}.

\bibitem{Martinovic:2021hzy}
K.~Martinovic, C.~Badger, M.~Sakellariadou, and V.~Mandic, ``{Searching for
  parity violation with the LIGO-Virgo-KAGRA network},''
  \href{http://dx.doi.org/10.1103/PhysRevD.104.L081101}{{\em Phys. Rev. D}
  {\bfseries 104} no.~8, (2021) L081101},
  \href{http://arxiv.org/abs/2103.06718}{{\ttfamily arXiv:2103.06718 [gr-qc]}}.

\bibitem{Callister:2023tws}
T.~Callister, L.~Jenks, D.~Holz, and N.~Yunes, ``{A New Probe of Gravitational
  Parity Violation Through (Non-)Observation of the Stochastic
  Gravitational-Wave Background},''  (12, 2023) ,
  \href{http://arxiv.org/abs/2312.12532}{{\ttfamily arXiv:2312.12532 [gr-qc]}}.

\bibitem{Cruz:2024esk}
N.~M.~J. Cruz, A.~Malhotra, G.~Tasinato, and I.~Zavala, ``{Measuring the
  circular polarization of gravitational waves with pulsar timing arrays},''
  (6, 2024) , \href{http://arxiv.org/abs/2406.04957}{{\ttfamily
  arXiv:2406.04957 [astro-ph.CO]}}.

\bibitem{Blanchet:1996pi}
L.~Blanchet, B.~R. Iyer, C.~M. Will, and A.~G. Wiseman, ``{Gravitational wave
  forms from inspiralling compact binaries to second postNewtonian order},''
  \href{http://dx.doi.org/10.1088/0264-9381/13/4/002}{{\em Class. Quant. Grav.}
  {\bfseries 13} (1996) 575--584},
  \href{http://arxiv.org/abs/gr-qc/9602024}{{\ttfamily arXiv:gr-qc/9602024}}.

\bibitem{Arun:2004ff}
K.~G. Arun, L.~Blanchet, B.~R. Iyer, and M.~S.~S. Qusailah, ``{The 2.5PN
  gravitational wave polarisations from inspiralling compact binaries in
  circular orbits},'' \href{http://dx.doi.org/10.1088/0264-9381/21/15/010}{{\em
  Class. Quant. Grav.} {\bfseries 21} (2004) 3771--3802},
  \href{http://arxiv.org/abs/gr-qc/0404085}{{\ttfamily arXiv:gr-qc/0404085}}.
  [Erratum: Class.Quant.Grav. 22, 3115 (2005)].

\bibitem{Regimbau:2012ir}
T.~Regimbau {\em et~al.}, ``{A Mock Data Challenge for the Einstein
  Gravitational-Wave Telescope},''
  \href{http://dx.doi.org/10.1103/PhysRevD.86.122001}{{\em Phys. Rev.}
  {\bfseries D86} (2012) 122001},
\href{http://arxiv.org/abs/1201.3563}{{\ttfamily arXiv:1201.3563 [gr-qc]}}.

\bibitem{Meacher:2014aca}
D.~Meacher, E.~Thrane, and T.~Regimbau, ``{Statistical properties of
  astrophysical gravitational-wave backgrounds},''
  \href{http://dx.doi.org/10.1103/PhysRevD.89.084063}{{\em Phys. Rev. D}
  {\bfseries 89} no.~8, (2014) 084063},
  \href{http://arxiv.org/abs/1402.6231}{{\ttfamily arXiv:1402.6231
  [astro-ph.CO]}}.

\bibitem{Meacher:2015iua}
D.~Meacher, M.~Coughlin, S.~Morris, T.~Regimbau, N.~Christensen, S.~Kandhasamy,
  V.~Mandic, J.~D. Romano, and E.~Thrane, ``{Mock data and science challenge
  for detecting an astrophysical stochastic gravitational-wave background with
  Advanced LIGO and Advanced Virgo},''
  \href{http://dx.doi.org/10.1103/PhysRevD.92.063002}{{\em Phys. Rev. D}
  {\bfseries 92} no.~6, (2015) 063002},
  \href{http://arxiv.org/abs/1506.06744}{{\ttfamily arXiv:1506.06744
  [astro-ph.HE]}}.

\bibitem{Phinney:2001di}
E.~S. Phinney, ``{A Practical theorem on gravitational wave backgrounds},''
  (7, 2001) , \href{http://arxiv.org/abs/astro-ph/0108028}{{\ttfamily
  arXiv:astro-ph/0108028}}.

\bibitem{Renzini:2024pxt}
A.~I. Renzini and J.~Golomb, ``{Projections of the uncertainty on the compact
  binary population background using popstock},''
  \href{http://dx.doi.org/10.1051/0004-6361/202451374}{{\em Astron. Astrophys.}
  {\bfseries 691} (2024) A238},
  \href{http://arxiv.org/abs/2407.03742}{{\ttfamily arXiv:2407.03742
  [astro-ph.CO]}}.

\bibitem{Kouvatsos:2023bgd}
N.~Kouvatsos, A.~C. Jenkins, A.~I. Renzini, J.~D. Romano, and M.~Sakellariadou,
  ``{Unbiased estimation of gravitational-wave anisotropies from noisy data},''
  \href{http://dx.doi.org/10.1103/PhysRevD.109.103535}{{\em Phys. Rev. D}
  {\bfseries 109} no.~10, (2024) 103535},
  \href{http://arxiv.org/abs/2312.09110}{{\ttfamily arXiv:2312.09110
  [astro-ph.CO]}}.

\bibitem{Conneely:2018wis}
C.~Conneely, A.~H. Jaffe, and C.~M.~F. Mingarelli, ``{On the Amplitude and
  Stokes Parameters of a Stochastic Gravitational-Wave Background},''
  \href{http://dx.doi.org/10.1093/mnras/stz1022}{{\em Mon. Not. Roy. Astron.
  Soc.} {\bfseries 487} no.~1, (2019) 562--579},
  \href{http://arxiv.org/abs/1808.05920}{{\ttfamily arXiv:1808.05920
  [astro-ph.CO]}}.

\bibitem{Akmal:1998cf}
A.~Akmal, V.~R. Pandharipande, and D.~G. Ravenhall, ``{The Equation of state of
  nucleon matter and neutron star structure},''
  \href{http://dx.doi.org/10.1103/PhysRevC.58.1804}{{\em Phys. Rev. C}
  {\bfseries 58} (1998) 1804--1828},
  \href{http://arxiv.org/abs/nucl-th/9804027}{{\ttfamily
  arXiv:nucl-th/9804027}}.

\end{thebibliography}\endgroup
